\documentclass[sigconf, nonacm]{acmart}

\AtBeginDocument{%
  \providecommand\BibTeX{{%
    \normalfont B\kern-0.5em{\scshape i\kern-0.25em b}\kern-0.8em\TeX}}}

\newcommand{\system}{DEFORM}

\usepackage{tabularx}
\usepackage{subcaption}
\usepackage[longend, ruled]{algorithm2e}
\usepackage{dsfont}
\usepackage{makecell}
\usepackage{cleveref}
\usepackage{siunitx}
\ifdefined\unit\else
  \ifdefined\NewCommandCopy
    \NewCommandCopy\unit\si
  \else
    \NewDocumentCommand\unit{O{}m}{\si[#1]{#2}}
  \fi
\fi
\usepackage{multirow}
\usepackage{amsmath}
\usepackage{pdfrender}
\usepackage{tablefootnote}
\usepackage{natbib}
\usepackage{enumitem}
\usepackage{balance}
\usepackage{changepage}

\begin{document}

\title{\system: A Practical, Universal Deep Beamforming System}

\author{Hai N. Nguyen}
\email{nguyen.hai@northeastern.edu}
\affiliation{%
  \institution{College of Computer Sciences, Northeastern University}
  \city{Boston}
  \state{Massachusetts}
  \country{USA}
}

\author{Guevara Noubir}
\email{g.noubir@northeastern.edu}
\affiliation{%
  \institution{College of Computer Sciences, Northeastern University}
  \city{Boston}
  \state{Massachusetts}
  \country{USA}
}

\renewcommand{\shortauthors}{Hai et al.}
\newcommand{\bfpara}[1]{\vskip 1ex \noindent \textbf{#1.}}
\newcommand{\ignore}[1]{{}}

\newcommand{\code}[1]{\texttt{#1}}
\definecolor{codegray}{gray}{0.9}




\begin{abstract}
We introduce, design, and evaluate a set of universal receiver beamforming techniques. Our approach and system \system{}, a Deep Learning (DL)-based RX beamforming achieves significant gain for multi-antenna RF receivers while being agnostic to the transmitted signal features (e.g., modulation or bandwidth). It is well known that combining coherent RF signals from multiple antennas results in a beamforming gain proportional to the number of receiving elements. However in practice, this approach heavily relies on explicit channel estimation techniques, which are link specific and require significant communication overhead to be transmitted to the receiver. \system{} addresses this challenge by leveraging Convolutional Neural Network to  estimate the channel characteristics in particular the relative phase to antenna elements. It is specifically designed to address the unique features of wireless signals complex samples, such as the ambiguous $2\pi$ phase discontinuity and the high sensitivity of the link Bit Error Rate.  The channel prediction is subsequently used in the Maximum Ratio Combining algorithm to achieve an optimal combination of the received signals. While being trained on a fixed, basic RF settings, we show that \system{}’s DL model is universal, achieving up to 3 dB of SNR gain for a two-antenna receiver in extensive experiments demonstrating various settings of modulations, bandwidths, and channels. The universality of \system{} is demonstrated through joint beamforming-relaying of LoRa (Chirp Spread Spectrum modulation) and ZigBee signals, achieving significant improvements to Packet Loss/Delivery Rates relatively to conventional Amplify-and-Forward (LoRa PLR reduced by 23 times and ZigBee PDR increased by 8 times).
\ignore{We also demonstrate \system{} as a universal beamforming-relay of LoRa communications and achieve substantial improvement compared to the conventional Amplify-and-Forward method.}
\end{abstract}

\maketitle

\section{introduction}

The success of wireless communications was accompanied by a dramatic 
crowding of the RF spectrum. Beamforming is a widely used spatial filtering technique to steer RF emissions towards/from other devices. It enables array and diversity gains of multiple-input-single-output (MISO) systems~\cite{goldsmith_2005, van2004optimum}, and canceling interference from unwanted sources. Beamforming was extensively investigated  in a variety of applications such as in radars, sonars, acoustics, astronomy, and even medical devices design~\cite{liu_beamforming_2010}. Beamforming and more advanced MIMO techniques are widely used in systems targeting high throughput and spectral efficiency such as cellular systems since the third generation 3GPP 3G, and IEEE 802.11n. 

Extensive research was conducted on beamforming~\cite{cardoso_93, zhang_2017, vijay_2010, masouros_2015, wu_2019}. However, beamforming in today's systems  requires explicit mechanisms such as sounding and feedback in  802.11,  Demodulation Reference Signal (DMRS) in 5G, training sequences, etc. This approach introduces  critical limitations: Firstly, it results in significant overhead to transmit reference signals and estimate the channel. Secondly, accurate channel estimation typically exhibits long delay that is undesirable in fast-changing channels. Thirdly, it requires compatibility between the transmitter and receiver (to agree on when, what, and how reference signals are transmitted).

Therefore, a \textit{universal beamforming} scheme is highly desirable as it not only reduces the overhead but can also be used as a technology-agnostic component on the RF front end, and enable new applications such as relaying arbitrary signals. A universal beamformer can support the increasing heterogeneity of wireless technologies operating over the same spectrum. An example of application is a drone equipped with a technology-agnostic relay enabling communications between devices without line-of-sight in scenarios of disaster recovery bringing connectivity to first-responders and other ad hoc communications. Another application is a universal beamforming-relay to extend the range and bridge IoT devices operating on the same frequency. Most IoT devices are low-cost and not equipped with beamforming capabilities. \ignore{On that account, a universal beamforming-relay can support LoRa~\cite{lora_semtech}, Zigbee~\cite{zigbee_csa}, and Z-Wave~\cite{zwave} on the \SI{900}{\unit{\mega\hertz}} ISM band.}\ignore{Universal beamforming can also be used to estimate direction of arrival of signals with a wide variety of applications in localization of emissions.}

Recently, Machine Learning techniques were explored in support of beamforming. For instance, ML schemes were developed for selecting a beamforming scheme out of two possibilities maximum ratio transmission (MRT) or zero-forcing (ZF) beamforming~\cite{KwonLC2019}. Beamforming with Deep Learning was also recently investigated specially for OFDM systems, and without our goals of universality~\cite{ye_2017}. Furthermore, the vast majority of prior is limited to analytical and simulations results. In this work, we introduce a set of techniques, and their extensive experimental evaluations demonstrating the feasibility and practicality of universal beamforming for wireless receivers using Deep Learning. Our approach and system denoted \system{}, is agnostic to the specifics of transmitted signals such as modulation, bandwidth or standard. \system{} is designed around a deep convolutional neural network (CNN) augmented with a Maximum Ratio Combiner (MRC). It is specifically designed to address the unique features of wireless signals complex samples, such as the $2\pi$ phase ambiguous discontinuity. Also, RF links targeting low Bit Error Rates (e.g., below $10^{-4}$) are sensitive to the typical variations and outliers in the continuous-valued estimations of neural network models~\cite{bishop_prml}. \system{} addresses these challenges successfully achieving the \textbf{\textit{optimal}} beamforming gain using only two receiving antennas even in multi-path environments. Existing systems addressing multi-path effects typically require more antennas, or extensive support from other sensing hardware. Furthermore, our universal beamforming-relay approach based on \system{} significantly outperforms the conventional Amplify-and-Relay, for enhancing LoRa (modulated by Chirp Spread Spectrum) and ZigBee. This work also opens the field to DL applications of multi-antenna systems for interference cancellation and multi-user communications. Our contributions can be summarized as follows:

\begin{itemize}[leftmargin=*]
    \item A design, model, and algorithm for a novel Deep Learning-based \textit{universal} receiver beamformer (\system{}). To the best of our knowledge, our work is the first in the literature that leverages DL to enable multi-antenna receivers for beamforming RF signals irrespective of modulations, bandwidths, or wireless channels (and not requiring any explicit mechanisms such as sounding).
    \item A two-antenna Software Defined Radio receiver prototype leveraging \system{} that supports arbitrary and unseen modulations, bandwidths and channels. The training of \system{} is performed only on a fixed basic RF settings (single modulation (BPSK) and bandwidth (\SI{1}{\unit{\mega\hertz}}) in a cable environment) and bolstered with efficient RF dataset collection process with phased augmentation to improve dataset diversity.
    \item \system{} is extensively evaluated (BPSK, QPSK, 8-PSK, 16-QAM, GMSK with various bandwidth settings), first using RF coaxial cables (for reproducibility) emulating channel effects, then in over-the-air setups. \system{} achieved the optimal $\SI{3}{\decibel}$ gain of a two antenna receiver.
    \item \system{} is demonstrated for relaying LoRa and ZigBee signals connecting two devices unable to establish direct communications\ignore{(100\% Packet Loss Rate)}. We show that \system{} achieves LoRa Packet Loss Rate of up to 23 times lower and ZigBee Packet Delivery Rate of 8 times higher than the single-antenna Amplify-and-Forward method.
\end{itemize}

\section{\ignore{Beamforming Overview}Problem and Approach}\label{sec:overview}
Receiver (RX) beamforming aims to optimally combine the received signals from multiple antennas to maximize the Signal-to-Noise Ratio (SNR). \ignore{Due to various dynamic, complex artifacts of wireless channels, it remains challenging to achieve for practical systems. }In this section, we identify and formulate the RX beamforming problem and present the key ideas of \system{}.

\begin{figure}
    \centering
    \includegraphics[width=\linewidth]{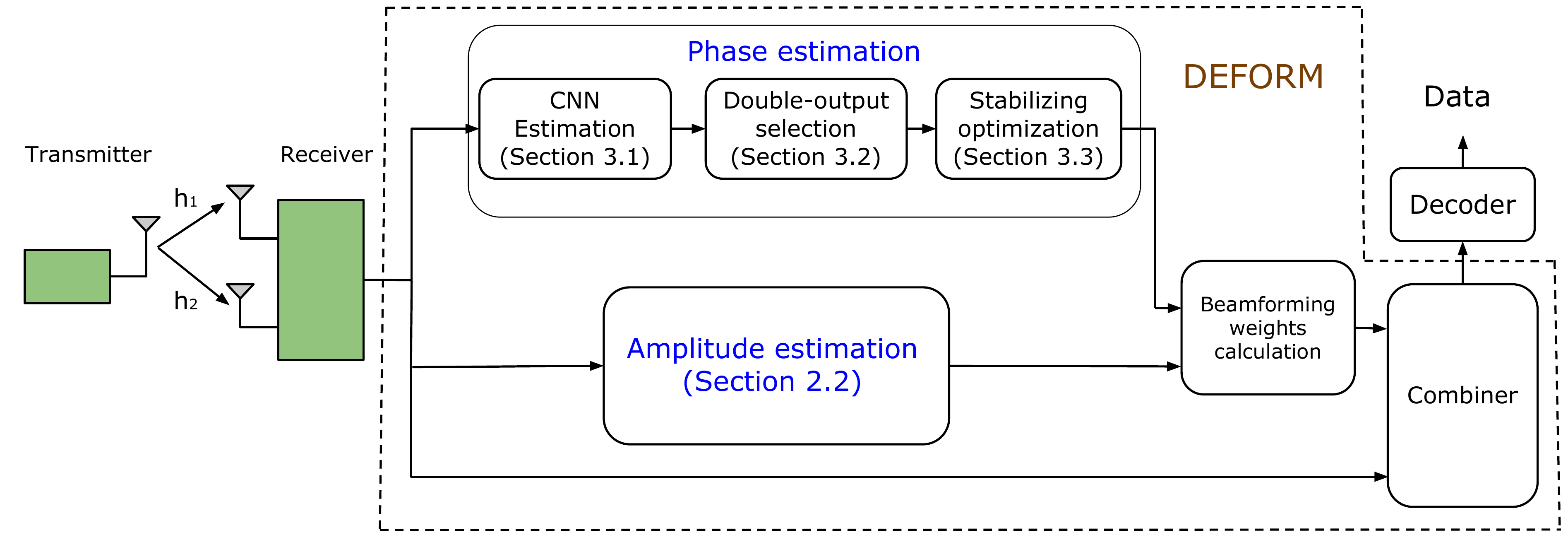}
    \caption{\system{}'s beamforming workflow}
    \label{fig:DEFORM_workflow}
\end{figure}

\subsection{Theory and Challenge}
Consider a transmitter and a receiver communicating through a slow fading channel. The receiver has $N$ antenna elements, where we model the signal $R_i$ received by antenna $i$ consisting of the transmitted signal $S$ adjusted by the channel gain $h_i$ and the additive Gaussian noise $N_i$:
\begin{equation}\label{eq:recv_sig}
    R_i = h_i S + N_i = s_i + N_i
\end{equation}
Receiver beamforming aims to leverage the diversity of the independent wireless channels between the transmitting and receiving antennas by combining the receiving branches with the adequate complex \textit{beamforming weights}:
\begin{equation}\label{eq:recv_bf}
    R_{\sum} = \sum_{i=1}^N a_i R_i = \sum_{i=1}^N (a_i s_i + a_i N_i)
\end{equation}
The \textit{beamforming weights} $a_i$ are chosen to maximize the combining Signal-to-Noise Ratio (SNR), which is given by:
\begin{equation}\label{eq:snr_bf}
    SNR_{\sum} = \frac{(\sum_{i=1}^N a_i s_i)^2}{N_0 \sum_{i=1}^N |a_i|^2}
\end{equation}
where we assume the noise in each branch is independent and has the same Power Spectral Density (PSD) $N_0$. The maximization of $SNR_{\sum}$ is solved using the Cauchy-Schwartz inequality~\cite{goldsmith_2005} and yields the optimal weights:
\begin{equation}\label{eq:opt_weight}
    \hat{a}_i = \frac{s_i^*}{\sum_{j=1}^N |s_j|} \; \; \forall j \in {1,...,N}
\end{equation}
where the denominator is the scaling factor for the weights. Substitute to~\Cref{eq:snr_bf}, we now have the combining SNR:
\begin{equation}\label{eq:snr_bf_new}
    SNR_{\sum} = \frac{\sum_{i=1}^N s_i^2}{N_0} = \sum_{i=1}^N SNR_i
\end{equation}
The optimal combining SNR is the sum of the SNRs from all receiving branches. In the best case scenario when all branches have the same SNRs, this beamforming technique (also known as Maximal-Ratio Combining \cite{goldsmith_2005}) can achieve a final SNR of $N$ times the SNR acquired from a single branch. For instance, with a two-antenna receiver we gain twice the SNR (corresponding to a 3 dB gain).

The main challenge to achieving the optimal combiner is how to estimate the optimal weight $\hat{a}_i$ for each receiving branch $i$. In \Cref{eq:opt_weight}, $\hat{a}_i$ is dependent on $s_i$ ($s_i=h_i S$), which is unknown to the receiver. Conventional techniques, which are still extensively used in OFDM or MIMO systems \cite{stuber2004broadband} , have to insert mutually-known data such as a training sequence or pilot symbols into the transmitted signals that causes significant  communications overhead. Meanwhile, phased-array systems \cite{van2004optimum} do not require channel estimation, but are especially inaccurate against multi-path effects which add up the channel gains from multiple paths and distort the channel characteristics of the direct path.

\subsection{Approach}\label{sec:approach}
We design the receiver beamforming system with the goal to estimate the optimal beamforming weights $\hat{a}_i$ accurately. We first reformulate $\hat{a}_i$ from \Cref{eq:opt_weight} in the polar representation:
\begin{equation}
    \hat{a}_i = \frac{|s_i|}{\sum_{j=1}^N |s_j|} e^{-j\theta_i} \; \; \forall j \in {1,...,N}
\end{equation}
To find $\hat{a}_i$, we need to estimate the amplitude $A_i =  \frac{|s_i|}{\sum_{j=1}^N |s_j|}$ and the phase $\theta_i$. First, we present a simple approach to estimate the amplitude. Out of $N$ receiving branches, we pick an arbitrary branch $k$, and approximate $A_i$ for every branch $i$ with the transformation:
\begin{equation}\label{eq:mrc_gain}
    A_i=\frac{\frac{|s_i|}{|s_k|}}{\sum_{j=1}^N{\frac{|s_j|}{|s_k|}}} \approx \frac{\frac{|R_i|}{|R_k|}}{\sum_{j=1}^N{\frac{|R_j|}{|R_k|}}}
\end{equation}
where the ratio $\frac{|s_i|}{|s_k|}$ is approximated by the amplitude ratio of received signals $\frac{|R_i|}{|R_k|}$  for every $i \neq k$ (If $i = k$, the ratio is $1$). 
It is easy to see that the approximation is correct when $|s_i| \approx |s_k|$. Nonetheless, as $|s_i|$ gets significantly bigger than $|s_k|$, the approximation error increases. The worst case scenario is when $|s_i| \gg |s_k|$ and at the same time, the signal power $|s_k|^2$ is close to the noise PSD $N_0$. Fortunately, there are two reasons that justify the approximation in the scenarios we consider. Firstly, $|s_i| \gg |s_k|$ is equivalent to  $|R_i| \gg |R_k|$, and both make $A_i$ very close to zero. In this extreme case, the beamforming can only provide the SNR as good as the best receiving branch, since the other branches have beamforming weights of zero and do not impact the combining signal $R_{\sum}$. Secondly, in practice, many wireless communications typically requires a sufficiently high SNR to operate, for example, at least \SI{20}{\decibel} is required for Wi-Fi technologies to achieve high data rates~\cite{SNR}. Moreover, as we will discuss in later sections, a SNR of \SI{10}{\decibel} is the minimum requirement for the RX to achieve a target Bit Error Rate of less than $10^{-4}$. With a sufficiently large SNR on each antenna (e.g., \SI{10}{\decibel}), the approximation error is negligible (less than 5\%) even when the SNR on the other antenna is substantially high (e.g., 30 dB). 

Estimating the phase $\theta_i$ is more challenging, especially in the absence of explicit information from the transmitter (e.g. reference signal or sounding mechanism). This is due to the effect of multi-path propagation, in which the constructive and destructive phase combining of multiple copies of the signal traversing the space is typically unpredictable. At the same time, estimating the phase is a critical requirement for the optimal beamforming weights to make the receiving branches \textit{\textbf{co-phased}} in the combined signal. Without this, the branch signals will not add up coherently and the combining signal will experience even further fading (similar to the behavior of multi-path) \cite{goldsmith_2005}. To address this, we present a new approach to achieve co-phasing and optimal beamforming weights. Instead of finding the \textit{\textbf{absolute}} signal phase $\theta_i$, we estimate the \textit{\textbf{relative}} signal phase $\Delta_{\theta_i}=\theta_i-\theta_k$ between the current branch $i$ and a pre-selected arbitrary branch $k$, resulting in the new weight:
\begin{equation}\label{eq:opt_weight_final}
    \bar{a}_i = A_i e^{-j \Delta_{\theta_i}}
\end{equation}
which makes the received signal from branch $i$ co-phased with the signal from branch $k$. As a result, all branch signals are co-phased in the combiner and we achieve the optimal SNR gain. 

To estimate the relative phase, we need to detach the signal component from the noise component. Phased-array systems \cite{van2004optimum} have been known for the capability to disentangle different components in the received signals. However, they typically perform poorly in the presence of multi-path fading \cite{rahamim2004source}. To address this problem, we propose a novel Deep Learning-based universal RX beamforming system (\system{}) that centers around an efficient, powerful Convolutional Neural Network (CNN). Inspired by the capability of CNN to filter and extract relevant low-level features from data in various areas such as text \cite{dcnn_language}, RF \cite{oshea16}, visual \cite{VGG-2014}, or speech \cite{cnn_speech_recog}, we develop a CNN model that precisely estimates the relative phase directly from the branch signals. The CNN is facilitated with optimization techniques to address the unique features of wireless signals complex samples, such as the misleading $2\pi$ phase discontinuity and the link Bit Error Rate sensitivity, bolstering \system{}'s universality. \system{} is implemented for a two-antenna receiver, and extensively evaluated in various RF settings of modulations, bandwidths, and channels to validate the universality in achieving the \textit{optimal} beamforming gain. Here, we emphasize that existing phase-array-based systems addressing multi-path typically require more antennas \cite{arraytrack} or multiple types of sensing hardware \cite{ubicarse}. Furthermore, they are limited by some assumptions on the communication channels, and therefore not universal. The operation workflow of \system{} is depicted in \Cref{fig:DEFORM_workflow}, where the branch signals are combined using the optimal beamforming weights, resulting in the output signal which is sent to the decoder to decode the data. The design and optimization of the most important module - \textit{phase estimation}, will be presented in details in \Cref{sec:design}.

\section{Phase Estimation For Universal RX Beamforming}\label{sec:design}

In this section, we present the design of \system{}'s phase estimation module, which leverages a Convolutional Neural Network (CNN) to accurately estimate the relative phases (which is critical for \system{} to calculate the beamforming weights), supported by optimization techniques specially designed to address the unique features of wireless signals complex samples. We have considered several neural network architectures and choose CNN because it is very powerful to extract relevant low-level features embedded in data of various types \cite{dcnn_language, oshea16, cnn_speech_recog, VGG-2014}. It is natural to leverage such capabilities to disentangle the signal components and provide accurate phase estimations.

\begin{figure}
    \centering
    \includegraphics[width=\linewidth]{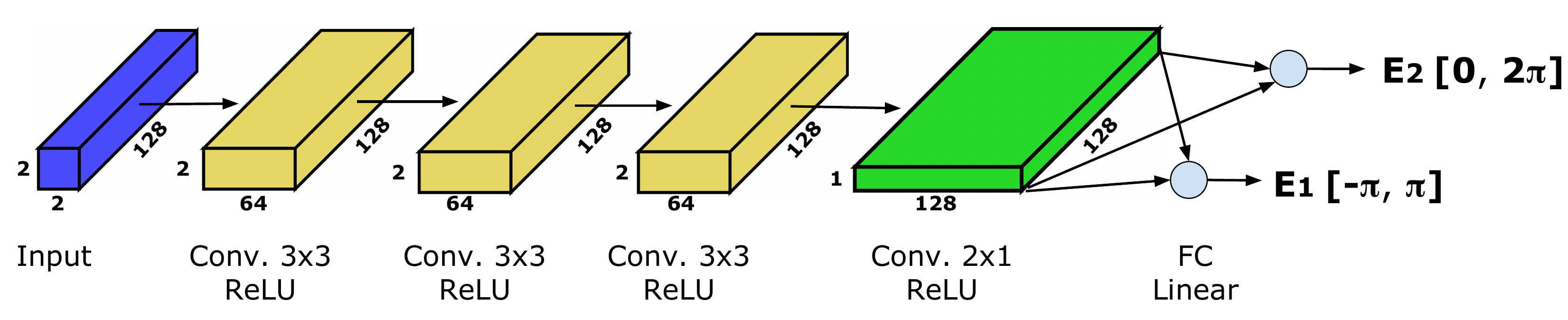}
    \caption{The CNN structure for \system{} with rotational double-output feature.}
    \label{fig:CNN_architecture}
\end{figure}

\subsection{Neural Network Design}\label{sec:cnn}
\bfpara{Goals} We define three goals for the design of the CNN model. Firstly, the model should be able to process the continuous stream of $I/Q$ data efficiently. Feeding a very long stream of samples to the CNN would significantly increase the size of the network as well as the computation cost. Secondly, the model should output the real-valued relative phase with the smallest estimation error as possible. Finally, the network estimation needs to be fast and computationally efficient. We note that there is a natural trade-off between the second and third goals, therefore we aim to find the best model that achieves optimal accuracy and speed throughout the training and validation processes.

\bfpara{CNN Architecture} To reduce the computational cost of the network, a long stream of RF samples is divided into equal chunks of $M$ samples ($M = 128$ in our implementation). As complex RF samples are composed of In-phase and Quadrature components, it is intuitive to view a block of M samples as a matrix of size $2 \times M$ where each entry is an In-phase ($1^{st}$ row) or Quadrature ($2^{nd}$ row) value. The matrices of $N$ antennas ($N=2$ in our current system) are stacked along the third dimension, which consequently form a $2 \times M \times N$ tensor of real-valued elements for each input data.

After having done a network structure search, we converge on an optimized structure that achieves good performance in both processing speed and estimation correctness. We find that as the network gets deeper, the estimation error generally decreases. However, after reaching a certain depth, the improvement becomes less significant. As one of our goals is estimation efficiency, we choose the fastest model that can achieve comparable performance with deeper models. We justify our selection by comparing with popular DL architectures (which is discussed in~\Cref{sec:model_compare}). The structure of our neural network is shown in \Cref{fig:CNN_architecture}. Preceding layers are three convolutional layers with kernel size of $3 \times 3$, followed by one $2 \times 1$ convolutional layer and the fully-connected output layer. $3 \times 3$ convolutional layers have been widely used in state-of-the-art DL architectures \cite{VGG-2014, resnet16} due to their capabilities of extracting low-level features appearing in local regions of the input data. \ignore{This is specially useful for RX beamforming where features about original signals embedded in multi-antenna samples are critical to estimate the relative phase offset.} On the other hand, the $2 \times 1$ convolutional layer aims to provide high-level semantics of angular distance with the sample-wise combining of $I/Q$ channels. The fully connected layer synthesizes the output from previous layers and makes prediction. Rectified Linear Unit (ReLU) activation is used for the convolutional layers because it is computationally efficient and more effective against the vanishing gradient problem \cite{relu_2011}, while linear activation is applied to the output layer. We utilize Batch Normalization \cite{batchnorm} for all convolutional layers to improve the training convergence and eliminate the needs for regularization. To avoid overfitting, we use Learning Rate Decay \cite{yogi} which lowers the learning rate if the validation error remains unimproved for a period of time (e.g., a few epochs). 

\subsection{Rotational Double-Output}\label{sec:double_output}
Theoretically, the neural network is required to provide one estimation for one relative phase between two receiving branches. Nonetheless, while investigating various models, we find that the phase estimation exhibits abrupt variations as the relative phase gets very close to the boundaries of phase values (i.e. upper and lower bounds of the ranges $[-\pi, \pi]$ or $[0, 2\pi]$). More specifically, the behavior is described as follows:
\begin{itemize}
    \item If the network estimates the phase in $[-\pi, \pi]$, then the estimation abruptly fluctuates when the true value is either in $[-\pi, -(\pi-\epsilon)]$ or $[\pi-\epsilon, \pi]$.
    \item If the network estimates the phase in $[0, 2\pi]$, then the estimation abruptly fluctuates when the true value is either in $[0, \epsilon]$ or $[2\pi-\epsilon, 2\pi]$.
\end{itemize}
where $\epsilon \approx 0.2\pi$ based on our investigation during the validation process. We believe that such behavior is related to the discontinuity of the phase when it travels beyond the boundaries, for example, above $\pi$ and below $-\pi$ for $[-\pi,\pi]$. Because of the rotational characteristics (i.e., $2\pi+\theta = \theta \mod 2\pi$), the phase will be shifted backward by an angle of $2\pi$. This confuses the estimation model because those values are far apart (by a distance of $2\pi$) in the numerical axis.

To overcome this problem, we enhance the CNN model with what we call a \textit{rotational double-output} feature, which incorporates two estimation outputs $E_1$, $E_2$ (as depicted in \Cref{fig:CNN_architecture}) for the relative phase converted in $[-\pi, \pi]$ and $[0, 2\pi]$, respectively. We note that the two estimations do not experience abrupt variations simultaneously. Therefore, if we know which output is experiencing errors and not usable, we can select the other output. The double-output selection is described in~\Cref{alg:output_selection}, where we define $SELECT_1$ and $SELECT_2$ to imply the correctness of the respective outputs $E_1$ and $E_2$. If both indicators are $True$, the algorithm selects the predictions with a smaller Standard Deviation (STD) in a history windows of $K=10$ elements. In the last case, the average value of two outputs is taken. We note that eventually, $E_2$ should be converted to $[-\pi, \pi]$ to avoid inconsistencies of the selections:
\begin{equation}\label{eq:convert_output}
    E_2 = \begin{cases}
        E_2 - 2\pi &\text{$\overline{E}_2 > \pi$}\\
        E_2 &\text{else}
\end{cases}
\end{equation}
where $\overline{E}_2$ is the mean of $K=10$ last predictions in $[0, 2\pi]$. With this feature, our CNN is trained to minimize the modified Mean Squared Error:
\begin{equation}\label{eq:mse}
    \mathcal{L}_{\theta} = (\Delta_{\theta_1}-E_1)^2 + (\Delta_{\theta_2}-E_2)^2
\end{equation}
where $\Delta_{\theta_1}$ and $\Delta_{\theta_2}$ are the conversions of the true relative phase $\Delta_{\theta}$ in $[-\pi, \pi]$ and $[0, 2\pi]$, respectively.

\begin{algorithm}[t]
\caption{Double-output selection}\label{alg:output_selection}
\SetAlgoLined
\SetKw{Or}{\hspace{0.3em}\itshape or \hspace{0.3em}}
\SetKw{And}{\hspace{0.3em}\itshape and \hspace{0.3em}}

\KwData{$E_1, E_2, STD_1, STD_2, \epsilon$}
\KwResult{$E_{cur}$}
\uIf{$\pi-\epsilon < E_2< \pi + \epsilon$}{
   $SELECT_2 \gets True$\
}
\Else{
   $SELECT_2 \gets False$\
}
\uIf{$-\epsilon < E_1 < \epsilon$}{
    $SELECT_1 \gets True$\
}
\Else{
   $SELECT_1 \gets False$\
}
Convert $E_2$ to $[-\pi, +\pi]$ by ~\Cref{eq:convert_output}\

\uIf{$SELECT_2 = True \And SELECT_1 = False$}{
    $E_{cur} \gets E_2$\
}
\uElseIf{$SELECT_2 = False \And SELECT_1 = True$}{
    $E_{cur} \gets E_1$\
}
\uElseIf{$SELECT_2 = True \And SELECT_1 = True$}{
    \uIf{$STD_1 < STD_2$}
    {
         $E_{cur} \gets E_1$\
    }
    \Else{
        $E_{cur} \gets E_2$\
    }
}
\Else{
    $E_{cur} \gets (E_1+E_2)/2$
}

\end{algorithm}

\subsection{Stabilizing The Estimations}\label{sec:model_optim}

When the \system{} beamforming system directly uses the continuous-valued phase estimations to compute the beamforming weights, it becomes susceptible to variations and outliers as typically seen in neural network models \cite{goodfellow_dl}. Meanwhile, practical wireless communication systems and standards require not only a high precision, but also the stability in the estimations as they target a Bit Error Rate in the orders of $10^{-4}$ for a proper operation. Short-term changes on some samples have significant and lasting impacts on the whole packet decoding and easily aggravate the Bit Error Rate. To address this problem, we propose two different methods for stabilizing the phase estimations, described as follows.

\bfpara{Temporal smoothing} Because wireless channels typically change with at a much slower rate than the incoming rate of RF samples ($128$ RF samples collected at 1Msamp/s is only $0.128$ ms long), we can stabilize the prediction at a given instant by combining with estimations from the recent past. We stabilize the estimation and improve the robustness of the RX beamforming by using the exponential smoothing function:
\begin{equation}\label{eq:smoothing}
    E^T = E_{cur} \lambda + E^{T-1} (1-\lambda)
\end{equation}
where the final phase estimation at the current time period $T$ is computed using the estimation from previous period $T-1$ and the current CNN output estimation acquired from \Cref{alg:output_selection}. Parameter $\lambda$ controls the smoothness of the result, and is chosen with the best value $\lambda=0.2$ through the validation process. It should be noted that if the offset between $E^{T-1}$ and $E_{cur}$ is significant (exceeding a certain threshold $\alpha$, i.e. $\alpha=1.5\pi$, we should select the instantaneous estimation $E_{cur}$ instead. This behavior is typically seen when the channel changes from being vacant to being occupied by the transmitter. In this case, a drastic change of the phase estimation indicates the beginning of a packet that we should account for.


\begin{figure}
    \centering
    \includegraphics[width=\linewidth]{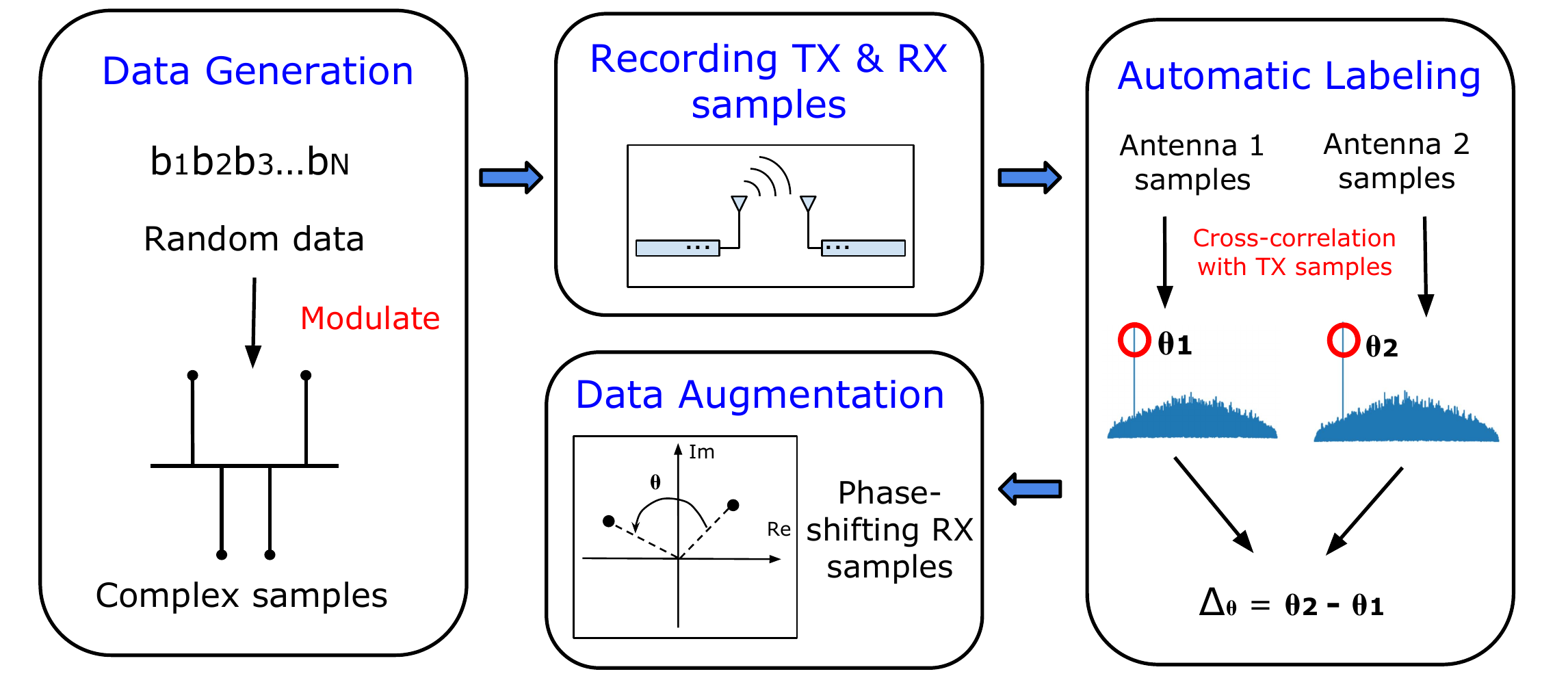}
    \caption{The procedure of building dataset for training the DL model of \system{}.}
    \label{fig:dataset_creation}
\end{figure}

\bfpara{Multi-trial averaging} Temporal smoothing requires multiple chunks of RF samples to achieve a stabilized prediction. In cases when the number of samples is limited but higher computation power is available (e.g., offline processing), we can instead stabilize the prediction by processing the same samples multiple times (thus \textit{multi-trials}). The algorithm is described in \Cref{alg:multitrial_prediction}. To avoid repeatedly getting the same estimation, in each trial, we augment the RF samples by artifically adjusting the phases with a random $\theta_{rand}$. We note that after being processed by the CNN, the current estimation $E_{cur}$ needs to be re-adjusted to account for the prior augmentation. To deal with possible drastic changes, we categorizes the estimations across $N$ trials into two clusters based on their values: We maintain the average estimations $E^T_{C_1}$ and $E^T_{C_2}$ for the clusters at time period $T$. An estimation $E_{cur}$ for any trial is categorized into cluster $C_1$ if $|E_{cur}-E^T_{C_1}| < \alpha$ where $\alpha=1.5\pi$ is the categorization threshold, otherwise cluster $C_2$. Finally, after $N$ trials, we count and compare the number of elements in each group, and choose the average estimation where the corresponding cluster has more elements, as the final estimation for current period.


\begin{algorithm}[t]
\caption{Multi-trial averaging}\label{alg:multitrial_prediction}
\SetAlgoLined
\SetKwFor{RepTimes}{repeat}{times}{end}
\KwData{$N$, RF samples in period $T$}
\KwResult{$E^T$}
\RepTimes{N}{
    Artificially adjust the phases with a random $\theta_{rand}$\;
    Compute instantaneous $E_{cur}$ using \Cref{alg:output_selection}\;
    Update $E_{cur}$ to original value with $\theta_{rand}$\;
    Categorize $E_{cur}$ into two clusters $C_1, C_2$\;
}
Count cluster elements $COUNT_{C_1}, COUNT_{C_2}$\;
Calculate average estimations $E^T_{C_1}, E^T_{C_2}$\;
\uIf{$COUNT_{C_1} > COUNT_{C_2}$}
{$E^T \gets E^T_{C_1}$}
\Else{
    $E^T \gets E^T_{C_2}$
}
\end{algorithm}

\subsection{Dataset Collection}\label{sec:dataset}

It is widely known that adequate, curated training data is critical to any Deep Learning approaches. For supervised learning, data is required to have sufficient high-quality labels. Unfortunately, data labeling is typically a manual process done by humans, requires domain knowledge, is slow, arduous, and can be very costly. Furthermore, open datasets of real RF emissions for RX multi-antenna beamforming remain absent. To address this, we devise an efficient multi-stage approach towards building a sufficiently large dataset for training our Deep Learning model, as depicted in \Cref{fig:dataset_creation}.

Our setup comprises a single-antenna transmitter (TX) and a multi-antenna receiver (RX). As a first step, we generate random complex samples and save them in the memory of both the TX and RX. Then, the TX transmits the saved samples and the RX collects the samples from receiving branches and saves to files. Because of the channel effects, the received samples will experience unknown phase shifts. To determine these phase shifts, we first chunk the received samples, which we subsequently cross-correlate with the transmitted samples already saved in the RX. The phase shift is calculated as the argument of the peak in the correlation output. The labels (relative phase $\Delta_{\theta}$) are obtained by taking the difference between the two phase shifts. When the channels are static, the acquired labels will have very little variance. This would negatively impact the training and bias the DL model towards a small range of values. To address this, and improve the diversity of the dataset, we employ a simple data augmentation technique. For each chunk of RF samples, we randomly shift the phases by a value in $[-\pi,\pi]$ and adjust the label accordingly. We emphasize that while this process is quite efficient, it is \textit{not necessary} for \system{} to repeat this process for each type of data (modulation or bandwidth). As we will show in later sections, our deep beamforming system is agnostic to channels, bandwidths, and modulations. Thanks to its universality, \system{} can be quickly deployed without a prior knowledge about the RF signal and channel parameters.

\section{Experimental Evaluation}
\label{sec:evaluation}
We conduct extensive experiments to validate the universality of \system{} for different modulations, bandwidths, and wireless channels. To train the DL model, we use the techniques described in \Cref{sec:dataset} and collect a dataset of over $167$ million complex RF samples transformed into $654,553$ real-valued tensors of size $2 \times 128 \times 2$, each corresponds to a total of $256$ samples collected by the two RX antennas. For this dataset, we use an Ettus USRP B210 software-defined radio (SDR) with the TX implemented using GNURadio~\cite{gnuradio} to transmit BPSK signals (SNR ranging from 0 to 35 dB) to the RX through two identical coaxial cables with a fixed TX bandwidth of 1 MHz. The RX bandwidth is maintained at 1 MHz for the whole dataset. The dataset is split into training, validation, and test sets with ratio $0.64: 0.16 : 0.2$, respectively. While being trained with a fixed set of RF settings, \system{} still performs very well on other unseen and more sophisticated settings of modulations, bandwidths, and channels. To the best of our knowledge, our work is the first universal RX beamforming system in the literature that is designed using Convolutional Neural Network (CNN) and extensively evaluated for practical, universal RF beamforming capabilities. 

\begin{figure}
    \centering
    \includegraphics[width=\linewidth]{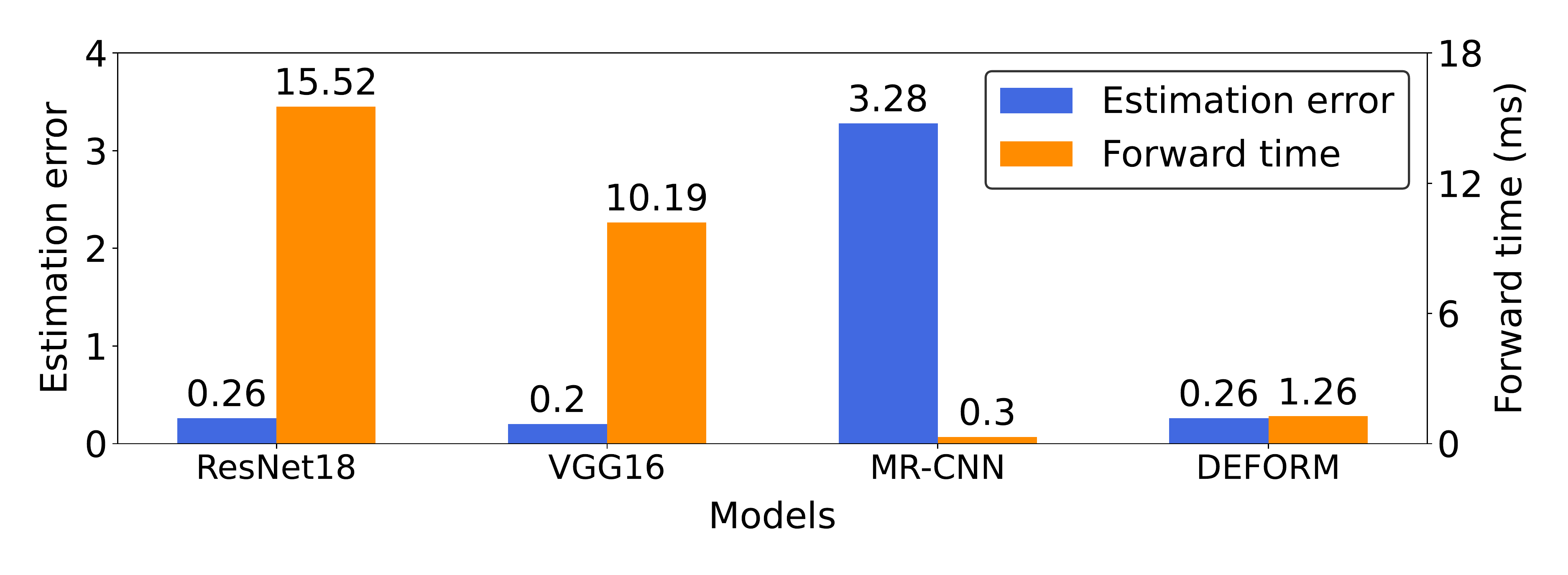}
    \caption{Comparison of the DL models}
    \label{fig:model_compare}
\end{figure}

\begin{figure}
    \centering
    \includegraphics[width=0.32\linewidth]{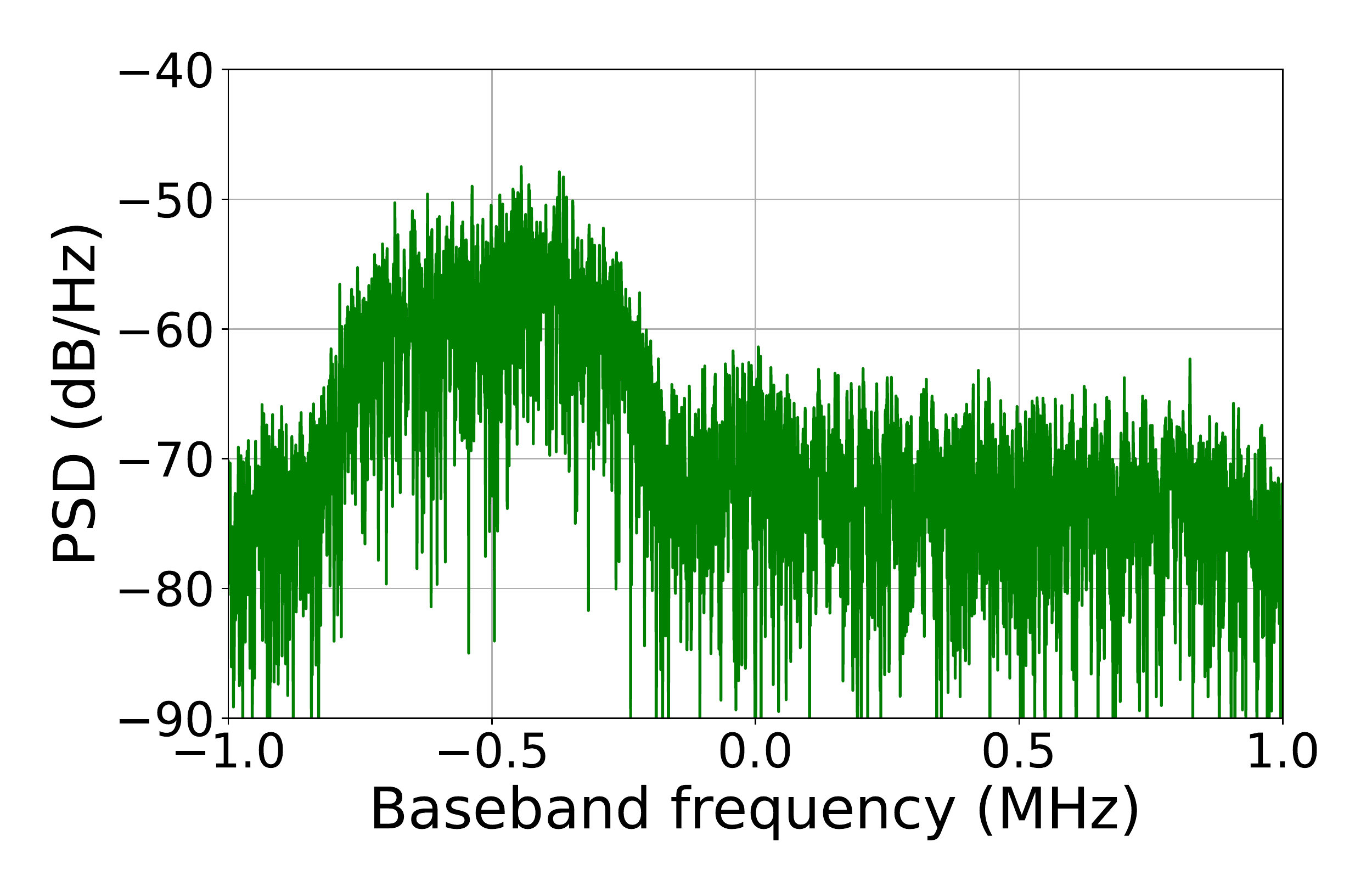}
    \includegraphics[width=0.32\linewidth]{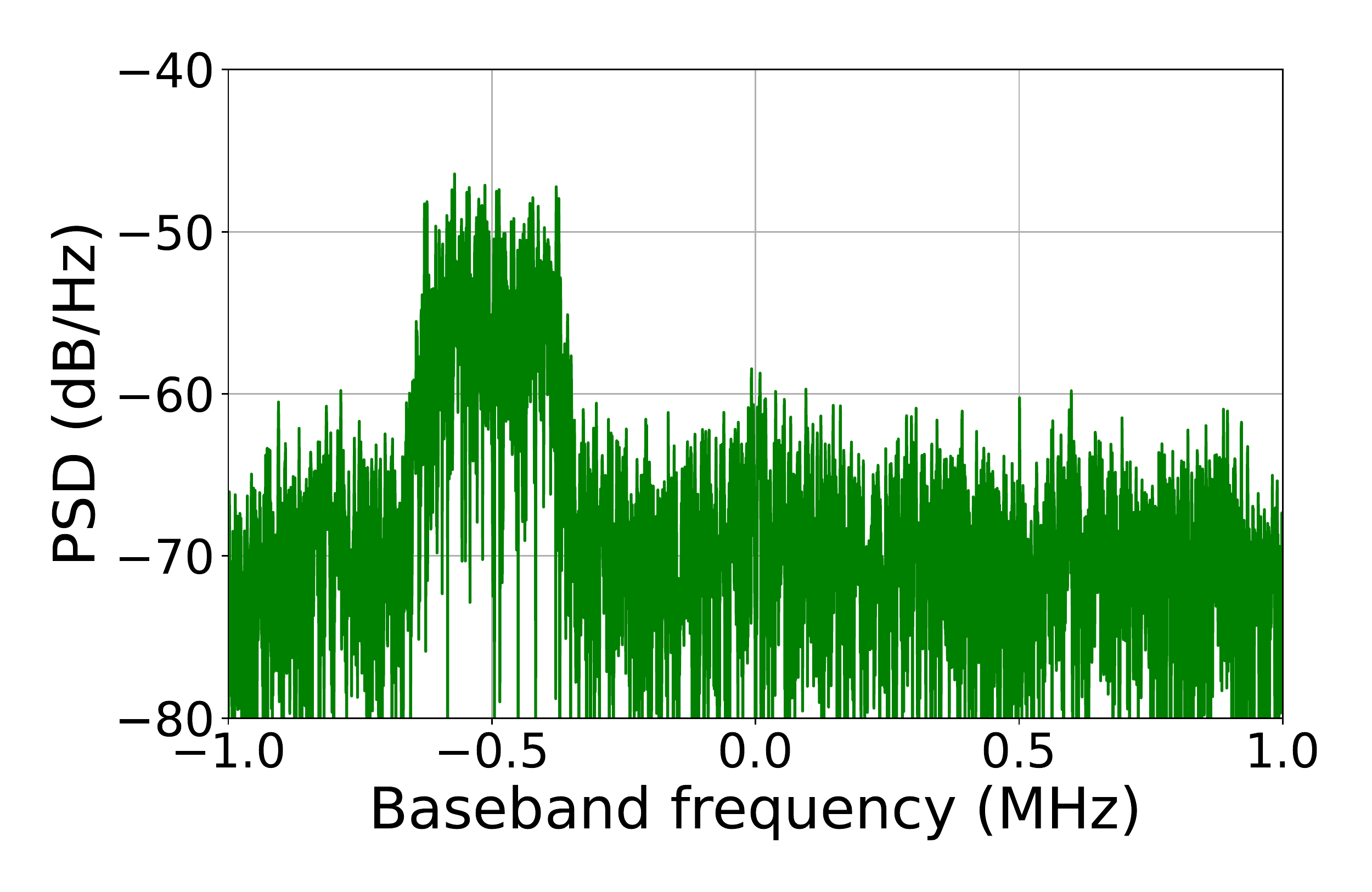} 
    \includegraphics[width=0.32\linewidth]{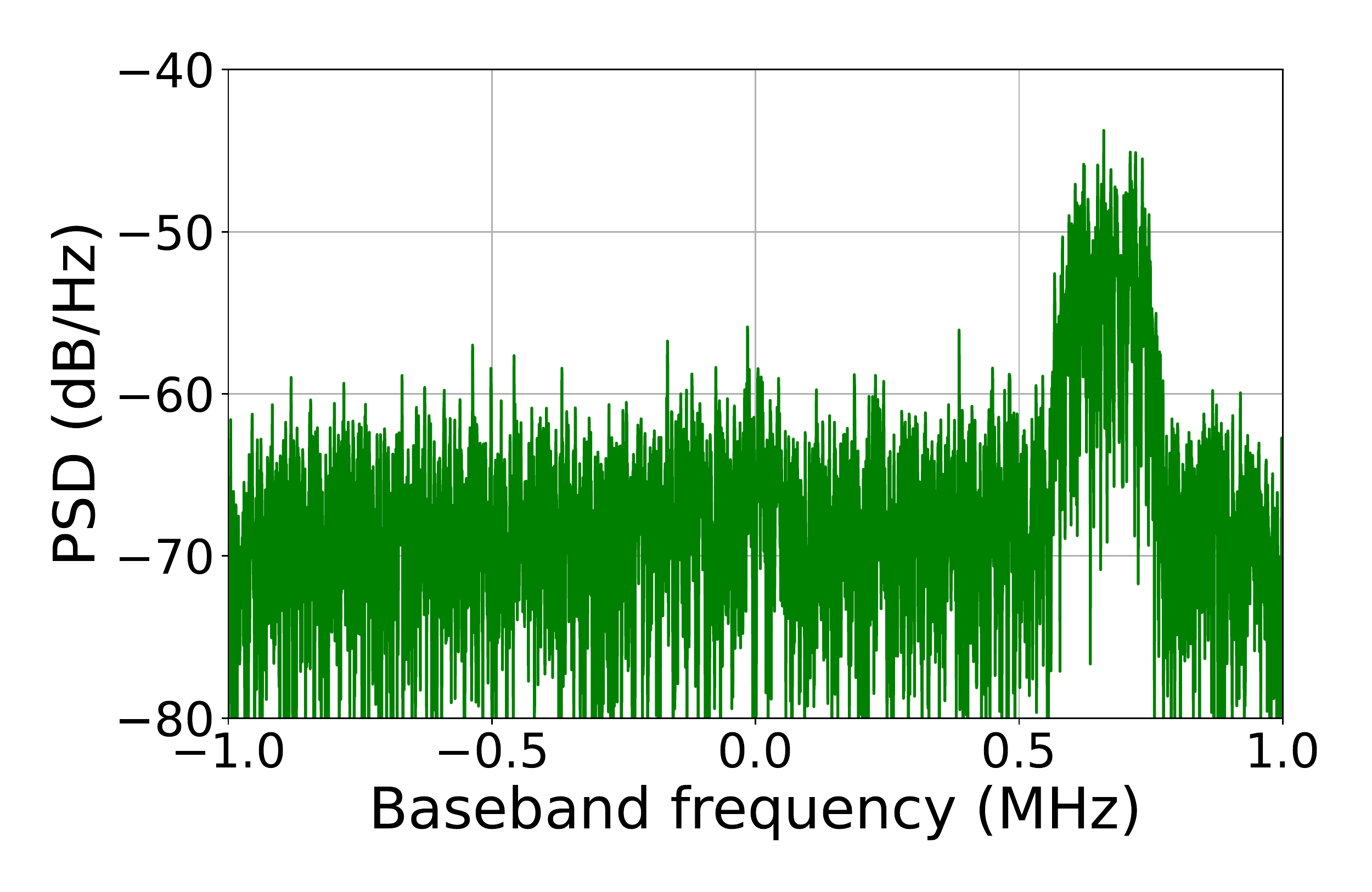}
    \caption{Power Spectral Density plot shows the center frequency offsets in the received wideband signals. It is noted that the y-axis scale is relative.}
    \label{fig:wideband_psd}
\end{figure}

\subsection{DL Model Comparison}\label{sec:model_compare}
To validate our design of neural network and highlight the benefits of our model for the specific task of phase estimation, we evaluate and compare the CNN model with three popular CNN architectures: VGG16 \cite{VGG-2014}, ResNet18 \cite{resnet16} and MR-CNN \cite{oshea16} using the test set of our beamforming dataset. All models are implemented using PyTorch library \cite{pytorch}, then trained and tested on a NVIDIA GeForce GTX 1080 GPU using Adam Optimizer \cite{kingma2017adam} and \colorbox{codegray}{\code{ReduceLROnPlateau}} Learning Rate Decay scheduler \cite{yogi} with initial learning rate $lr=0.005$. The evaluation metrics are estimation error and network forward time. Mean Square Error loss function (\Cref{eq:mse}) is used to calculate the estimation error. We use \colorbox{codegray}{\code{torch.cuda.synchronize}} from PyTorch to synchronize CUDA operations before and after the neural network propagation function and calculate the elapsed time of such function accurately. \Cref{fig:model_compare} compares the estimation error of models on the test set and the network forward time. It is clear that in terms of estimation correctness, \system{} outperforms MR-CNN with a test error of 12.6 times lower (0.26 compared to 3.28). Moreover, \system{}'s CNN is 8 times faster than VGG16 and 12 times faster than VGG16 while having comparable estimation error (equal to ResNet18, and only 0.06 lower compared to VGG16). Compared with these two models, \system{} is lightweight, more computationally efficient, and can quickly provide a phase estimation with high precision, which makes it more suitable for real-time and embedded systems.

\subsection{Over-the-cables Evaluation}\label{sec:cable_eval}
First, we evaluate \system{}'s performance in a relatively idealistic environment where the RF signals propagate through coaxial cables, where multi-path and other fading effects are absent. Data packets are sent from the TX to the two analog inputs of RX (devices are Ettus USRP B210 with SDR) through a pair of identical cables (So the received signals have similar SNRs). TX signal is modulated using differential BPSK, QPSK, 8-PSK, GMSK and 16-QAM techniques, and with a fixed TX bandwidth of \SI{1}{\unit{\mega\hertz}} and center frequency of 795 MHz.\ignore{To evaluate for different SNR levels, we adjusted the RX gain equally on both channels, so that two received signals will have the same SNR. } We assess \system{}'s wideband capability by using different values of RX bandwidths, with a random shift of RX center frequency when the bandwidth is larger than 1 MHz (\Cref{fig:wideband_psd}).

\begin{figure}
    \centering
    \includegraphics[width=\linewidth]{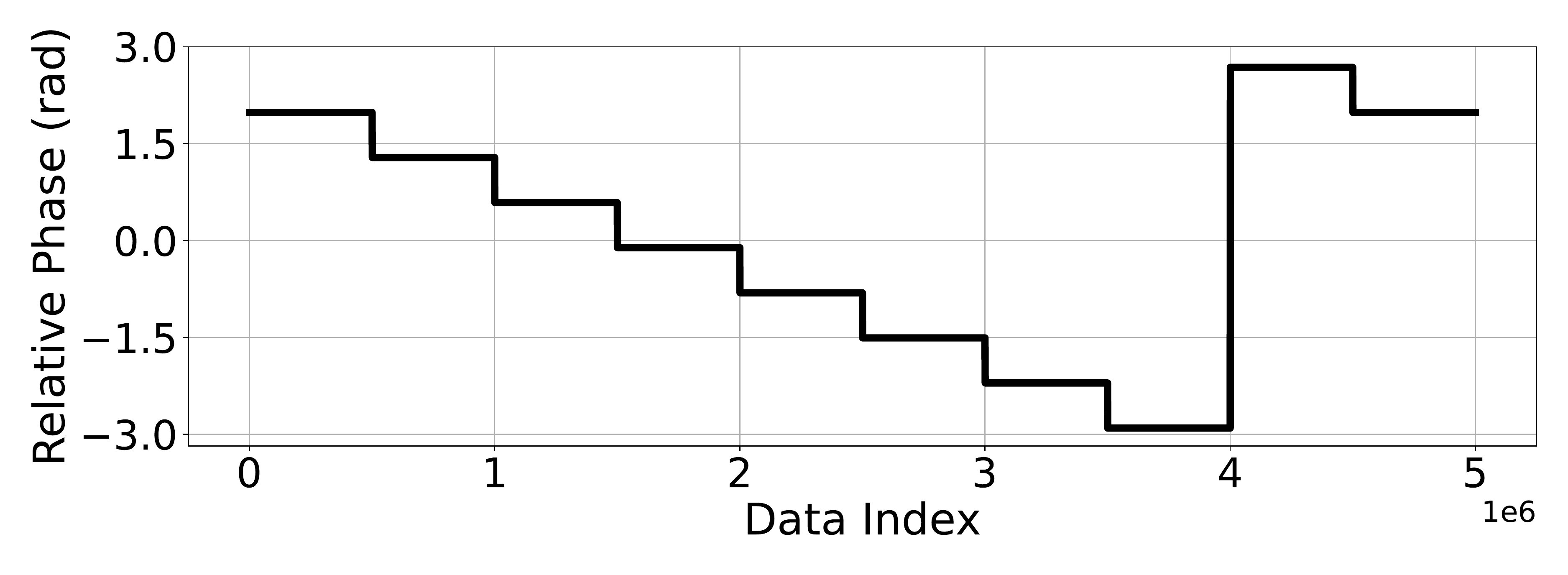}
    \caption{Emulated fading pattern for over-the-cables evaluation. The initial relative phase is randomly chosen within $[-\pi,\pi]$, then slowly changing throughout the experiment for a total amount of $2\pi$ radian.}
    \label{fig:fading_pattern}
\end{figure}

\begin{figure*}
    \centering
    \subcaptionbox{$\Delta_{\phi}=\frac{\pi}{10}$, 1MHz bandwidth \label{fig:symcab_comp_a}}{
        \includegraphics[width=.32\linewidth]{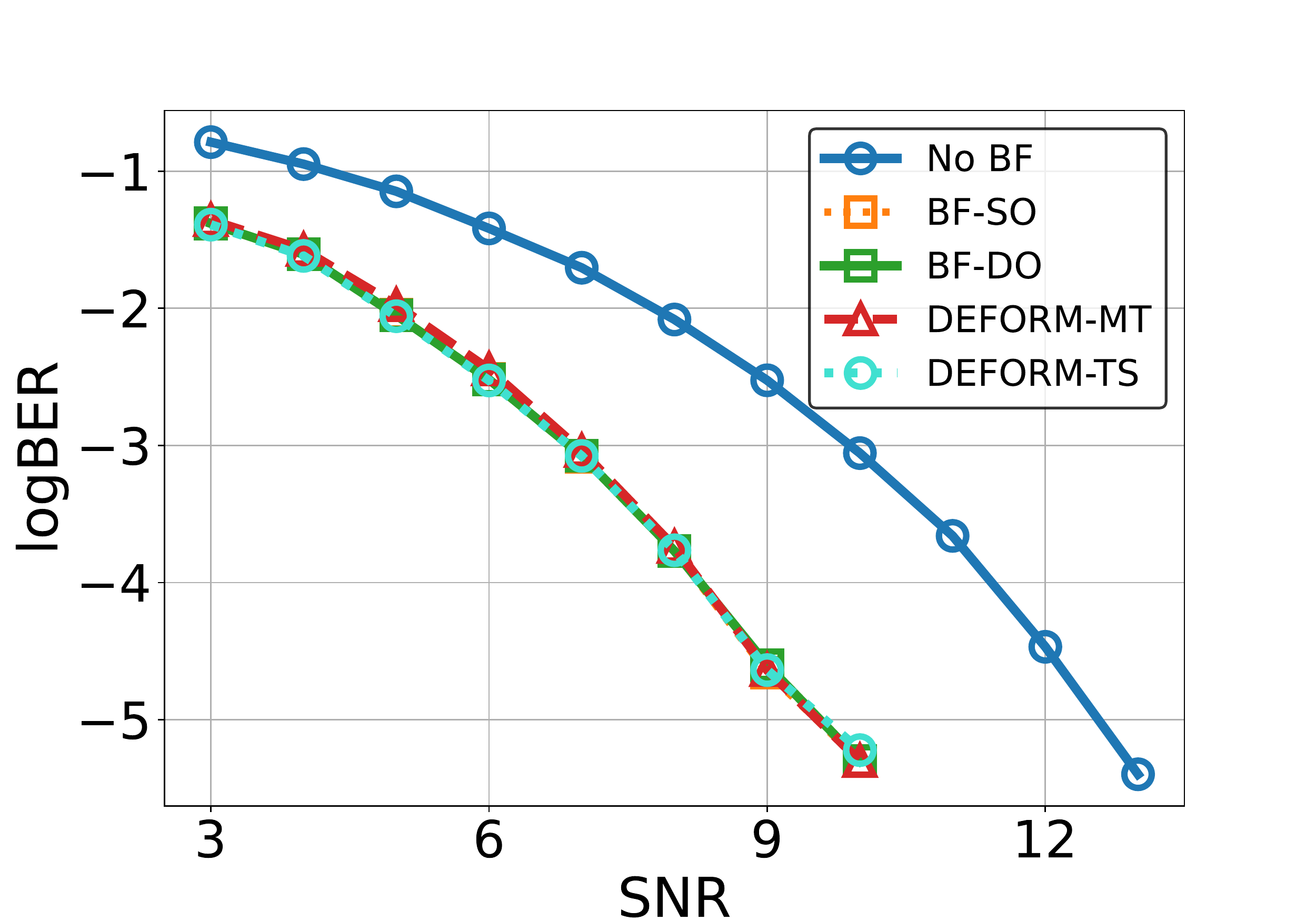}
    }
    \subcaptionbox{$\Delta_{\phi}=\frac{9\pi}{10}$, 1MHz bandwidth  \label{fig:symcab_comp_b}}{
        \includegraphics[width=.32\linewidth]{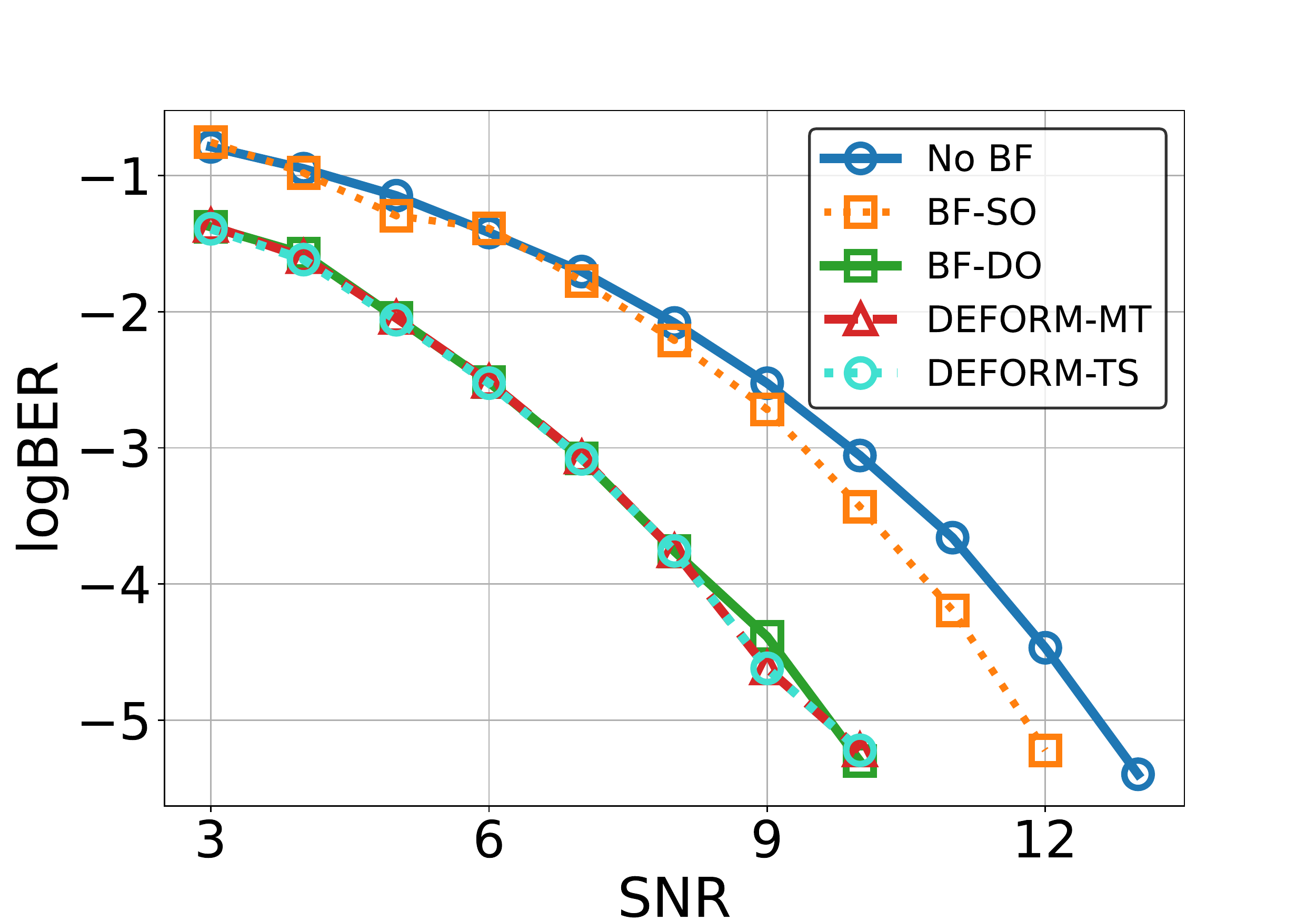}
    }
    \subcaptionbox{$\Delta_{\phi}=\frac{\pi}{2}$, 6MHz bandwidth  \label{fig:symcab_comp_c}}{
        \includegraphics[width=.32\linewidth]{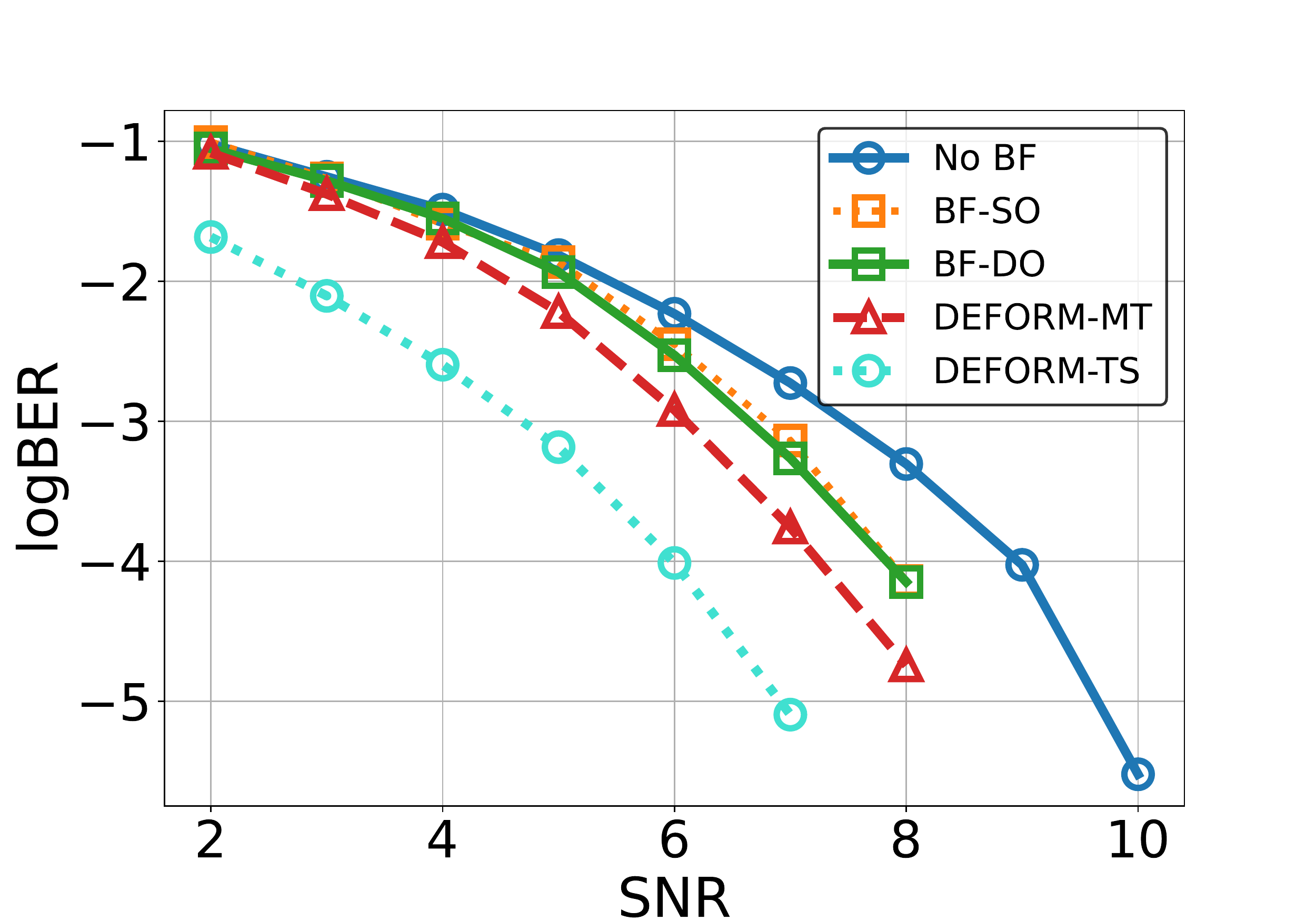}
    }
    \caption{BER comparison between non-beamforming, and beamforming with the baseline CNNs and \system{}'s optimizations.}
    \label{fig:symcab_opt_comp}
\end{figure*}

\bfpara{Evaluation for Model Optimizations} To highlight the impact of the optimization techniques, we measure and analyze the received Bit Error Rate (BER) in five cases: (1) When beamforming is turned off and one of the two received signals is selected for decoding (Note that in our over-the-cables experiments, received signals have similar SNRs), (2) when beamforming is used with the single-output CNN (BF-SO), or (3) with the rotational double-output CNN (BF-DO), and when \system{} is enabled with double-output CNN and (4) temporal smoothing (\system{}-TS) or (5) multi-trial averaging (\system{}-MT) optimization. To calculate BER, we record TX and RX signals at the SDRs and transfer them to the host computer to compare and count the bit errors.~\Cref{fig:symcab_opt_comp} illustrates the evaluation results, where we compare the approaches with BPSK transmissions. For each scenario, RX signal features (relative phase, bandwidth) are artificially adjusted to show the effects of model optimizations: For~\Cref{fig:symcab_comp_a}, because $\Delta_{\theta}$ is far from the phase boundaries, all beamforming settings including the baseline single-output CNN achieve 3 dB gain compared to non-beamforming. When $\Delta_{\phi}=\frac{9\pi}{10}$ which is very close to $\pi$ (where the estimation of the single-output CNN estimating phase in $[-\pi,\pi]$ experiences abrupt variations -~\Cref{fig:symcab_comp_b}), the efficiency of BF-SO decreases to less than 1 dB gain, while BF-DO and further optimization settings maintain the 3 dB gain. When we increase RX bandwidth to 6 MHz (~\Cref{fig:symcab_comp_c}), \system{}-TS and \system{}-MT outperform the baseline CNNs with 3 dB and 2 dB gains compared to non-beamforming, respectively. This justifies the efficiency of \system{}'s optimizations to address the increasing variations and outliers of the baseline CNNs introduced by communication settings of wider bandwidths.

\begin{figure*}
    \centering
    \begin{subfigure}{\linewidth}
        \includegraphics[width=0.19\linewidth]{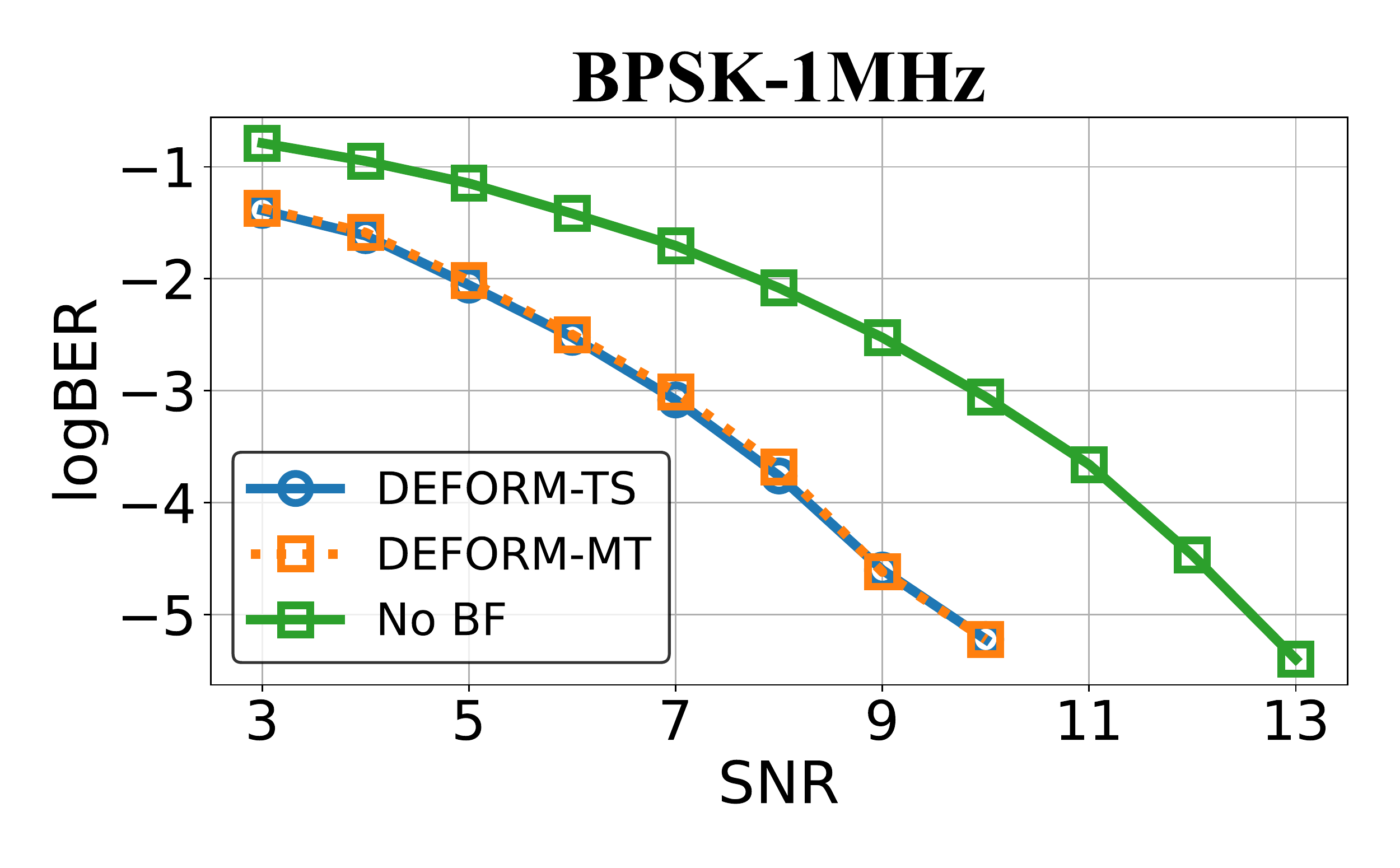}
        \includegraphics[width=0.19\linewidth]{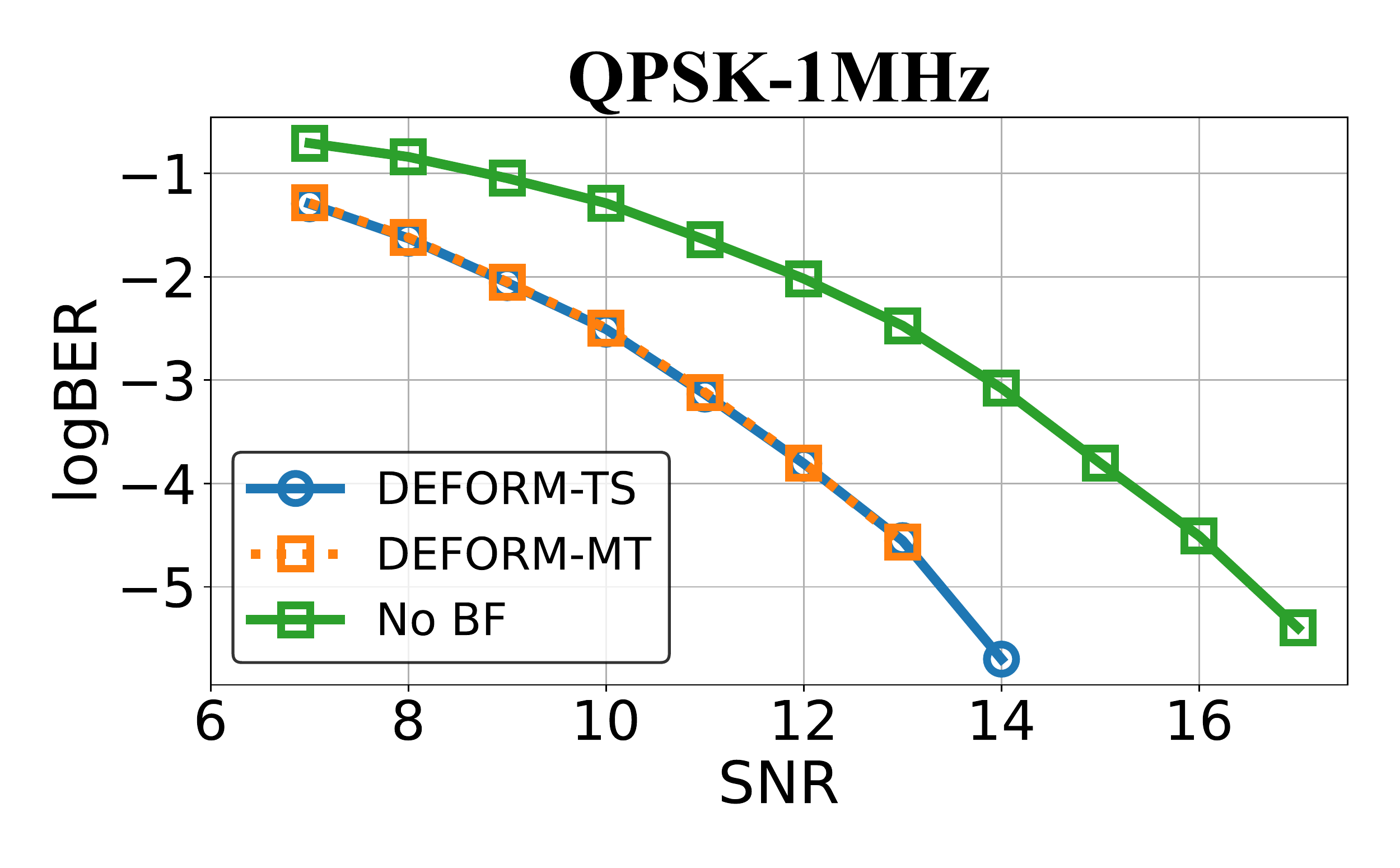}
        \includegraphics[width=0.19\linewidth]{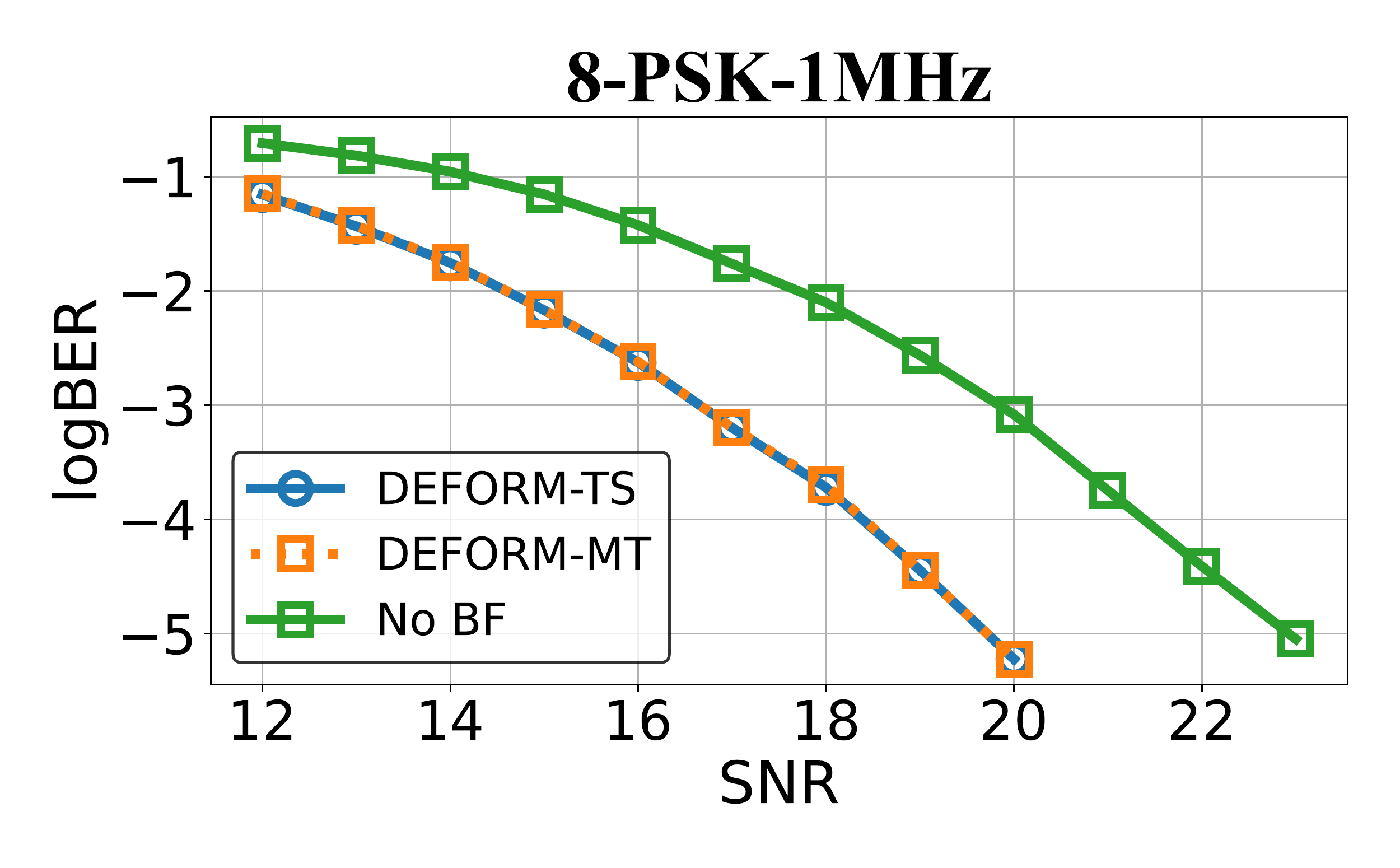}
        \includegraphics[width=0.19\linewidth]{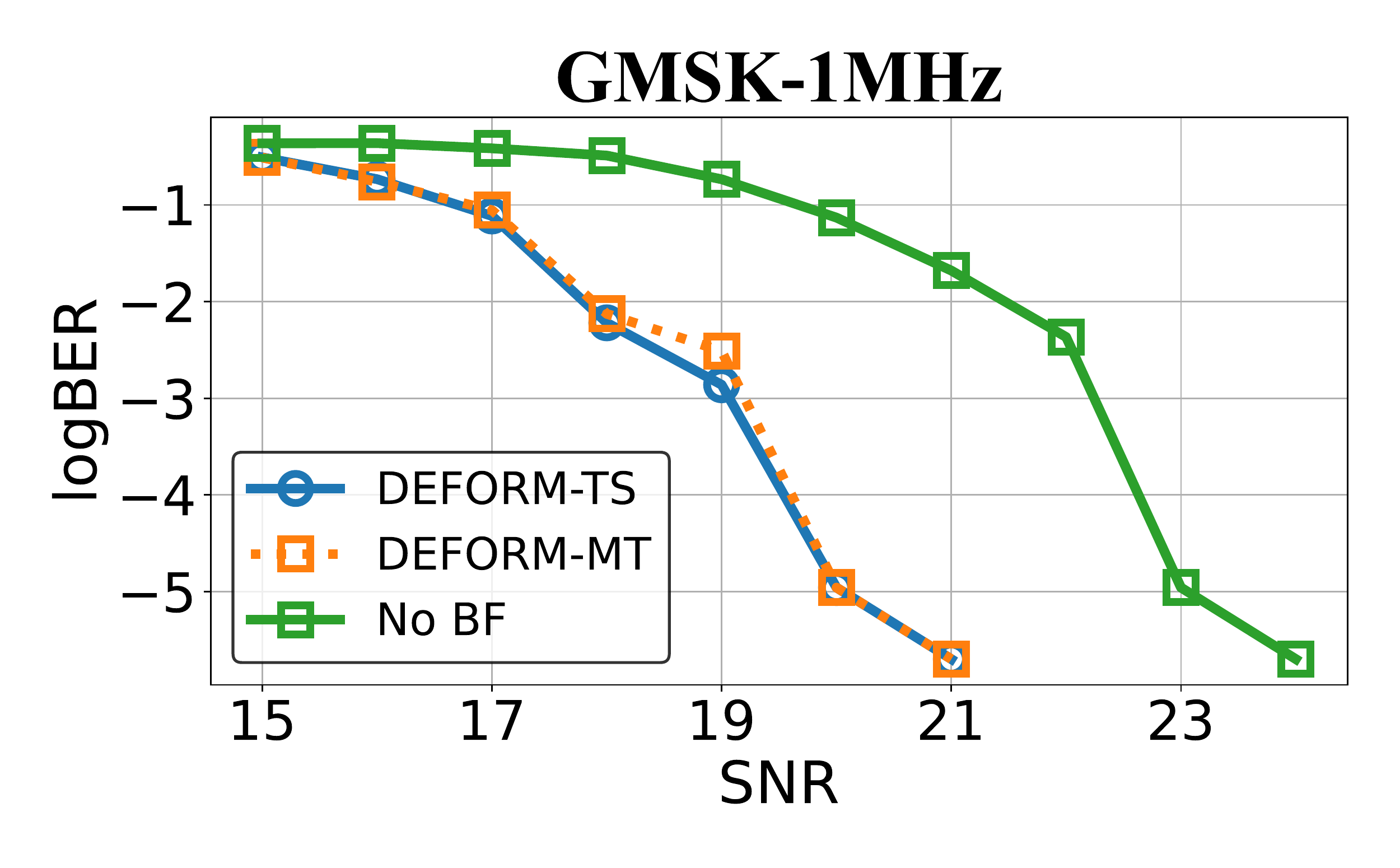}
        \includegraphics[width=0.19\linewidth]{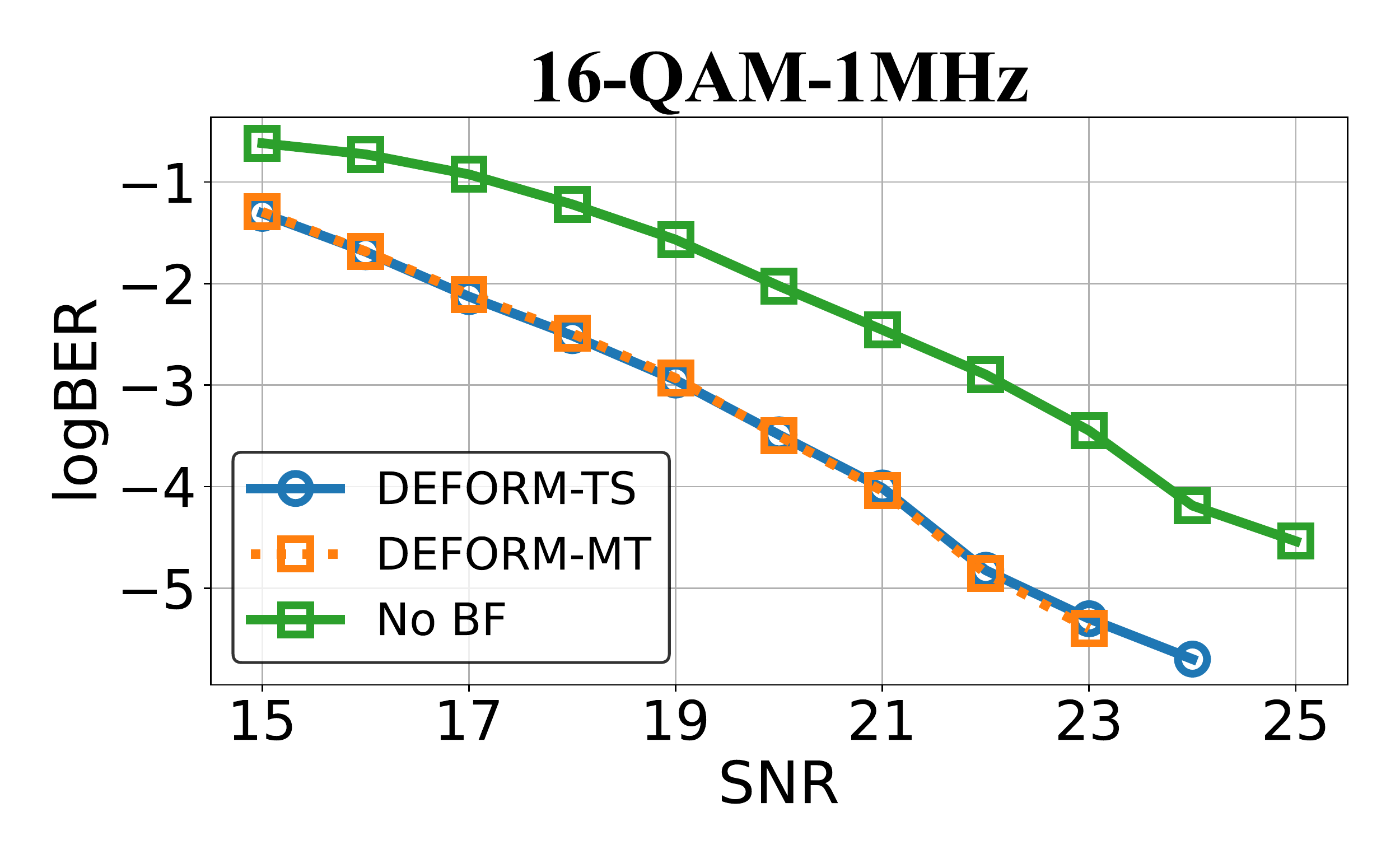}
    \end{subfigure}
    \begin{subfigure}{\linewidth}
        \includegraphics[width=0.19\linewidth]{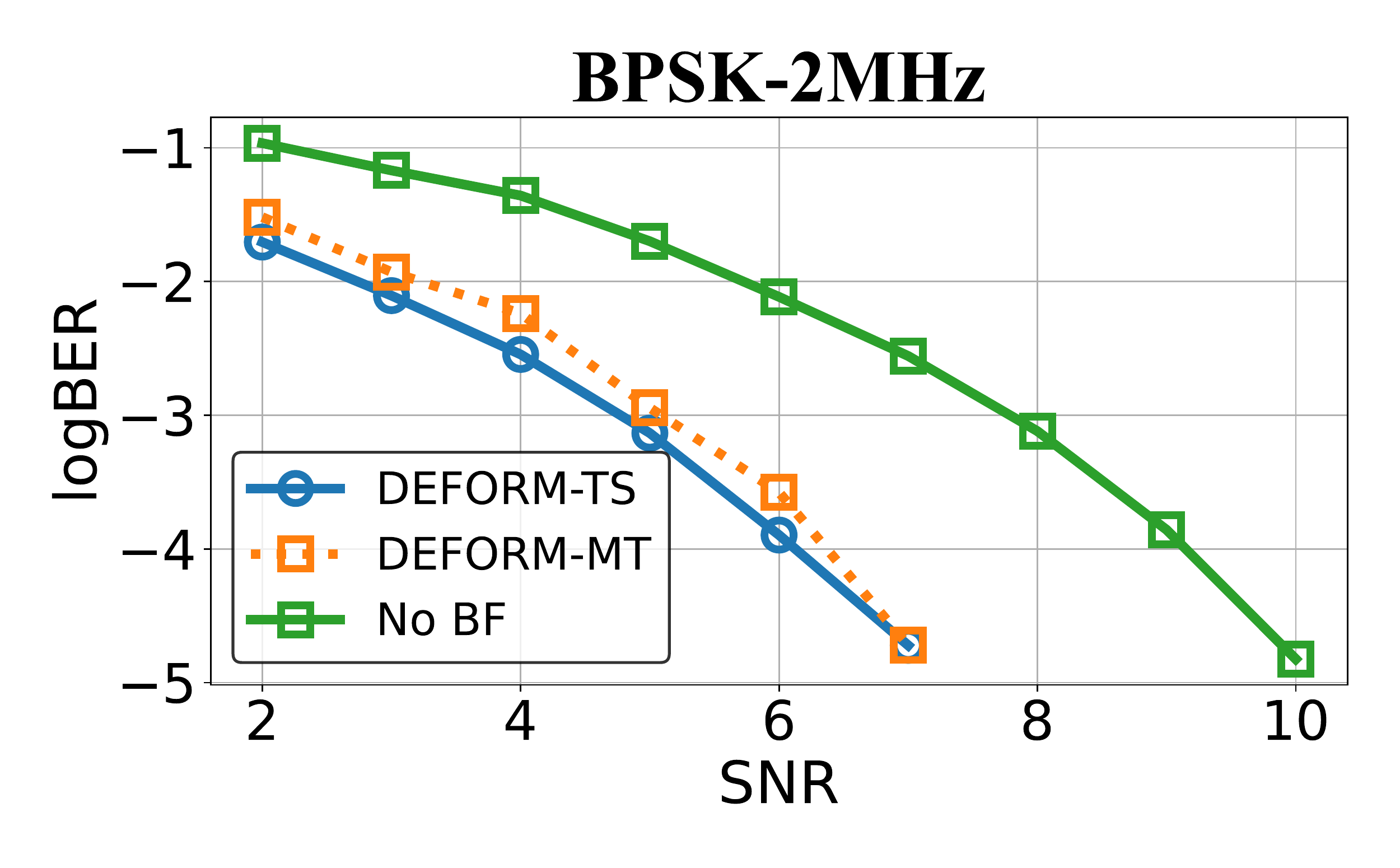}
        \includegraphics[width=0.19\linewidth]{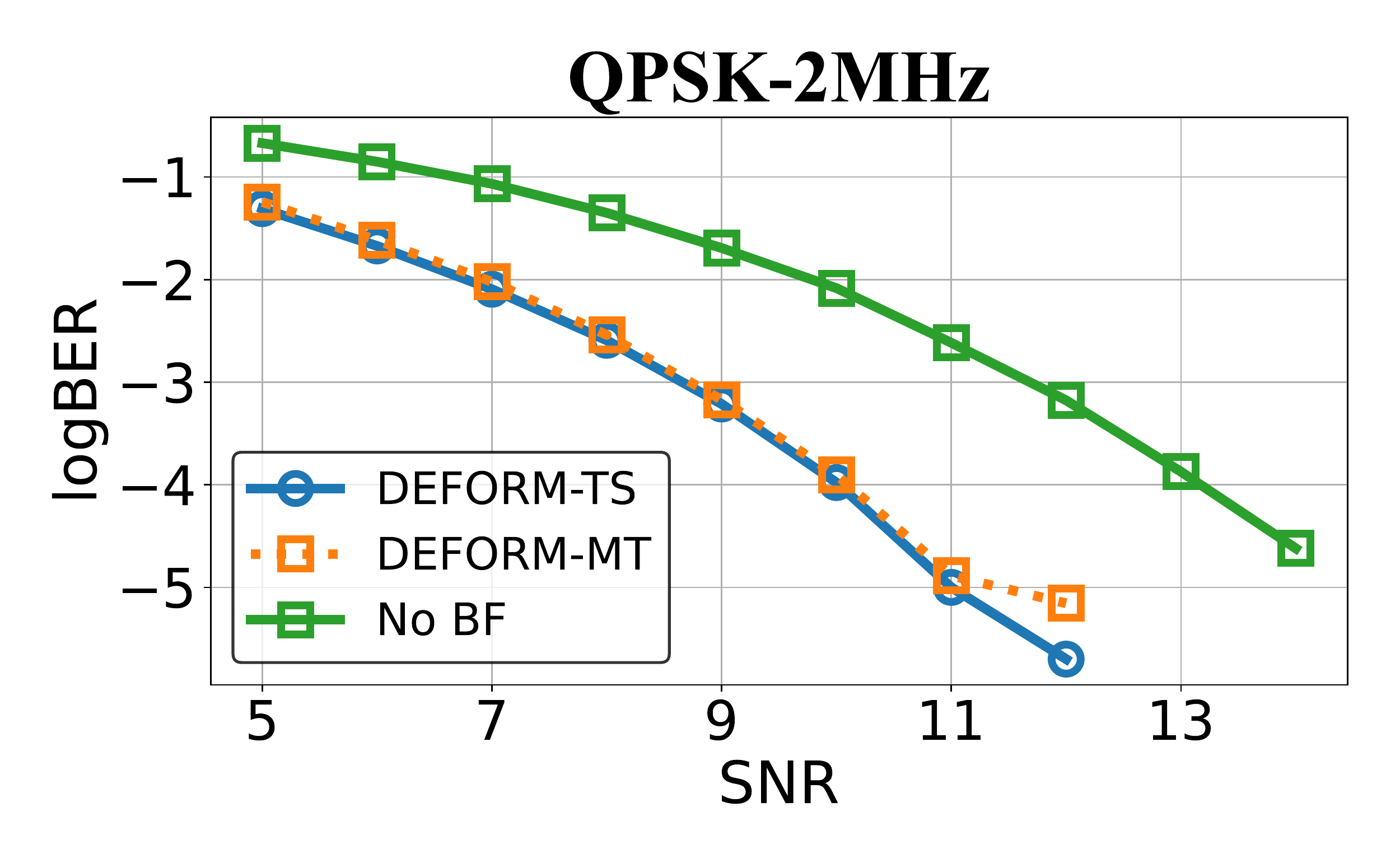}
        \includegraphics[width=0.19\linewidth]{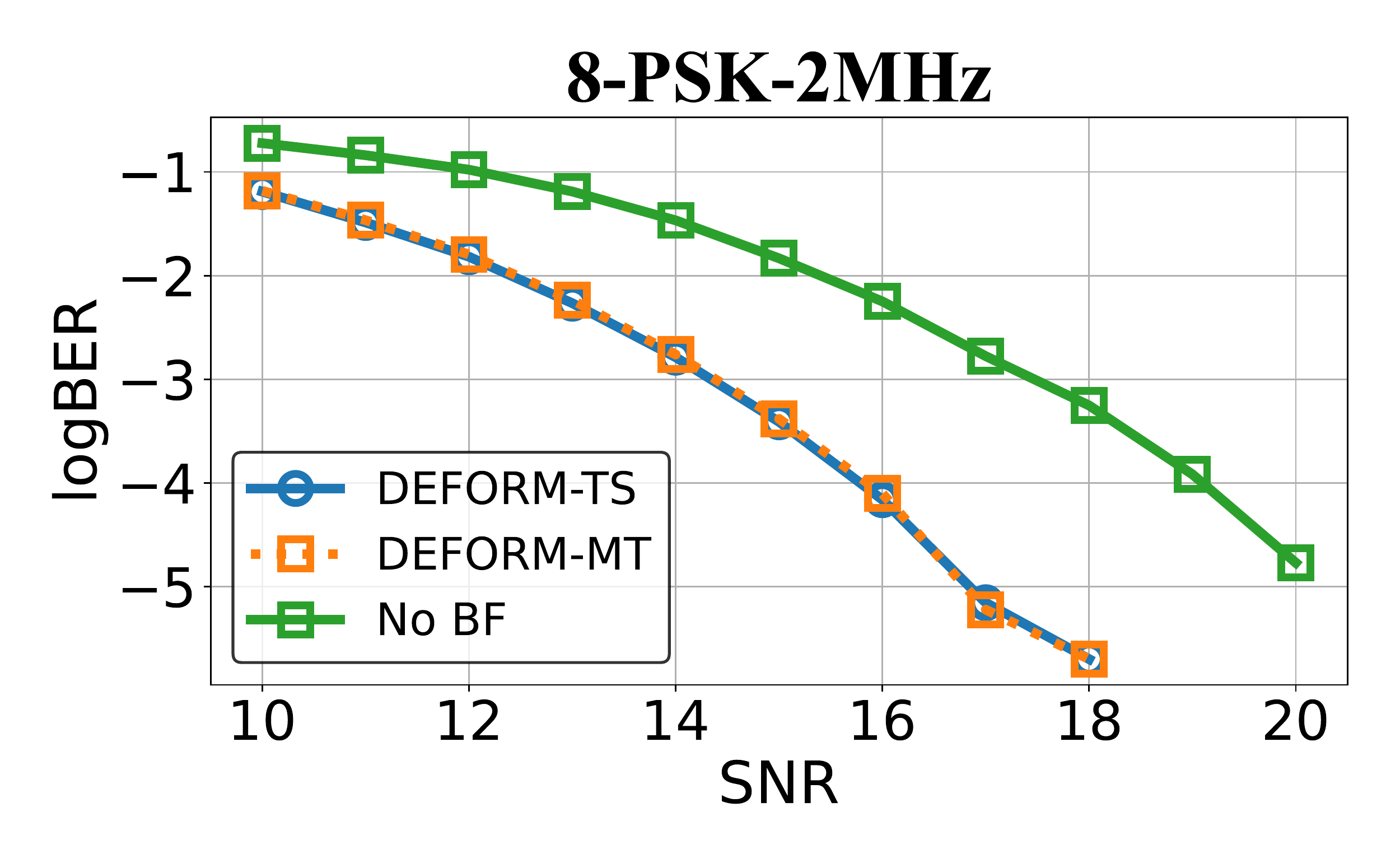}
        \includegraphics[width=0.19\linewidth]{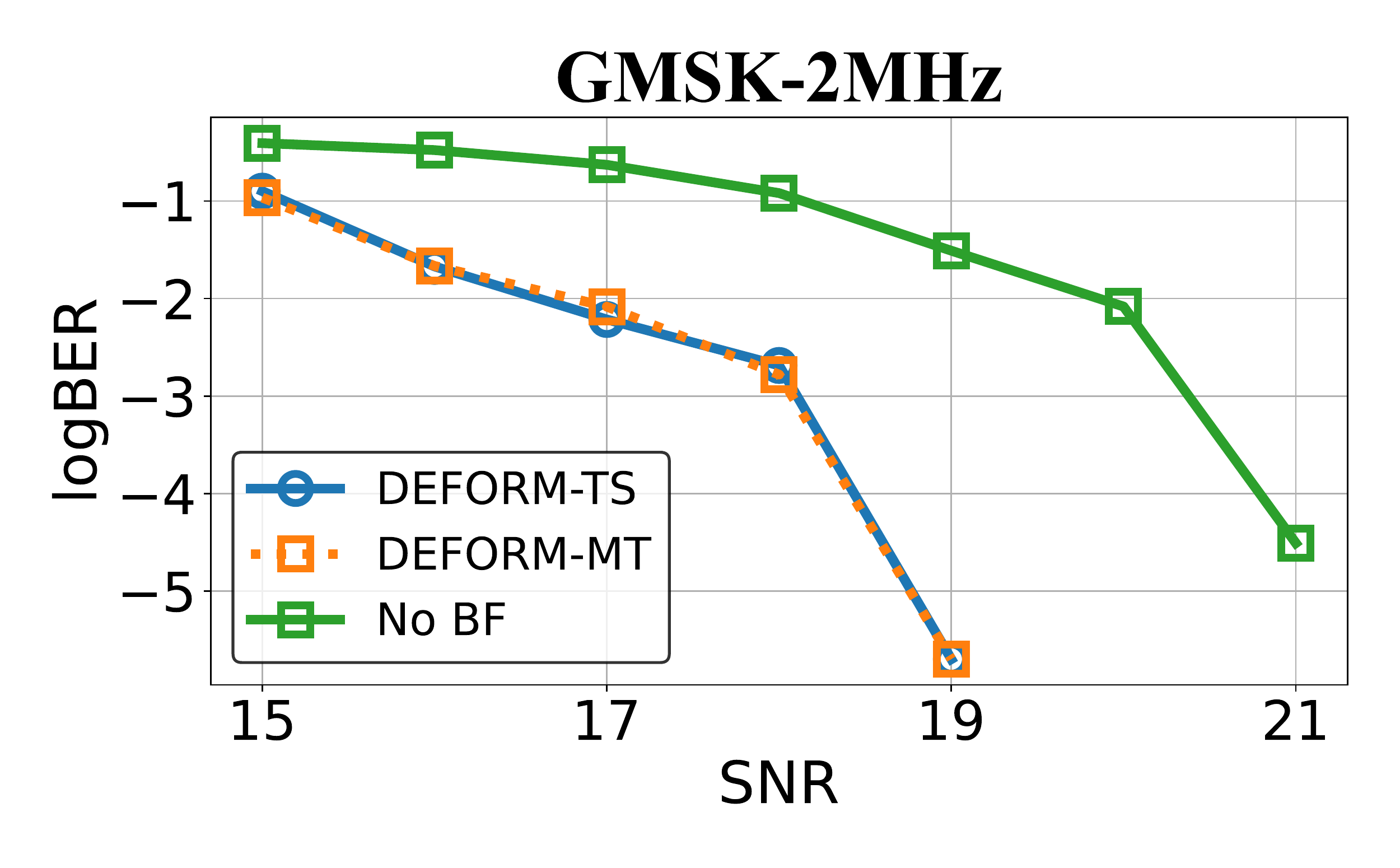}
        \includegraphics[width=0.19\linewidth]{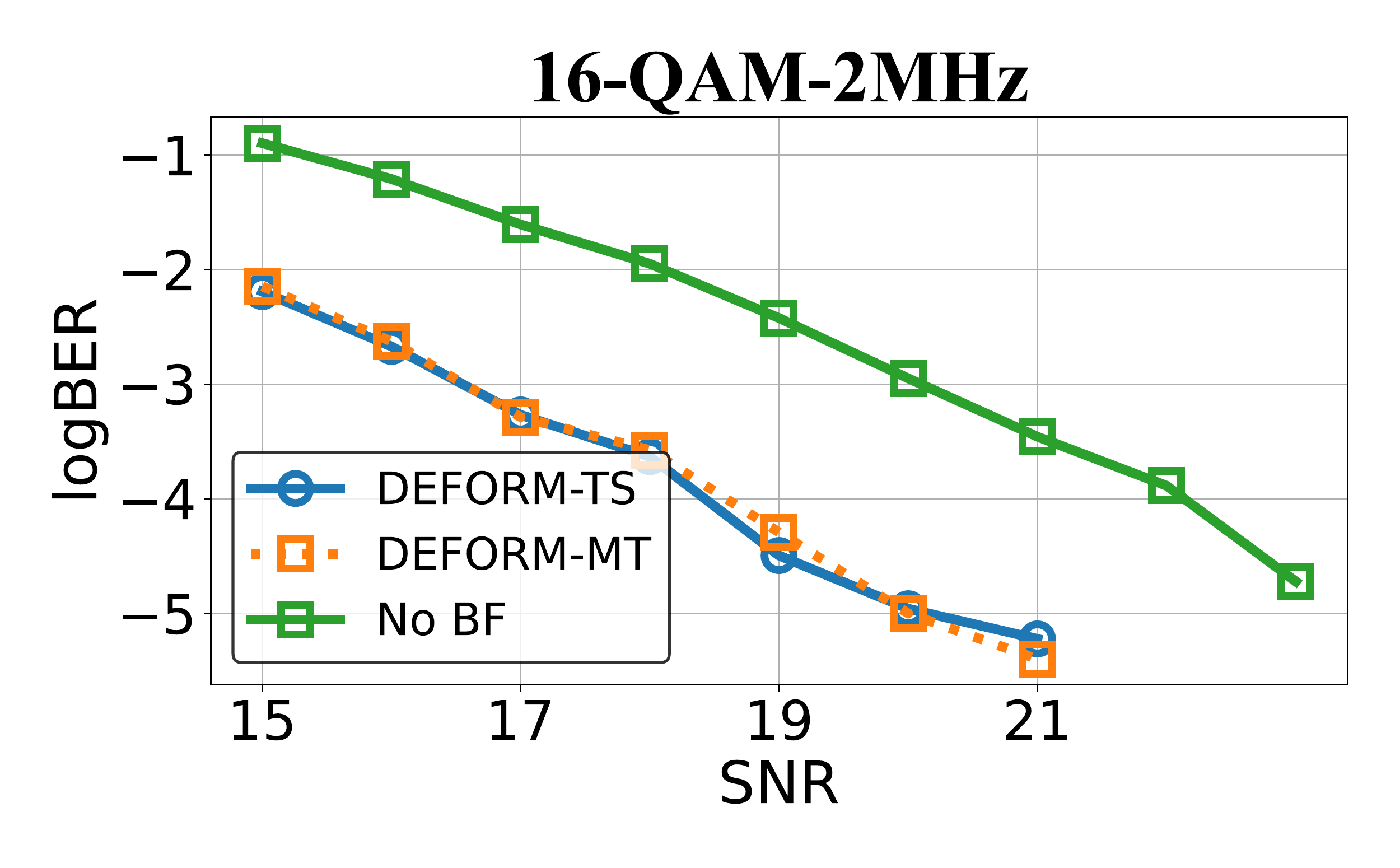}
    \end{subfigure}
    \begin{subfigure}{\linewidth}
        \includegraphics[width=0.19\linewidth]{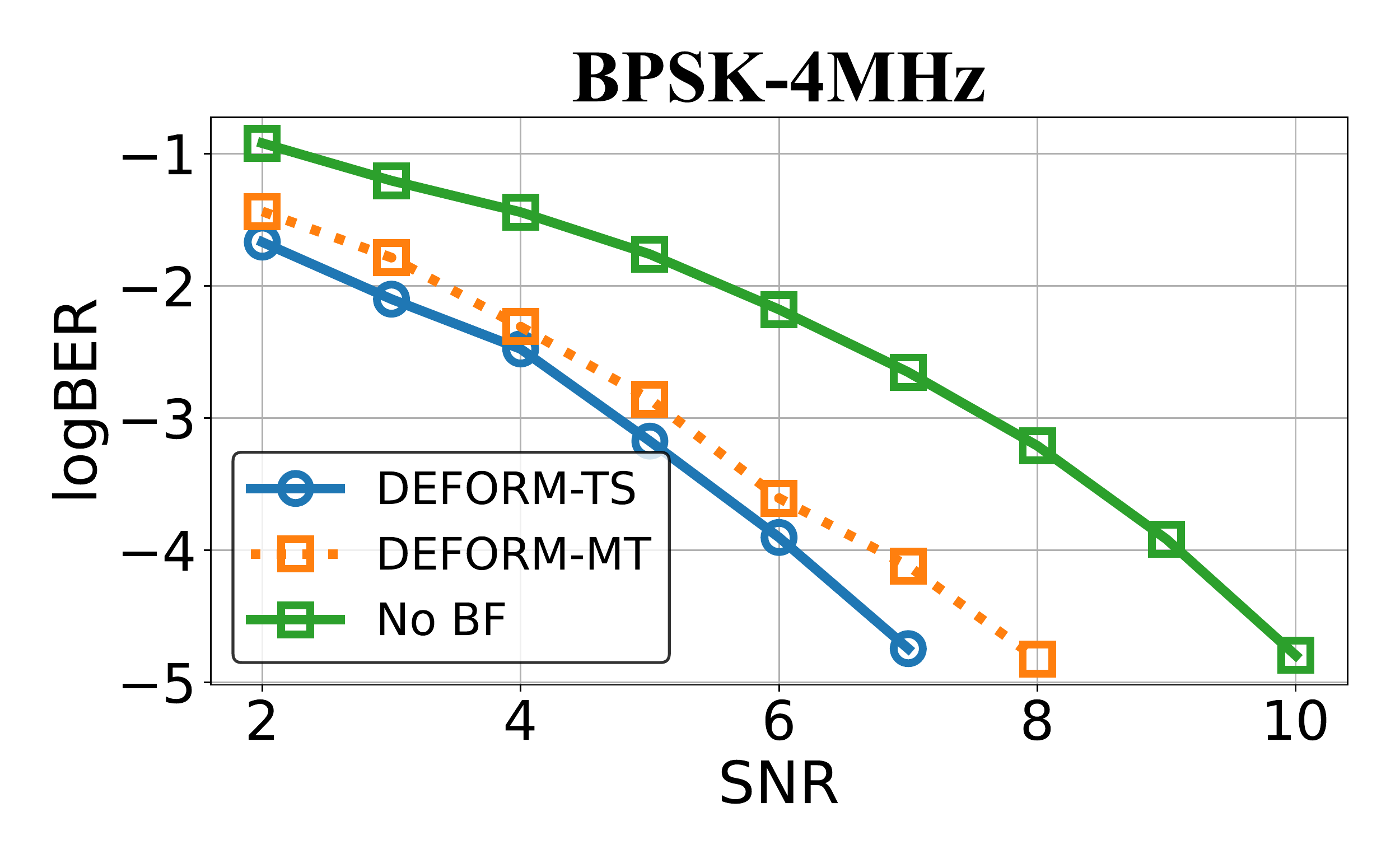}
        \includegraphics[width=0.19\linewidth]{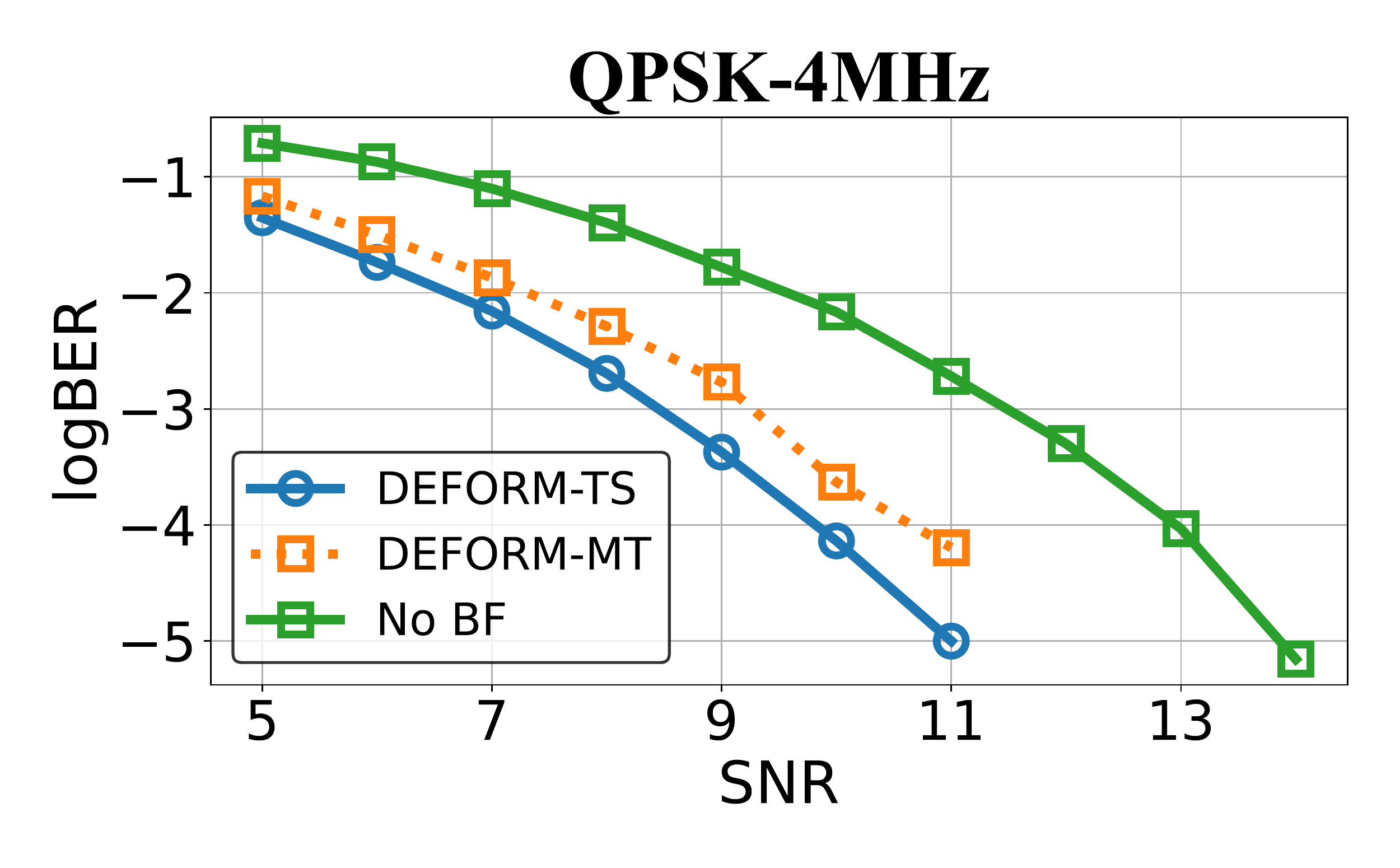}
        \includegraphics[width=0.19\linewidth]{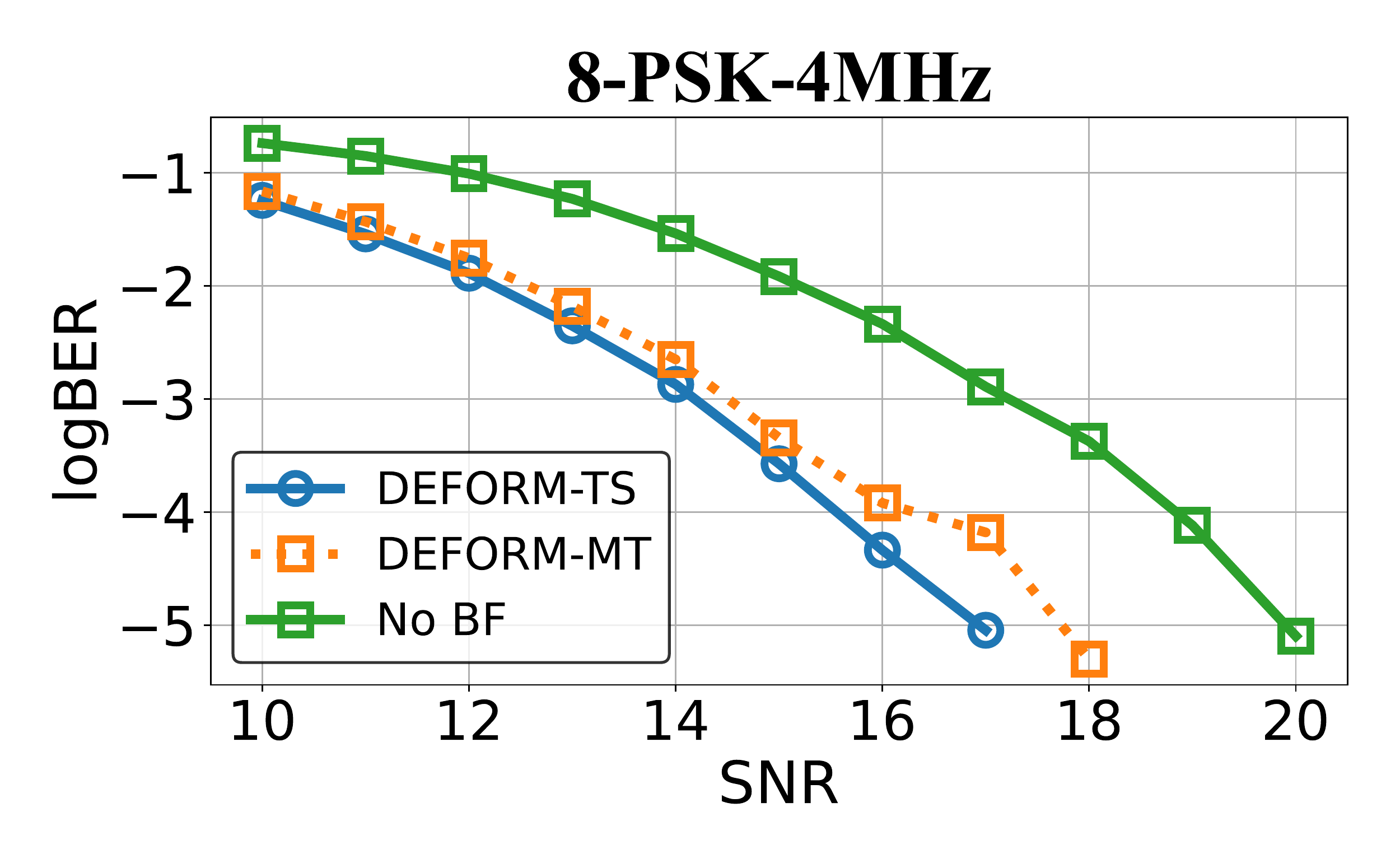}
        \includegraphics[width=0.19\linewidth]{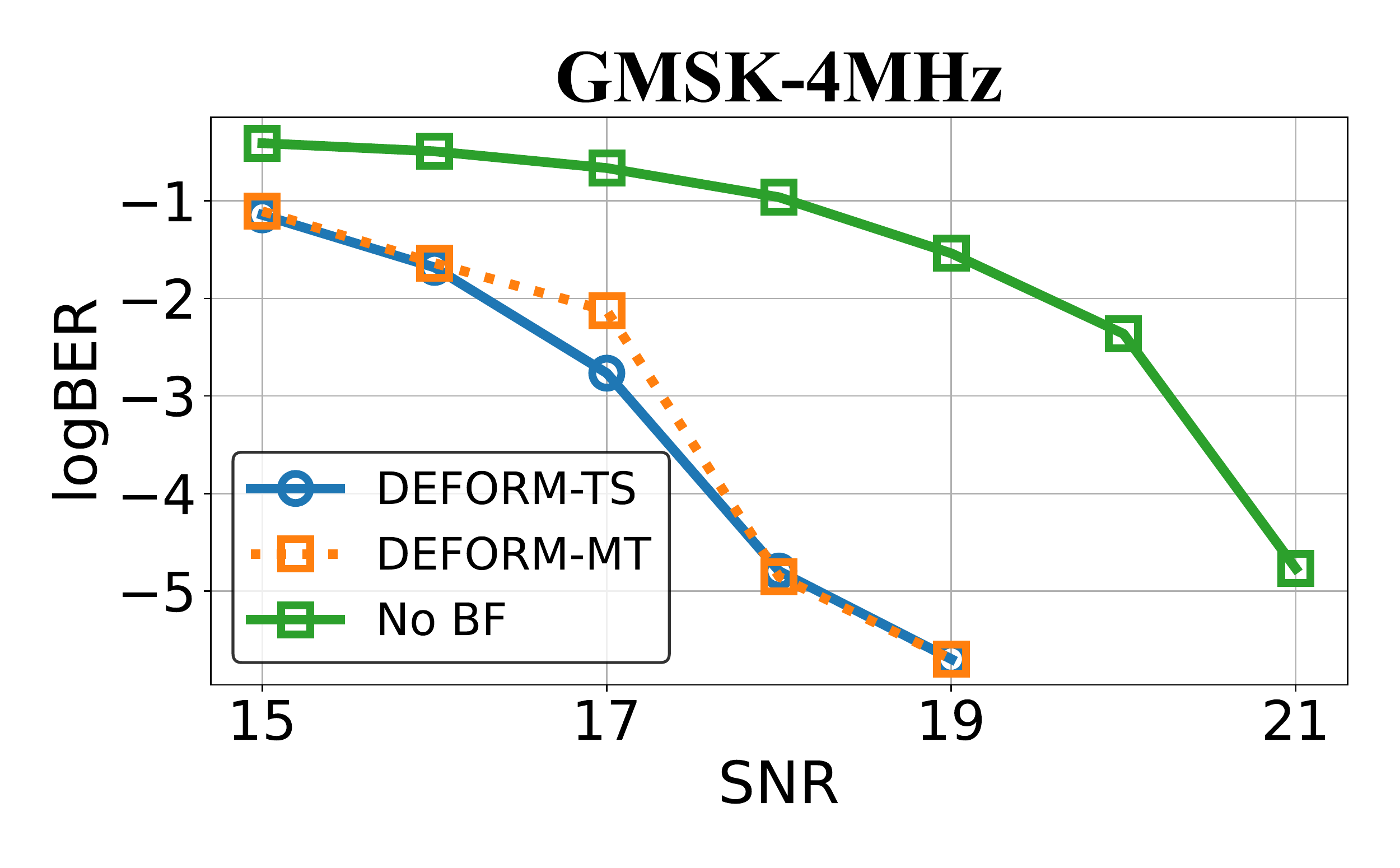}
        \includegraphics[width=0.19\linewidth]{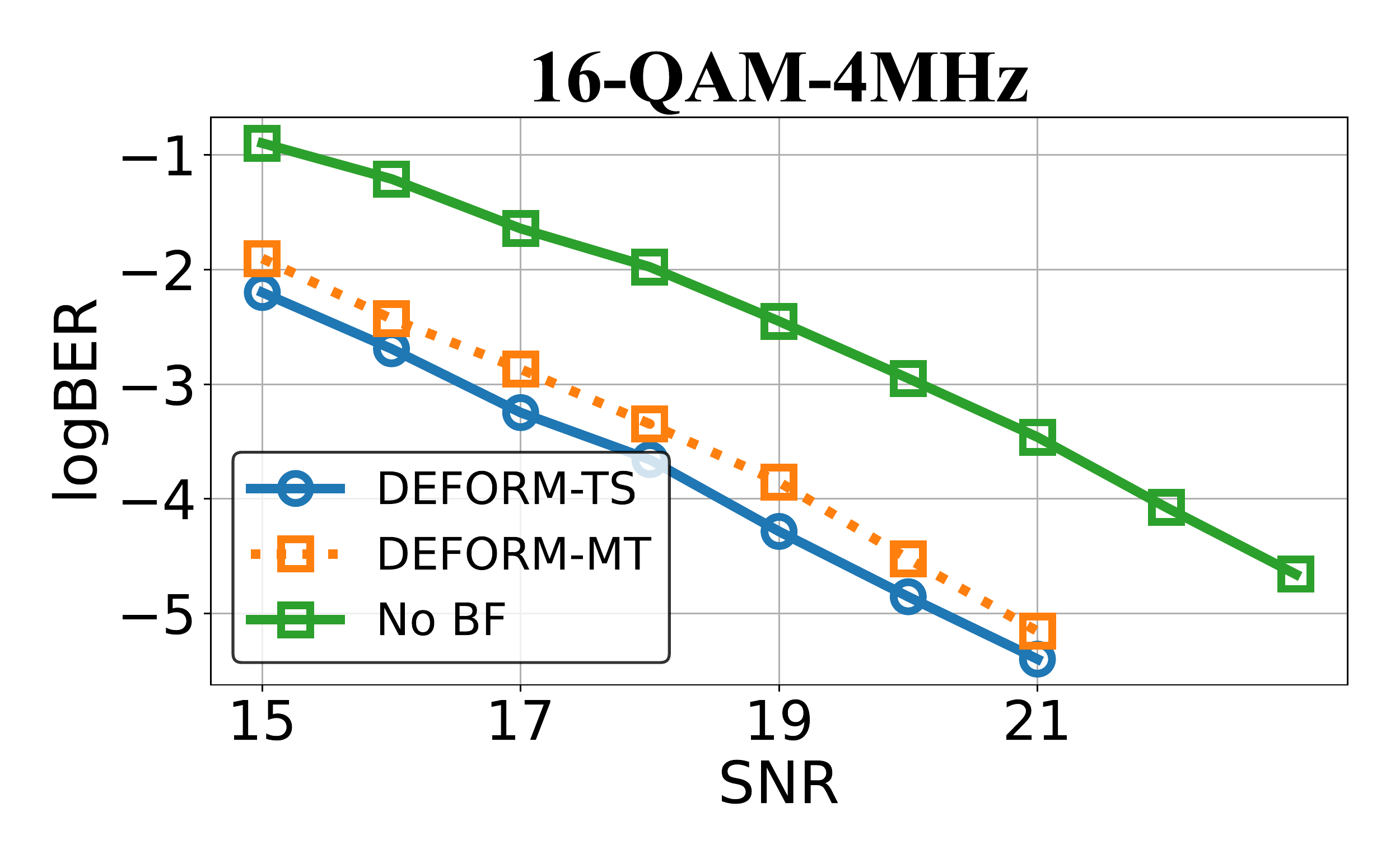}
    \end{subfigure}
    \begin{subfigure}{\linewidth}
        \includegraphics[width=0.19\linewidth]{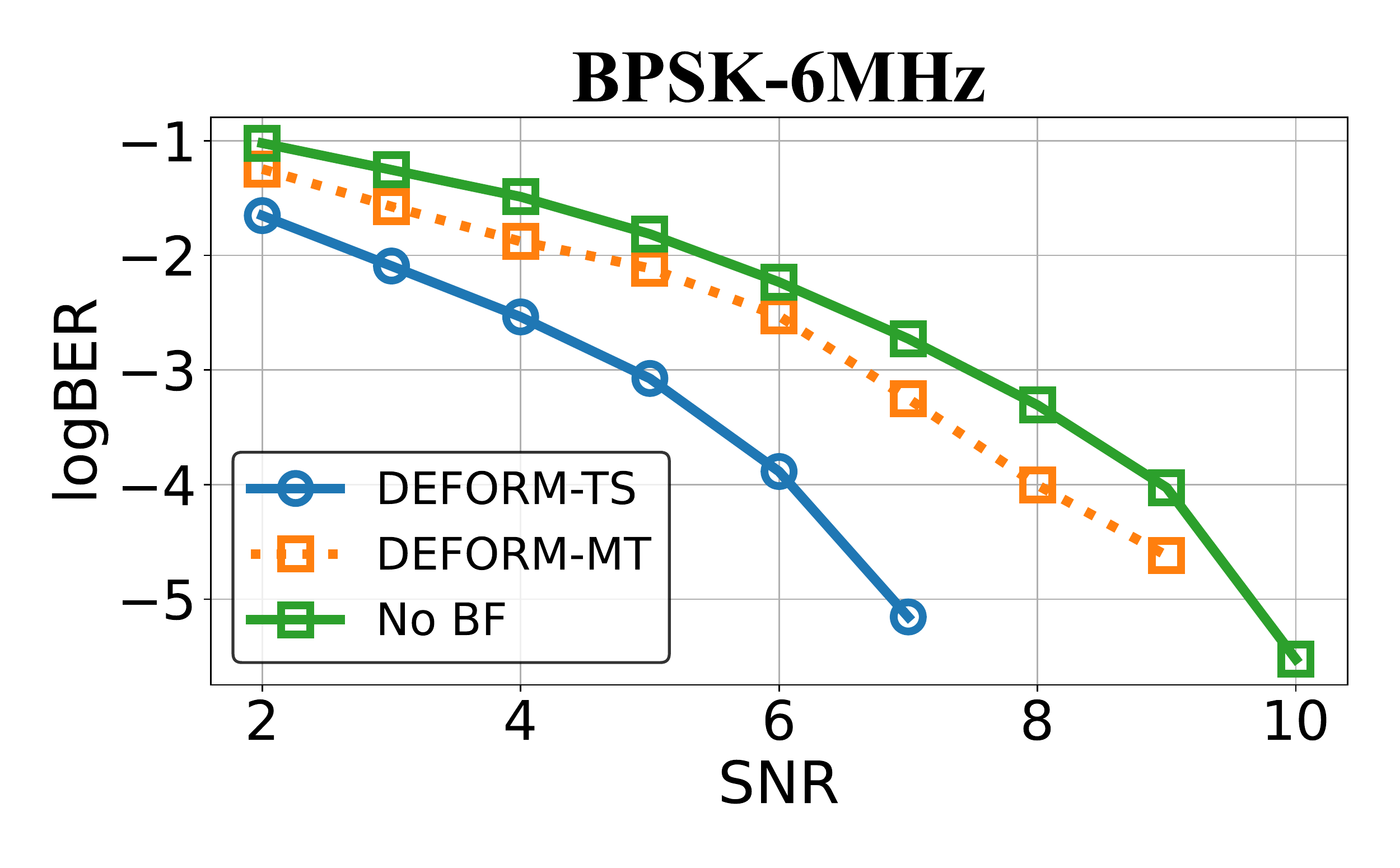}
        \includegraphics[width=0.19\linewidth]{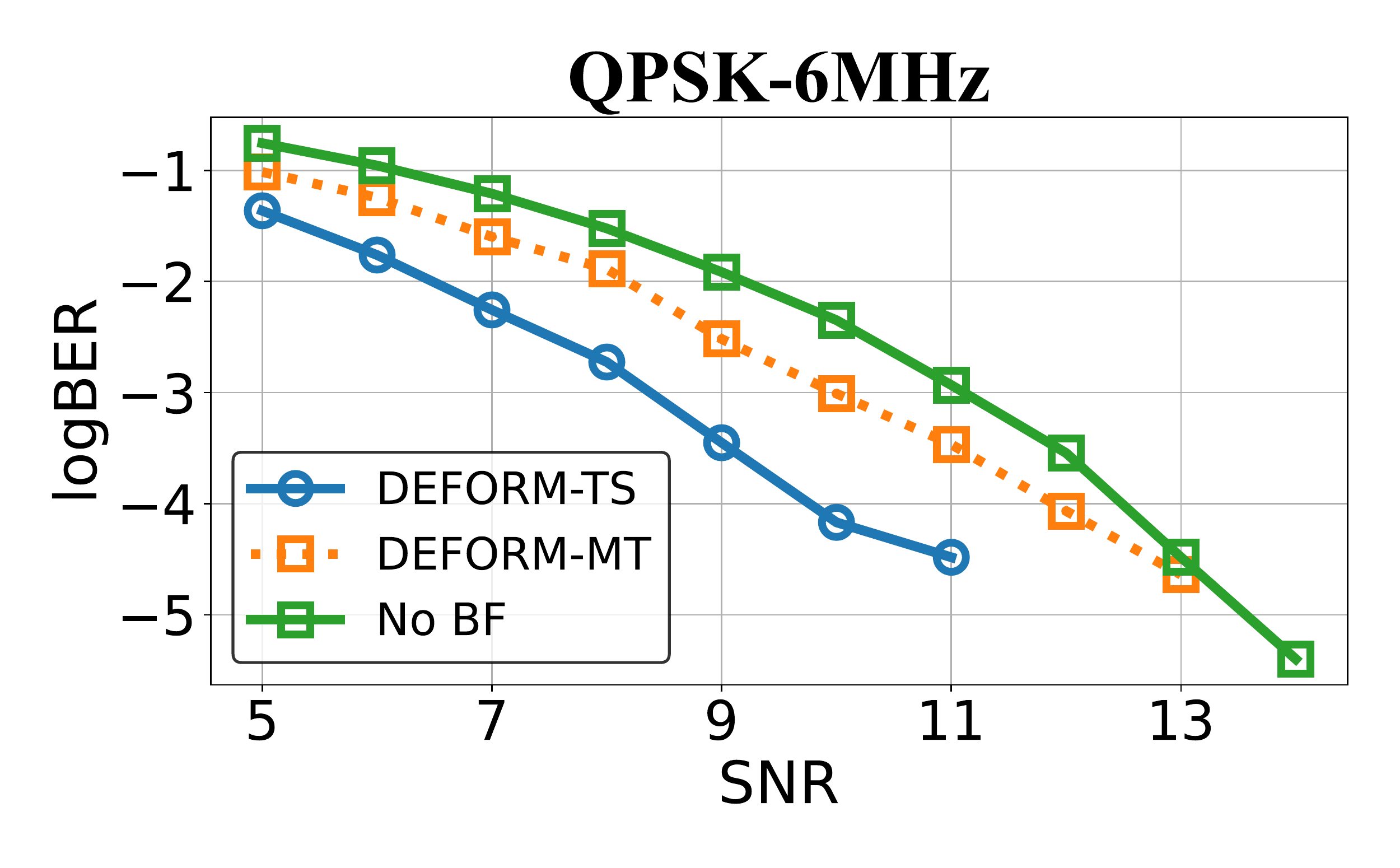}
        \includegraphics[width=0.19\linewidth]{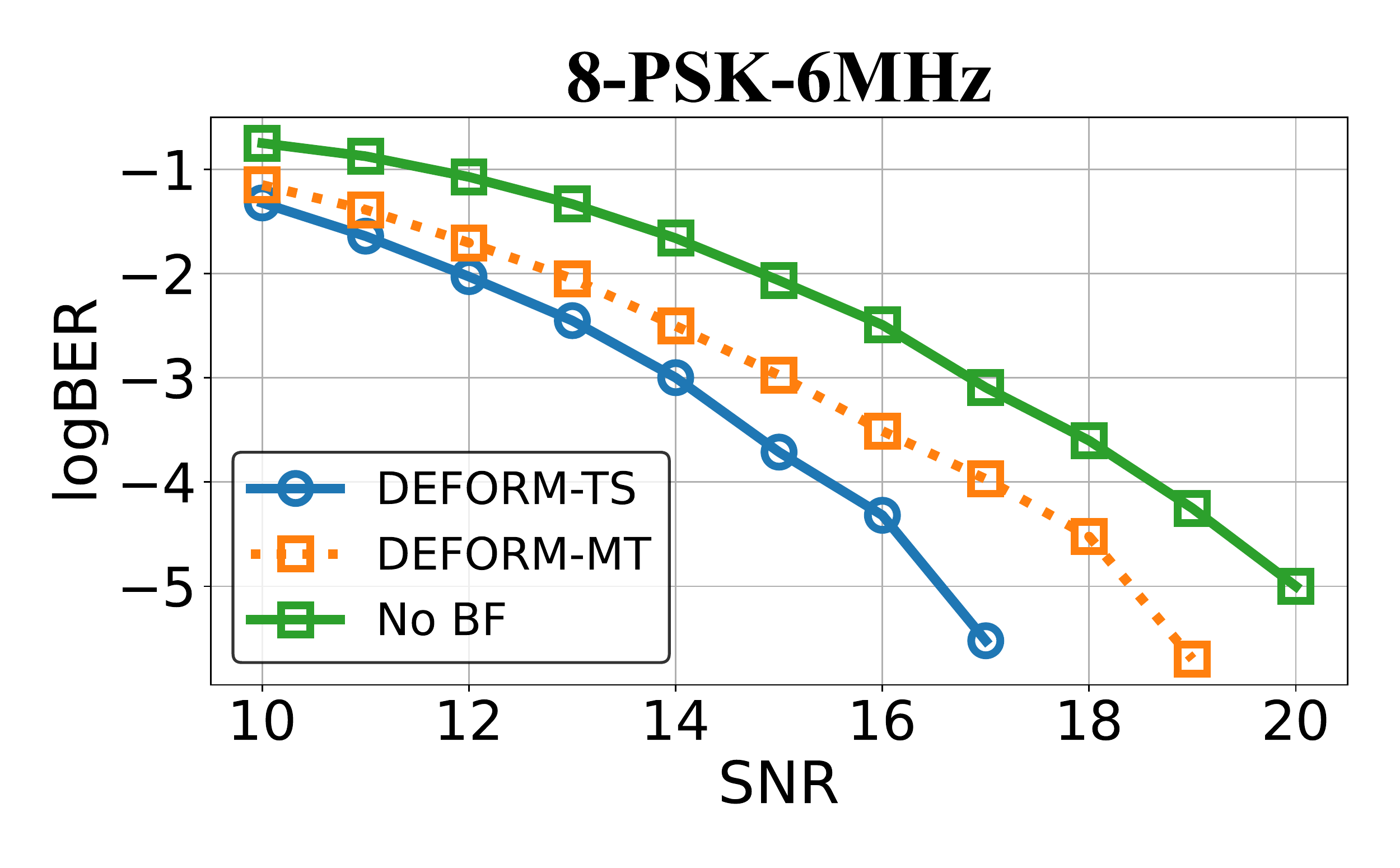}
        \includegraphics[width=0.19\linewidth]{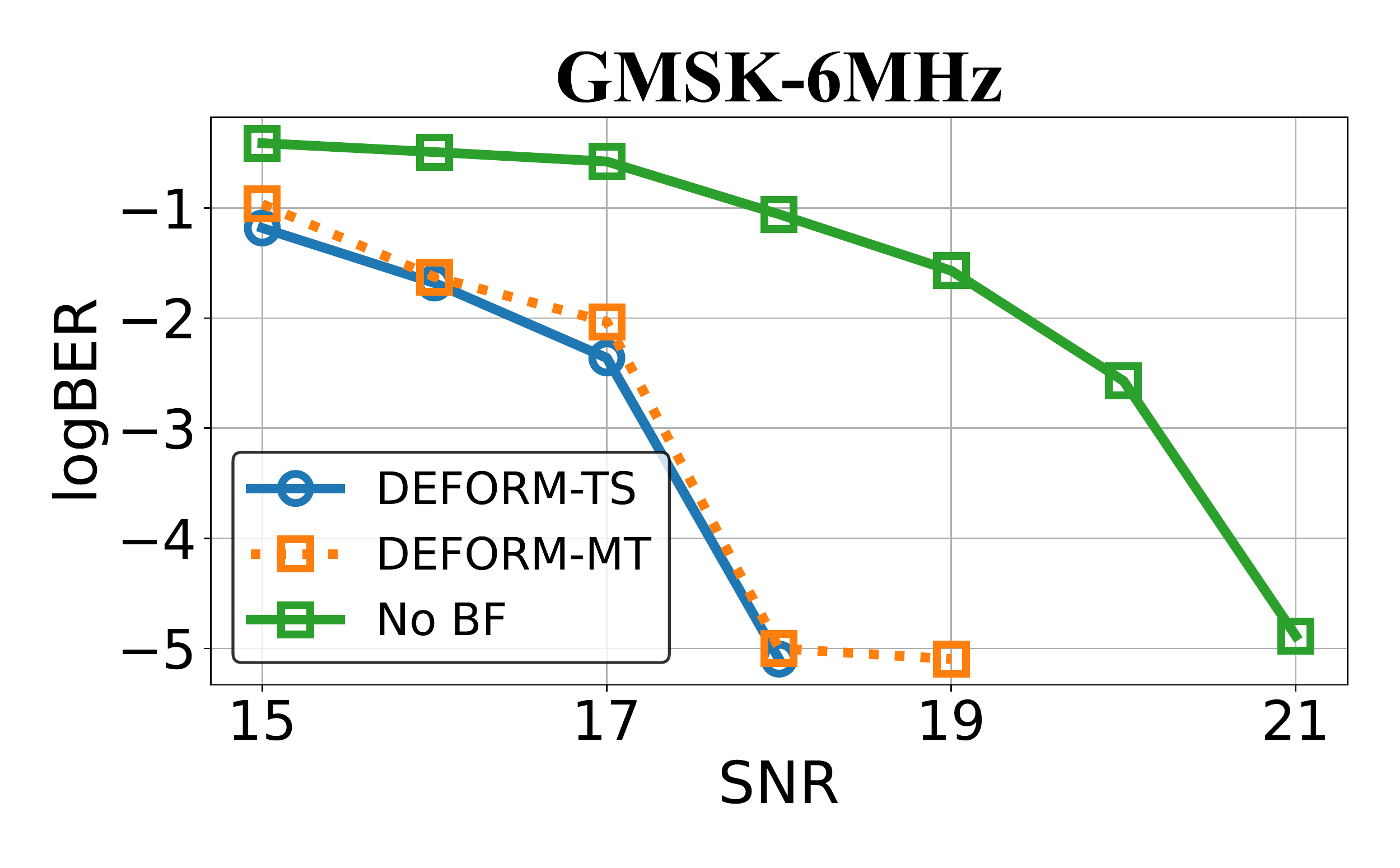}
        \includegraphics[width=0.19\linewidth]{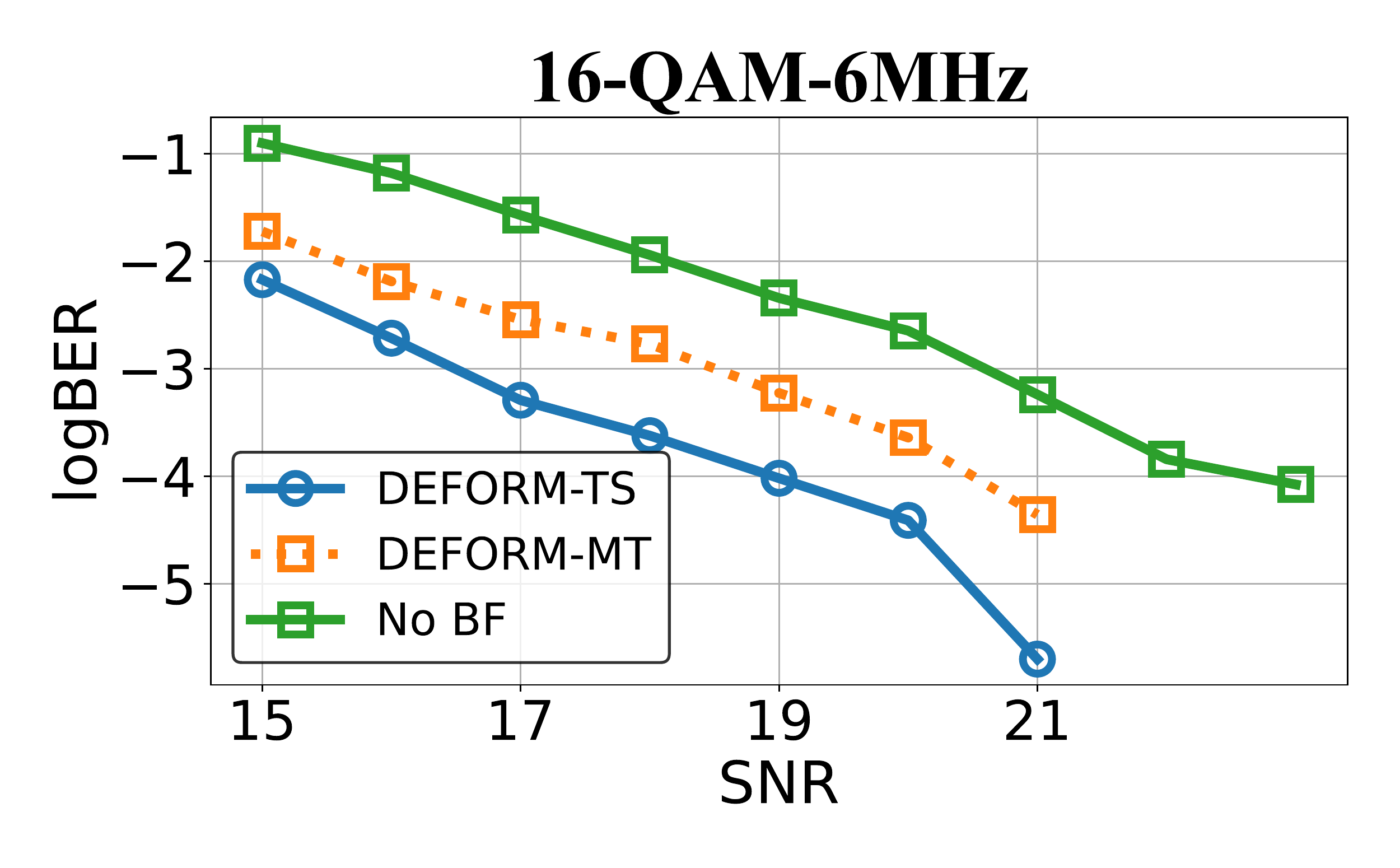}
    \end{subfigure}
    \caption{BER analysis of over-the-cables communications for different modulation schemes and RX bandwidths. \system{} is evaluated with Temporal Smoothing (\system-TS) and Multi-Trial Averaging (\system{}-MT).}
    \label{fig:symcab_multimod}
\end{figure*}

\bfpara{Evaluation for Signal Features Universality} 
As mentioned above, we consider various settings of modulations and bandwidths to evaluate the universality of \system{}.  We emulate the slow fading effect of the real wireless channels by artificially adjusting the phases of received signals following a stair-step pattern for the relative phase $\Delta_{\theta}$ as illustrated in ~\Cref{fig:fading_pattern} to cover the whole $2\pi$ phase range. \Cref{fig:symcab_multimod} shows the Bit Error Rate (BER) analysis where \system{} is evaluated with both \system{-TS} and \system{}-MT approaches, and compared with non-beamforming. It is evident that for all combinations of modulation and RX bandwidth, \system{}-TS can provide a 3 dB SNR gain compared to using a single RX branch, with the only exception of GMSK-2MHz where the gain for BER $=10^{-5}$ is approximately 2.5 dB. Meanwhile, \system{}-MT also achieves 3 dB gain for 1 and 2 MHz RX bandwidths, and about 2-3 dB gain for 4 MHz bandwidth. However, with 6 MHz bandwidth, the SNR gain of \system{}-MT declines to 1 dB for BPSK and QPSK at BER $\geq10^{-4}$, and to 2 dB for 8-PSK and 16-QAM. Interestingly, it still performs equally well with \system{}-TS for GMSK in 6 MHz RX bandwidth, both having a 3 dB gain for all BER levels.

\begin{figure}
\begin{minipage}[t]{0.61\linewidth}
    \vspace{0pt}
    \centering
    \includegraphics[width=\linewidth]{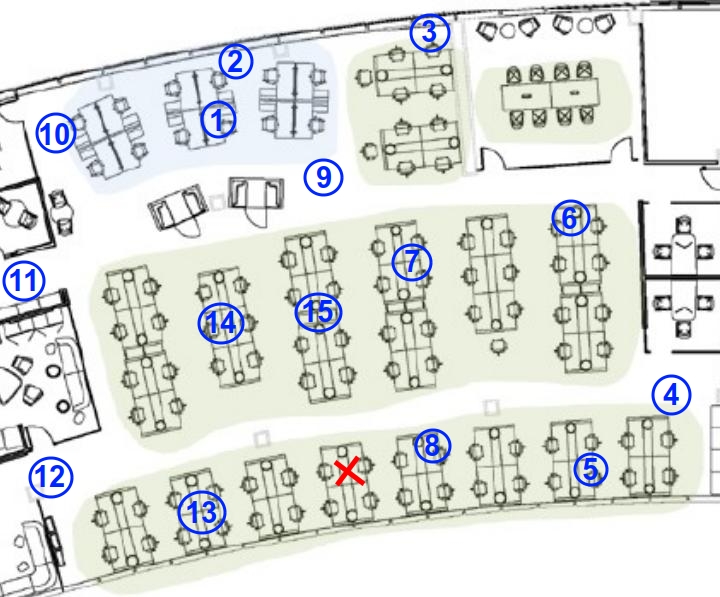}
\end{minipage}
\hspace{1em}
\begin{minipage}[t]{0.23\linewidth}
\vspace{0pt}
\footnotesize
    
    \begin{tabular}{m{1pt}m{1.2cm}}
    1 & GMSK 1M\\
    2 & 8-PSK 4M \\
    3 & QPSK 6M \\
    4 & QPSK 1M \\
    5 & QPSK 2M \\
    6 & 8-PSK 4M \\
    7 & GMSK 6M \\
    8 & 16-QAM 2M\\
    9 & QPSK 4M\\
     10 & BPSK 2M\\ 
     11 & BPSK 4M\\
    12 & GMSK 4M\\
    13 & GMSK 2M\\
    14 & 16-QAM 6M\\
    15 & QPSK 4M\\
    \end{tabular}
\end{minipage}
    \caption{Non-LOS over-the-air testbed environment in a 50 ft. $\times$ 100ft. lab office (left) and RF experiment settings (right).  TX locations are marked by the \textcolor{blue}{blue numbered circles}, while the fixed location of RX is marked by the \textcolor{red}{red cross}.}
    \label{fig:non_los_environment}
\end{figure}

\subsection{Over-the-air Evaluation}
\begin{figure*}
    \centering
    \begin{subfigure}{\linewidth}
        \includegraphics[width=0.19\linewidth]{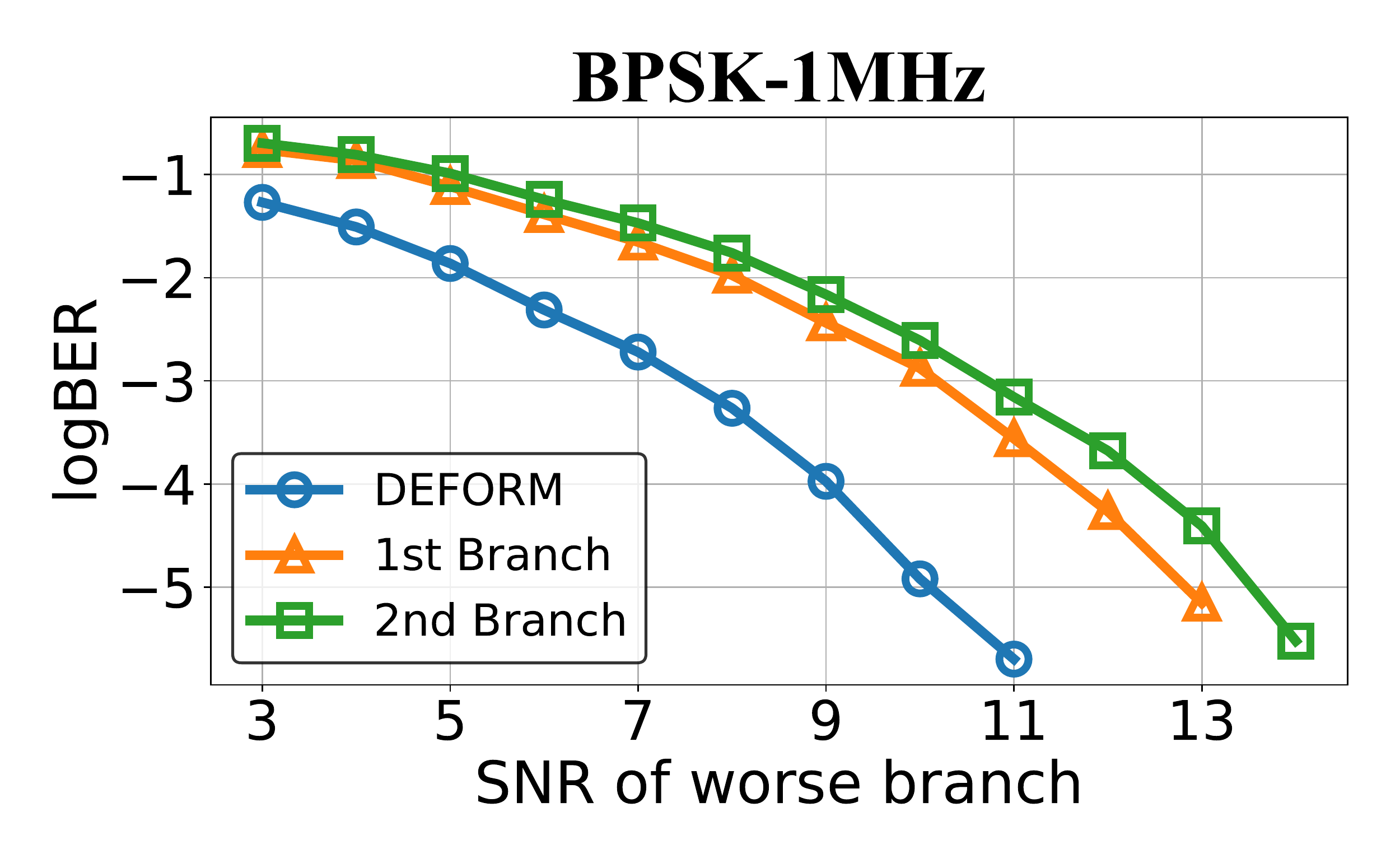}
        \includegraphics[width=0.19\linewidth]{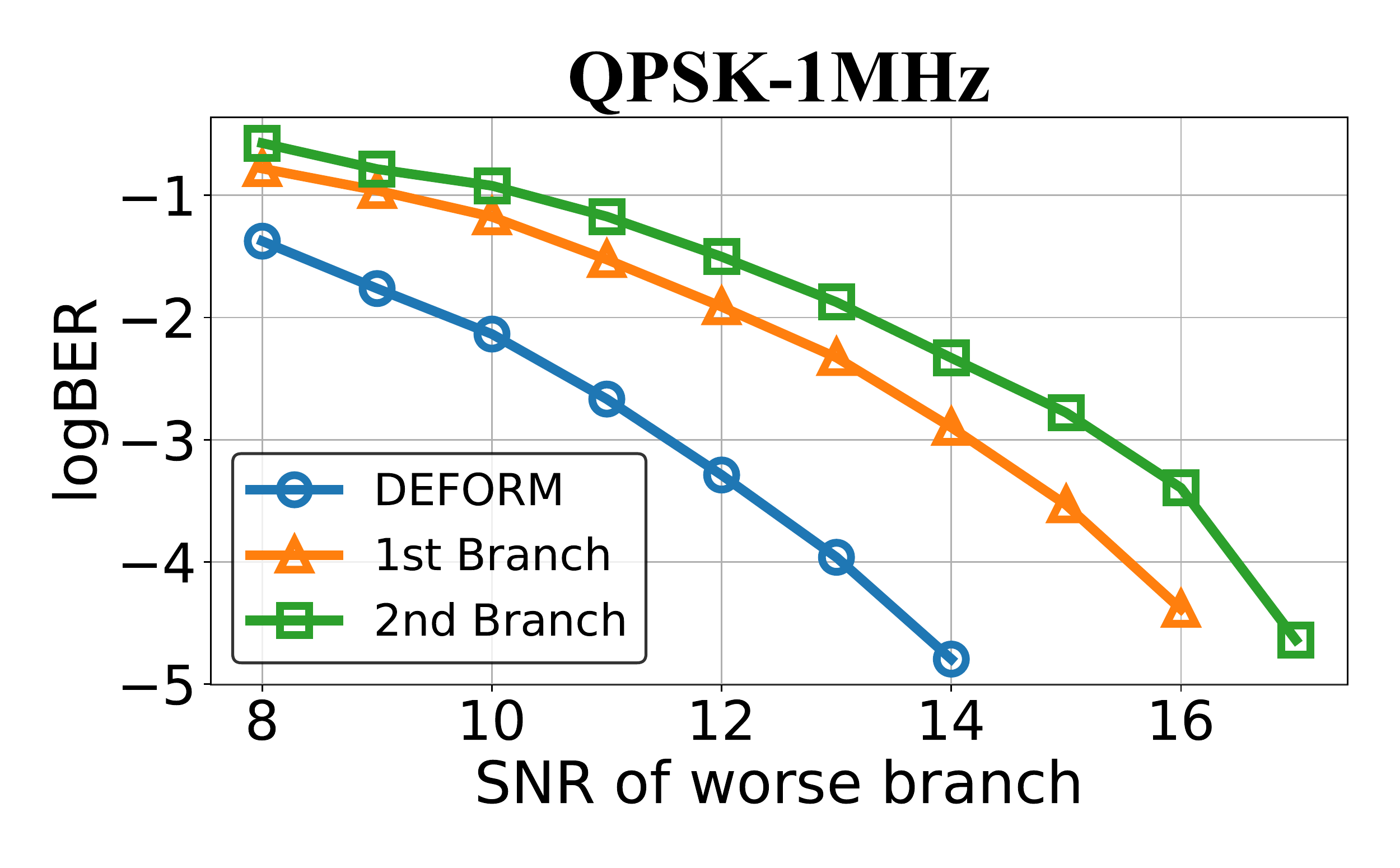}
        \includegraphics[width=0.19\linewidth]{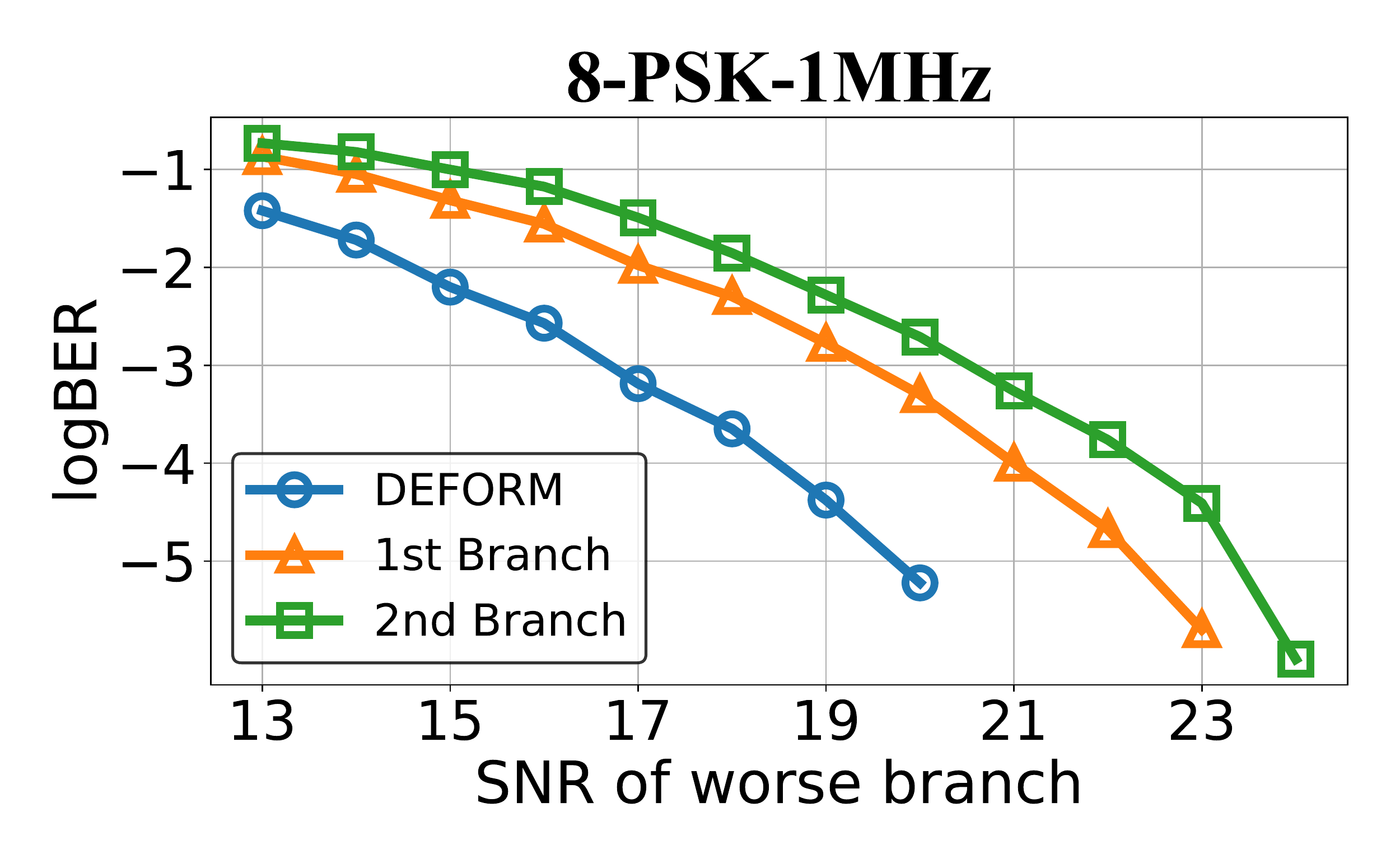}
        \includegraphics[width=0.19\linewidth]{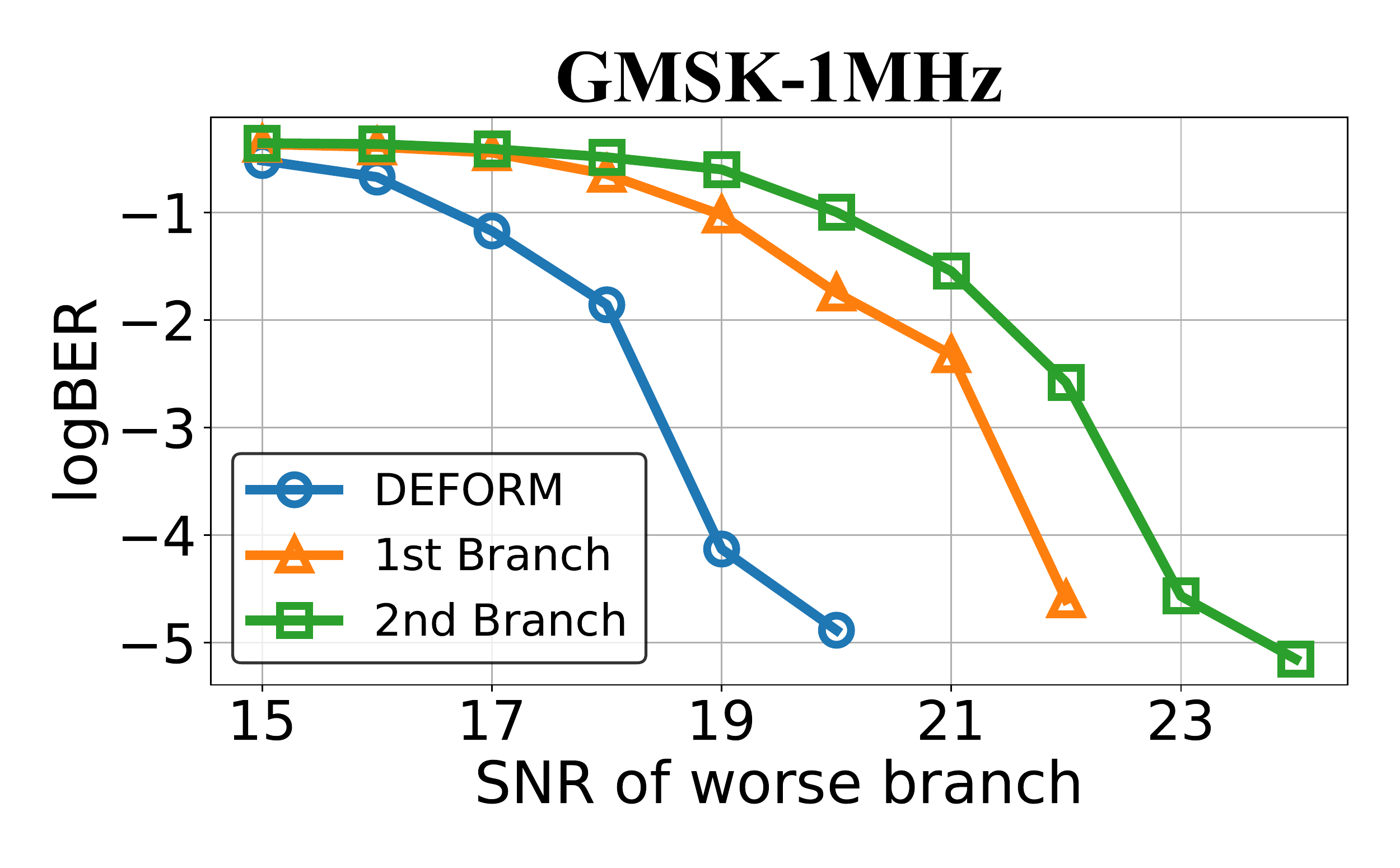}
        \includegraphics[width=0.19\linewidth]{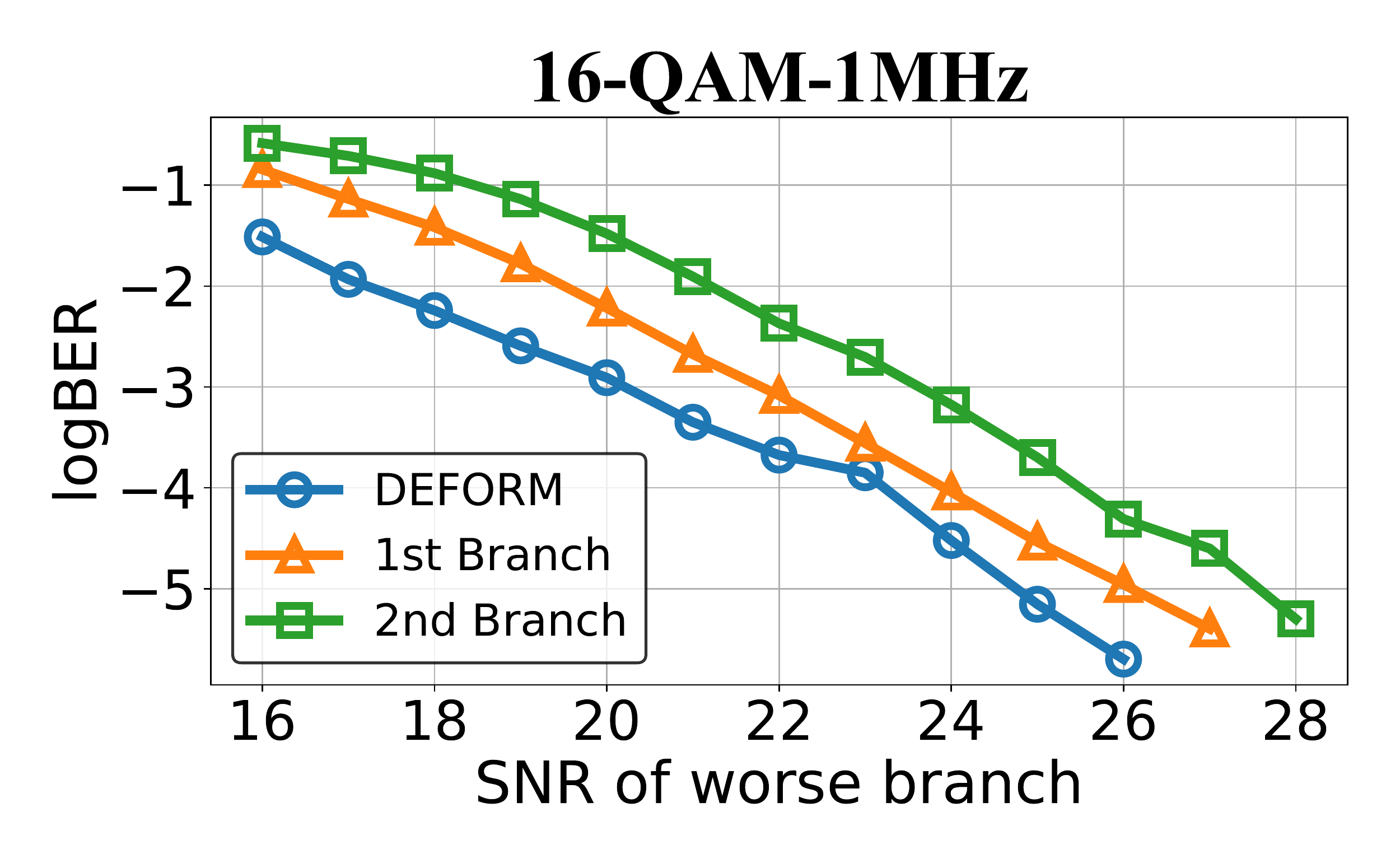}
    \end{subfigure}
    \begin{subfigure}{\linewidth}
        \includegraphics[width=0.19\linewidth]{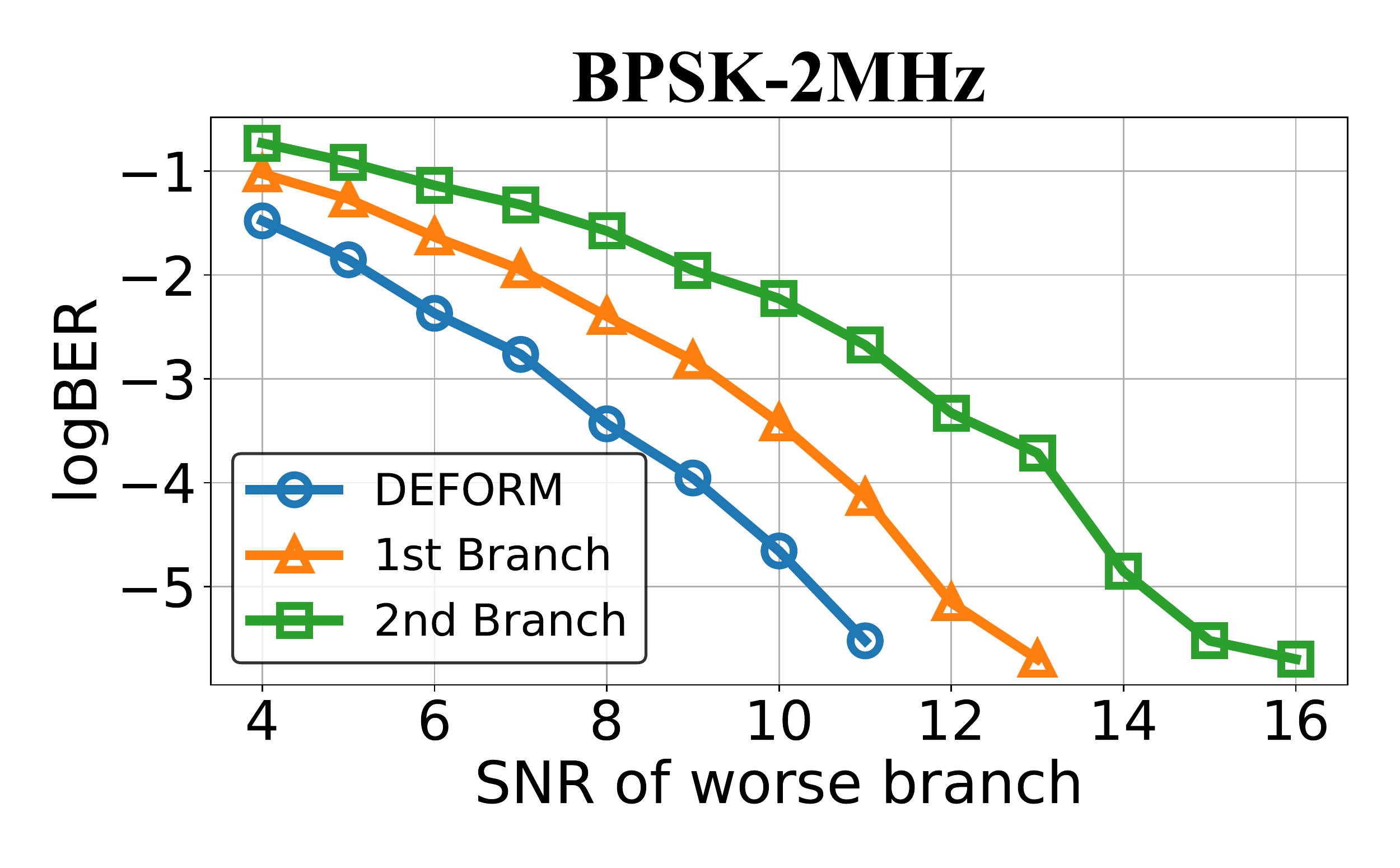}
        \includegraphics[width=0.19\linewidth]{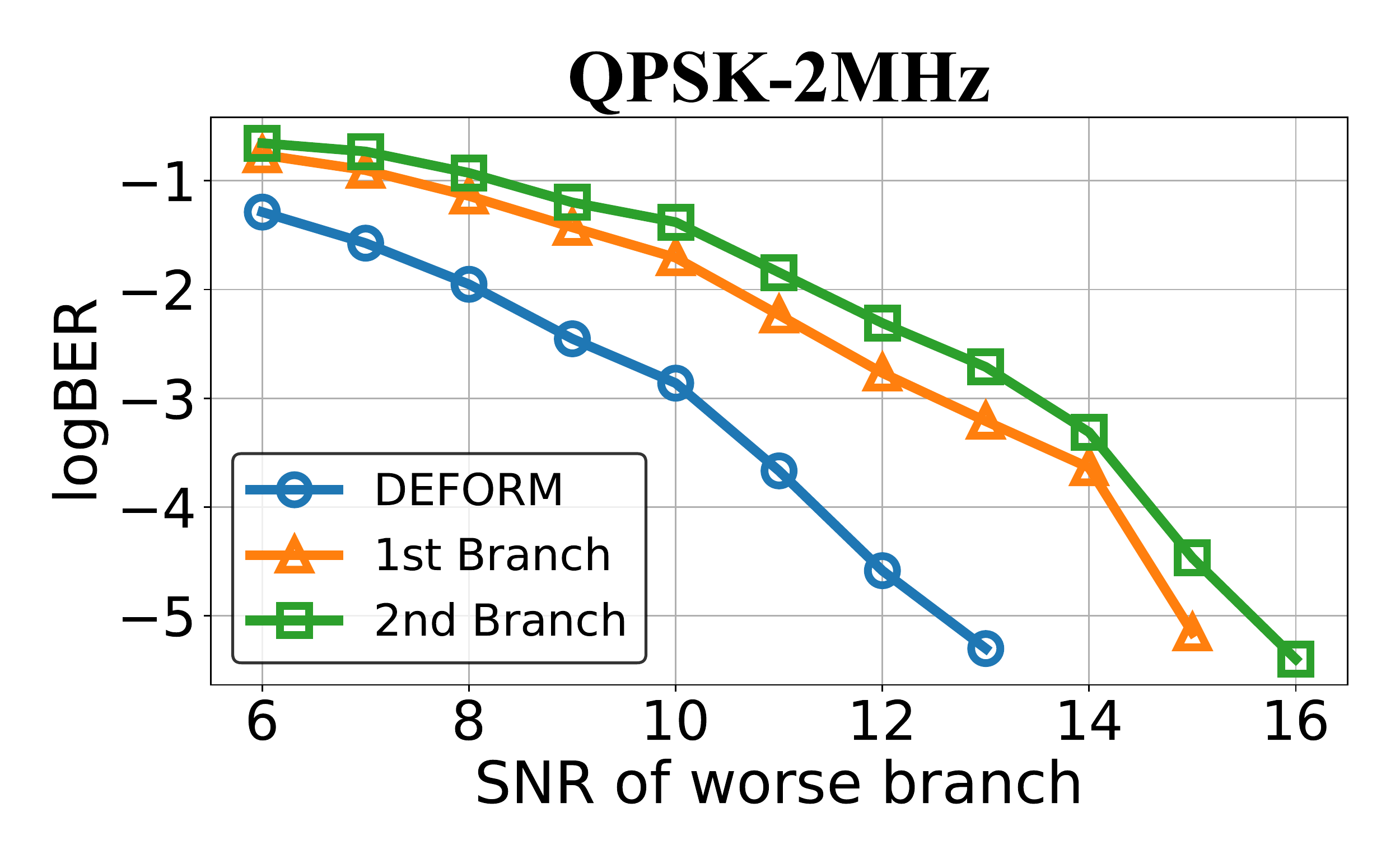}
        \includegraphics[width=0.19\linewidth]{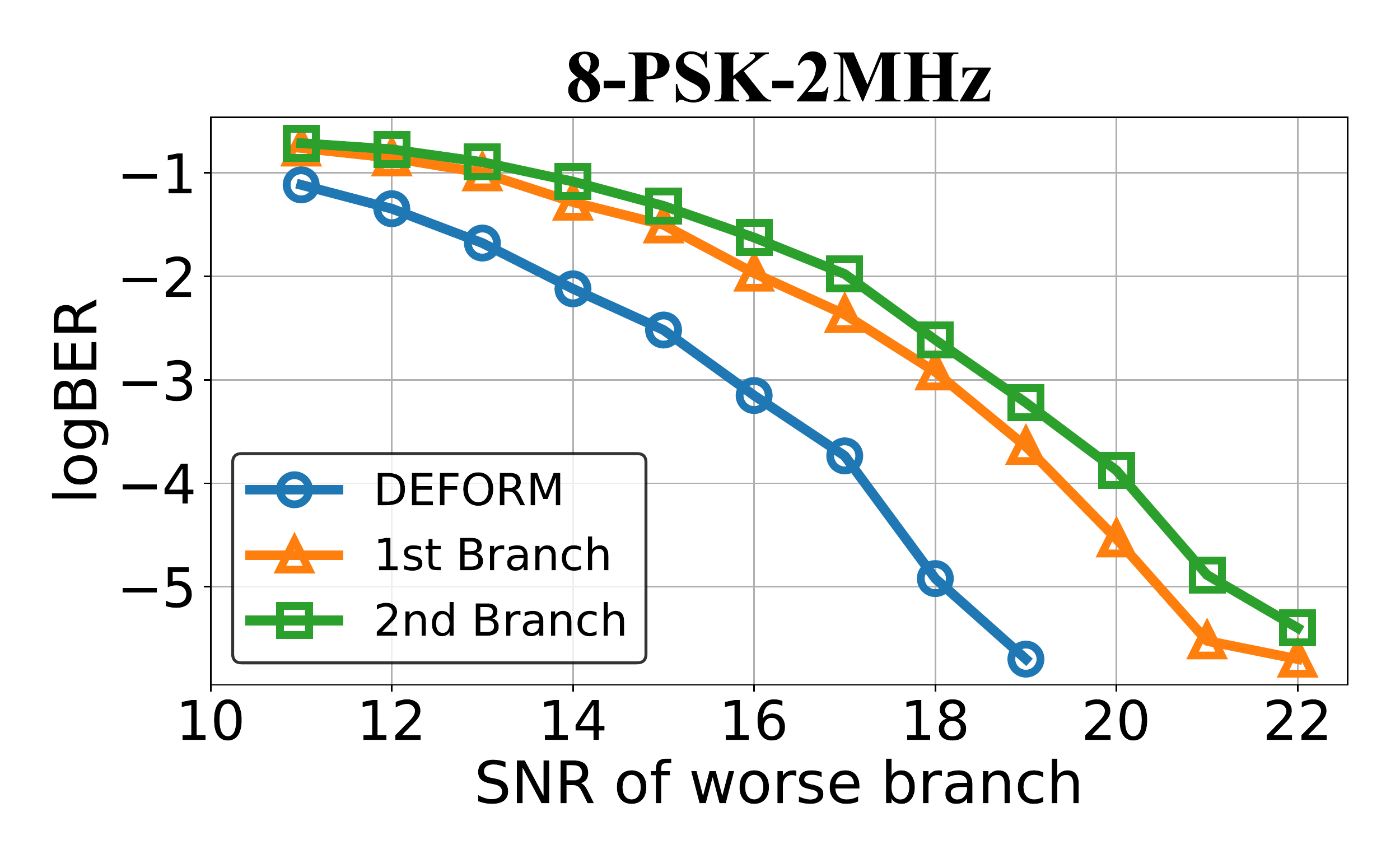}
        \includegraphics[width=0.19\linewidth]{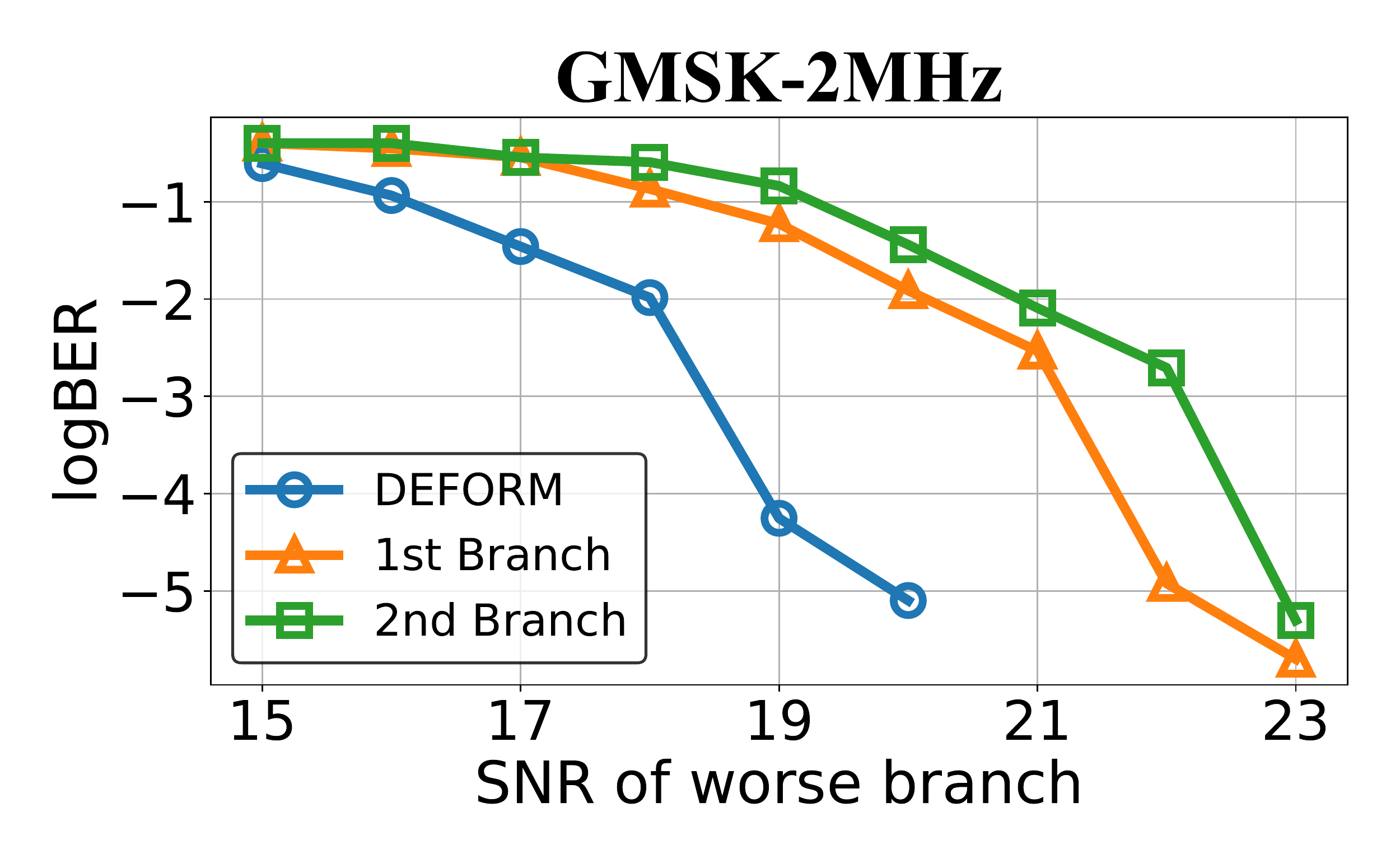}
        \includegraphics[width=0.19\linewidth]{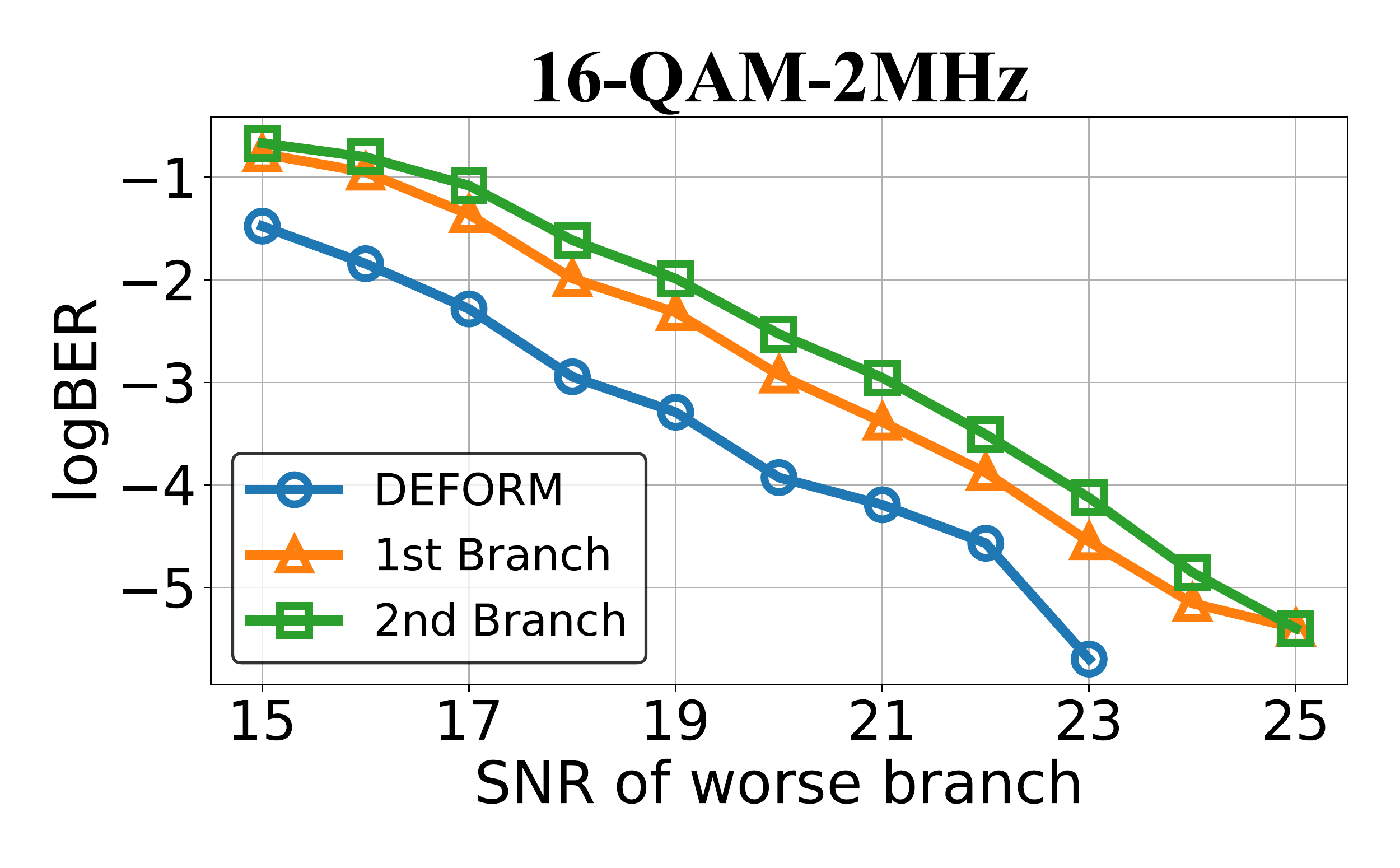}
    \end{subfigure}
    \begin{subfigure}{\linewidth}
        \includegraphics[width=0.19\linewidth]{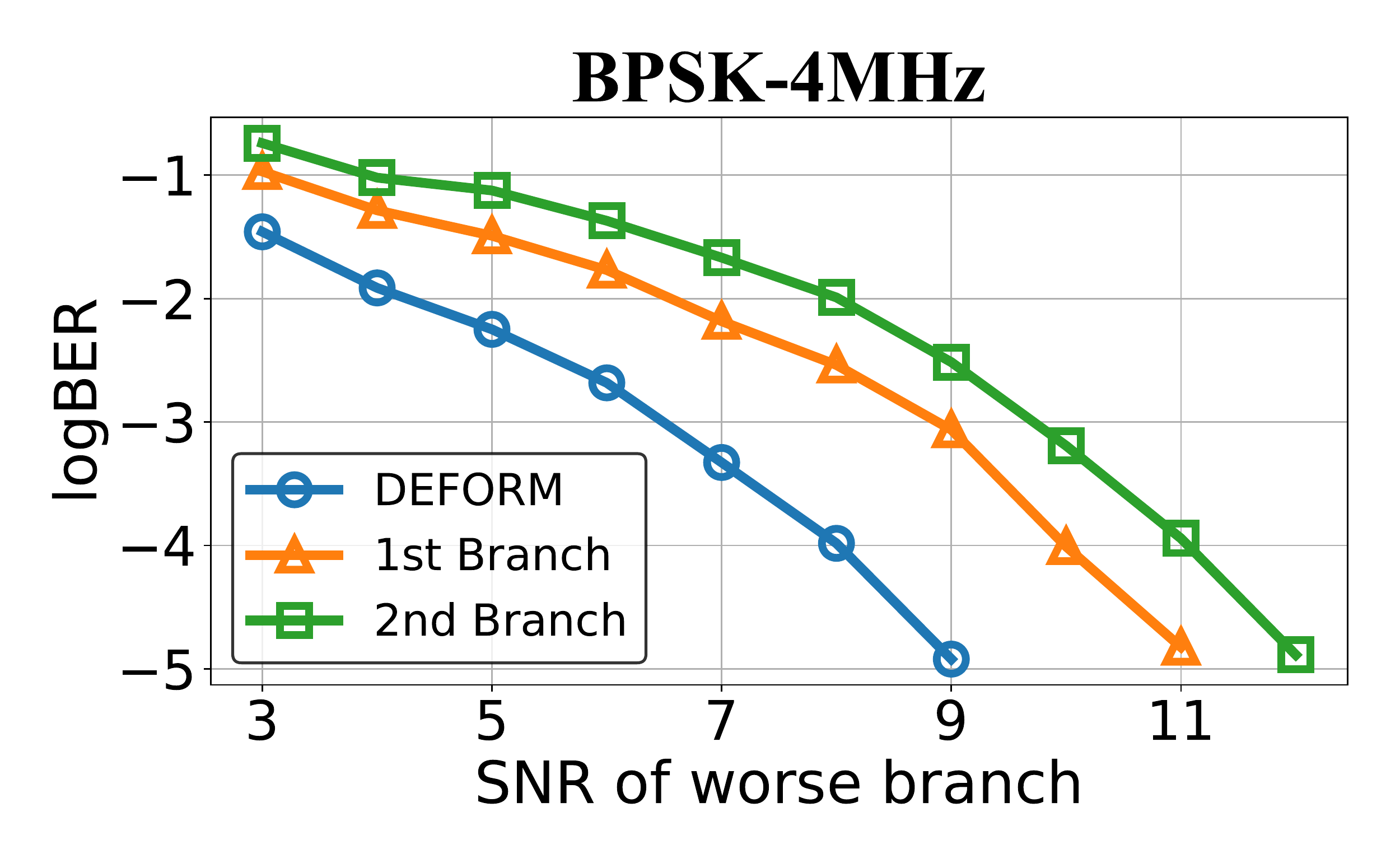}
        \includegraphics[width=0.19\linewidth]{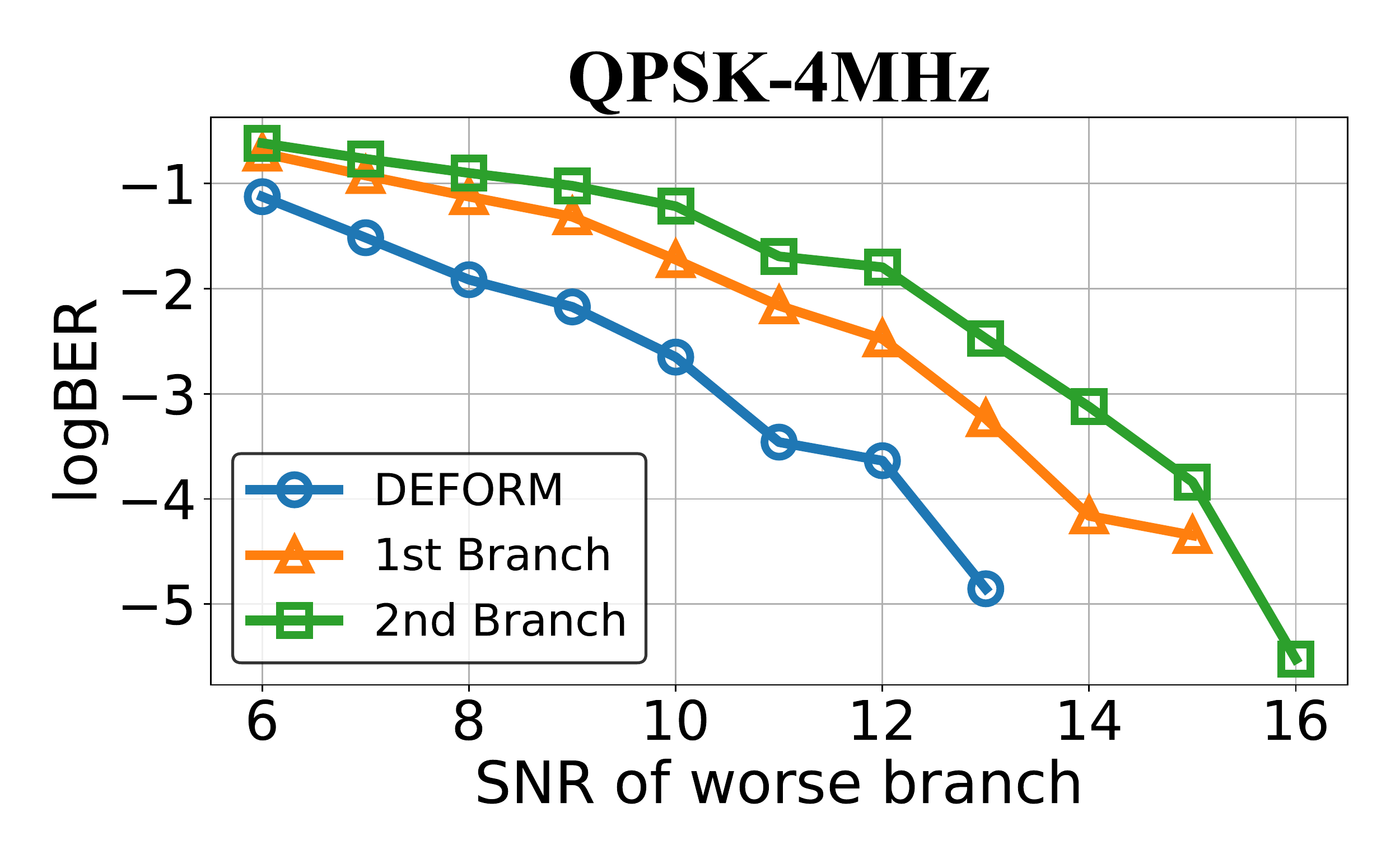}
        \includegraphics[width=0.19\linewidth]{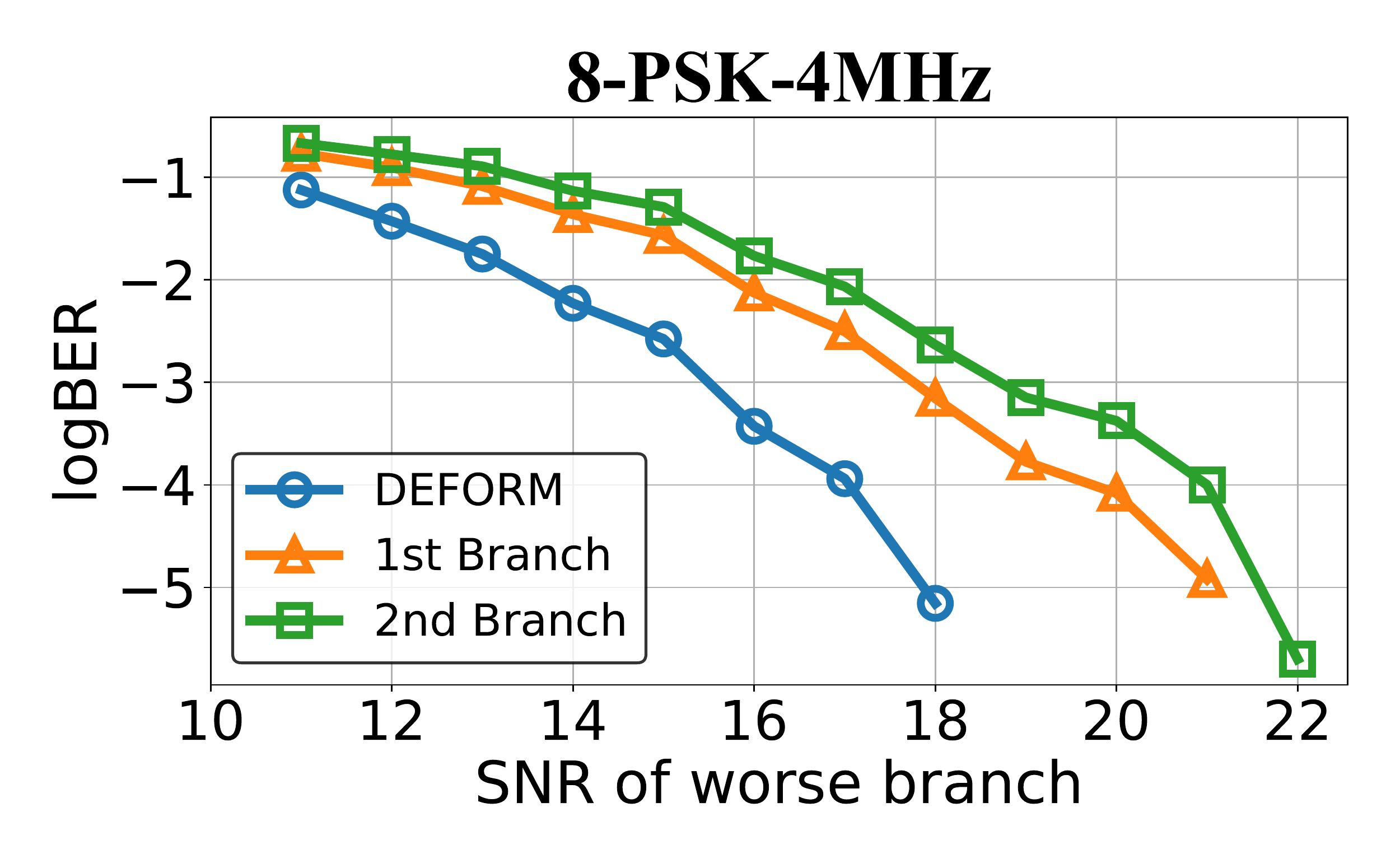}
        \includegraphics[width=0.19\linewidth]{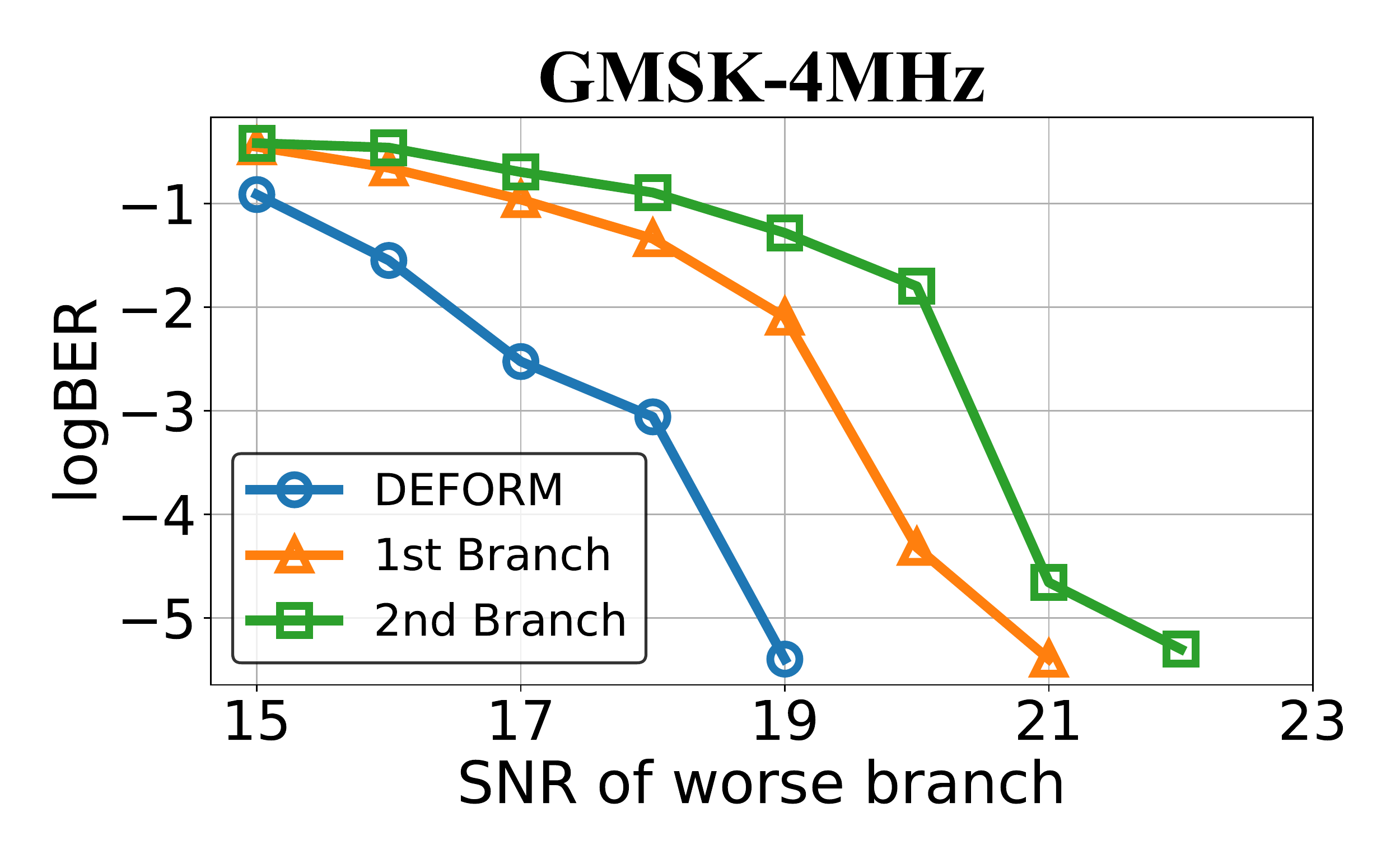}
        \includegraphics[width=0.19\linewidth]{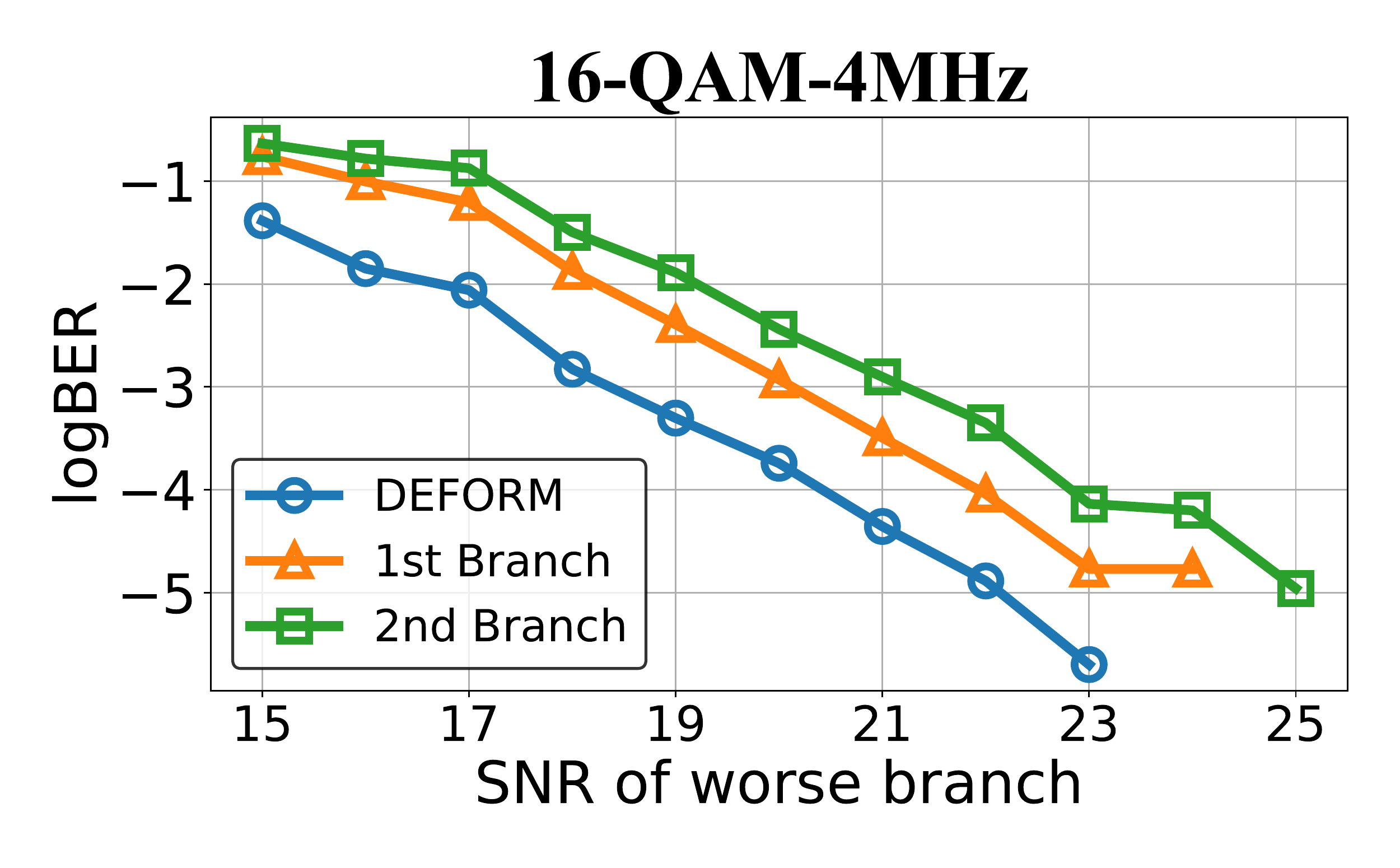}
    \end{subfigure}
    \begin{subfigure}{\linewidth}
        \includegraphics[width=0.19\linewidth]{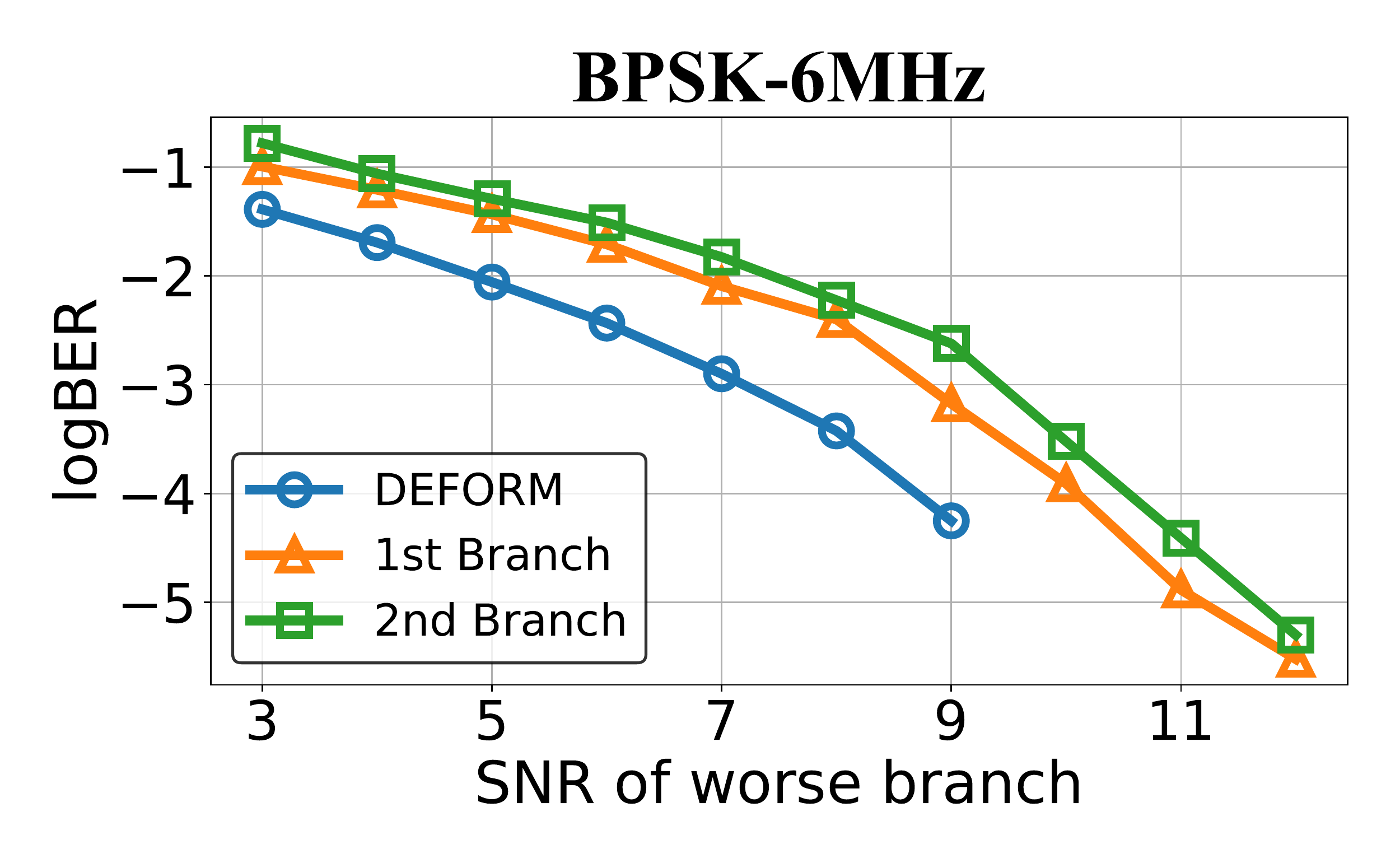}
        \includegraphics[width=0.19\linewidth]{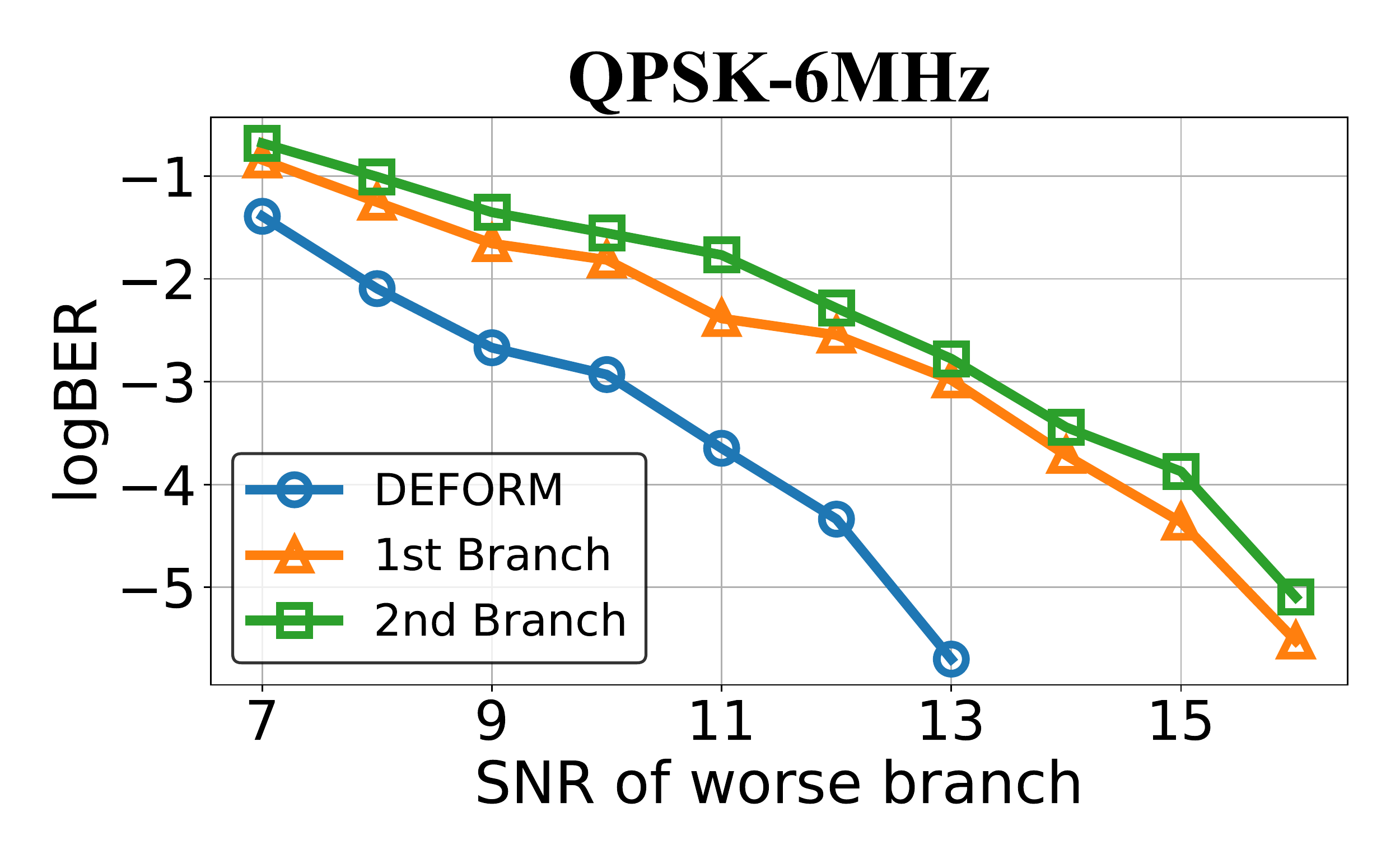}
        \includegraphics[width=0.19\linewidth]{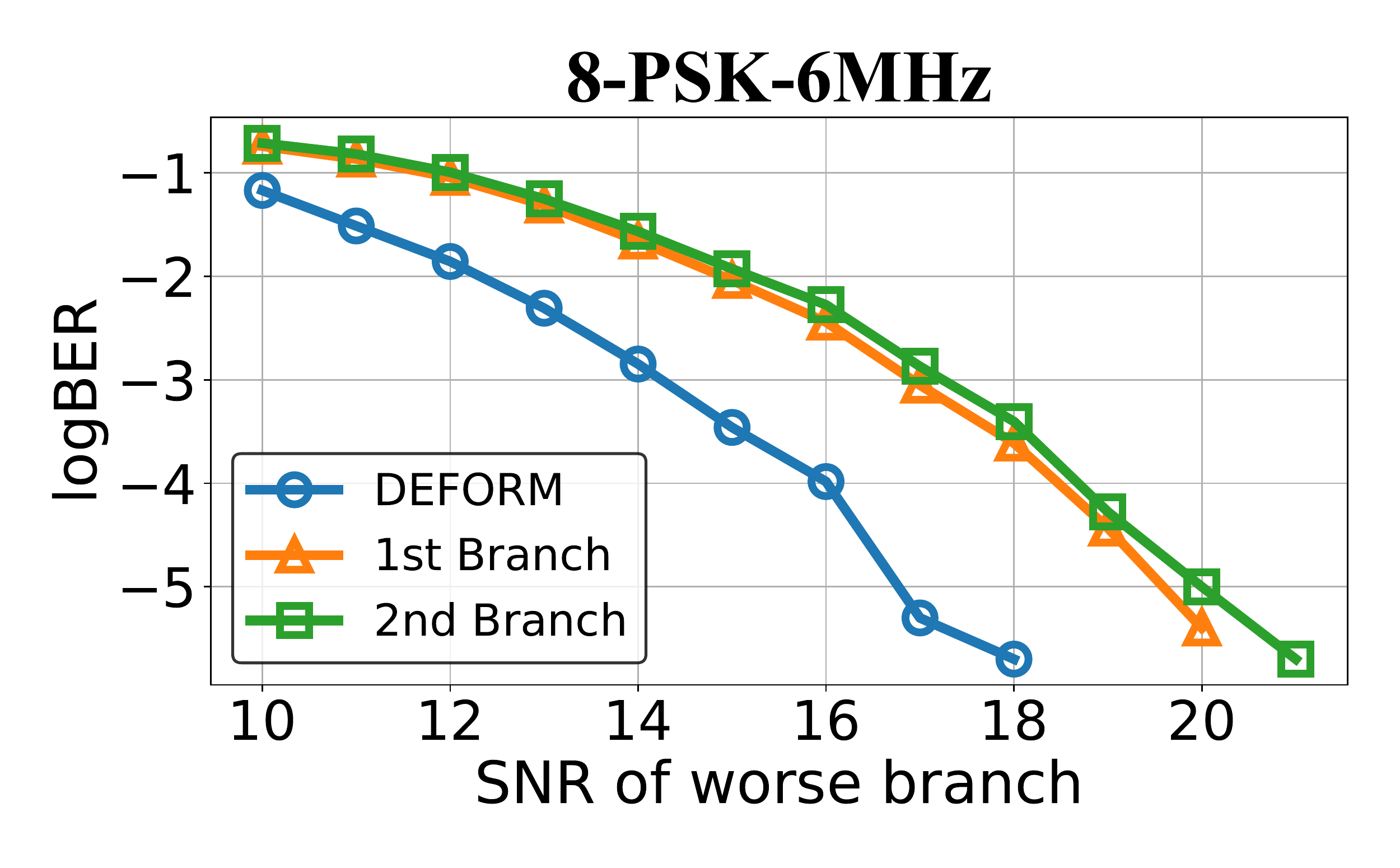}
        \includegraphics[width=0.19\linewidth]{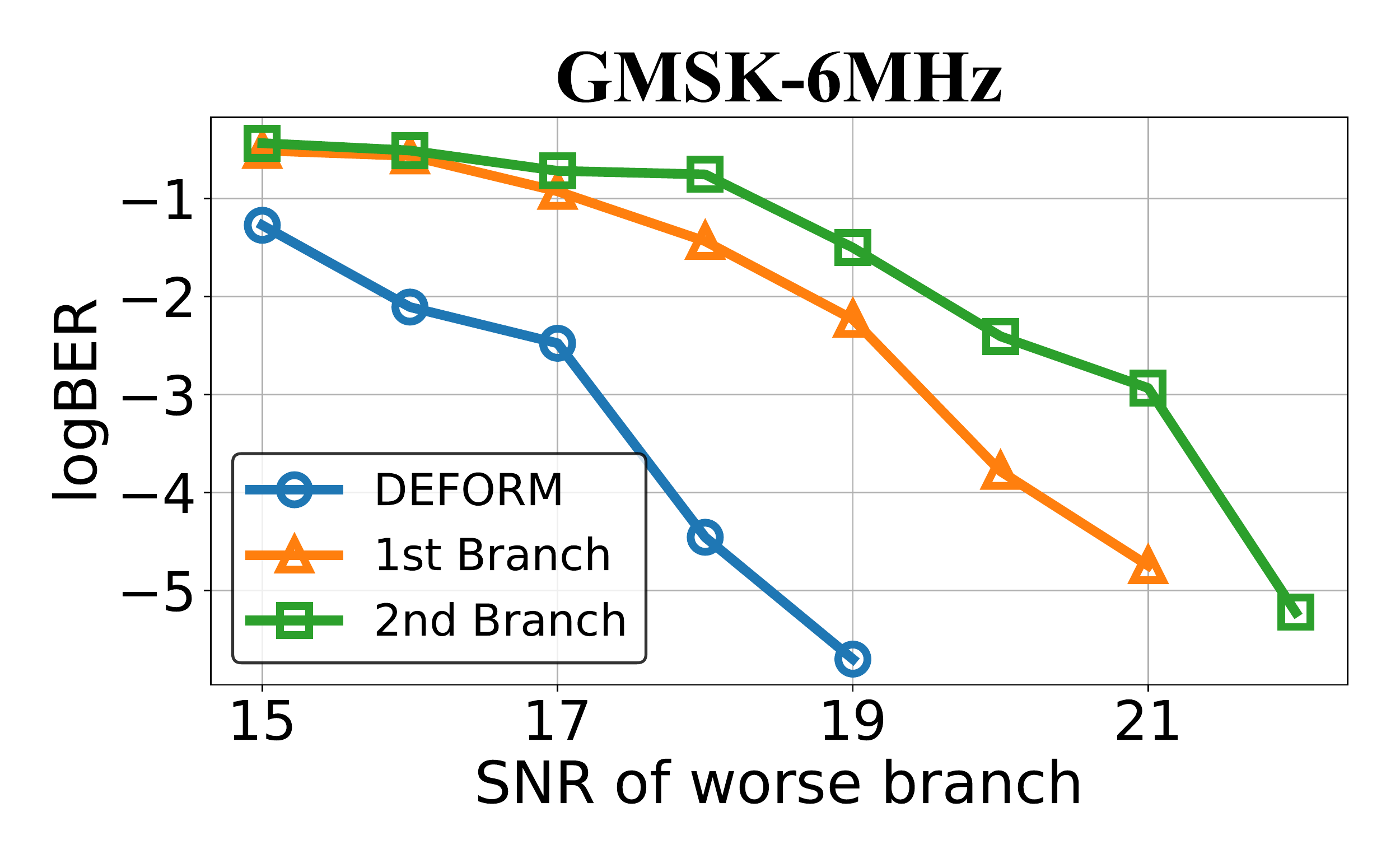}
        \includegraphics[width=0.19\linewidth]{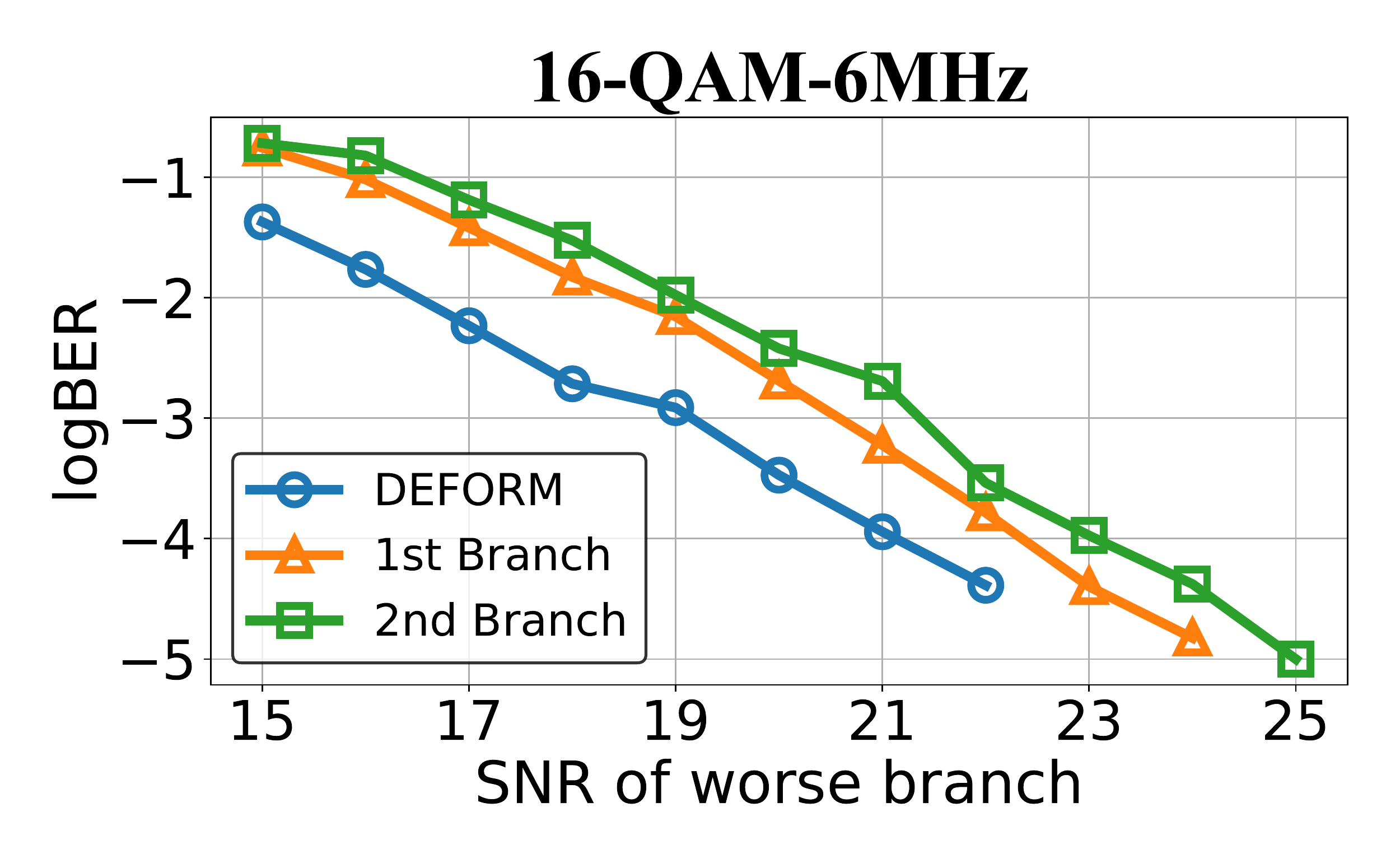}
    \end{subfigure}
    \caption{BER analysis of over-the-air communications with inherent LOS on different modulations and RX bandwidths.}
    \label{fig:ota_los_multimod}
\end{figure*}

We conduct experiments to assess \system{}'s performance in two over-the-air communication scenarios: When Line of Sight (LOS) is inherent and when LOS is blocked by various types of objects.
\bfpara{Over-the-air with inherent LOS} We create the LOS by positioning the TX and RX devices to ensure no blocking objects in the direct path. We use similar TX settings as in over-the-cables experiments. The RX has two receiving branches tuned for the RX bandwidths of 1, 2, 4, 6 MHz with random center frequency shifting mentioned in \Cref{sec:cable_eval}. As over-the-air communications do not guarantee the same SNR for all of the RX branches, we observe the SNR of the worse receiving branch and adjust the RX gain for the measurements. Temporal Smoothing is used in \system{} for over-the-air evaluations because it performs better than Multi-Trial Averaging for large RX bandwidths. It is clear to see from the results in~\Cref{fig:ota_los_multimod} that in most cases, \system{} can achieve 2 dB of SNR gain when compared to the better receiving branch. Interestingly, in some cases the gain is approximately 3 dB, such as for QPSK-6MHz, or GMSK-6MHz at BER $=10^{-2}$. Moreover, when comparing with the worse branch, the gain can reach up to 4 dB, such as in QPSK-1MHz, 8PSK-1MHz and 8PSK-4MHz at BER $\geq 10^{-4}$. 


\bfpara{Over-the-air with no LOS} In the second over-the-air experiment, we place the multi-antenna RX at the location of the red cross in our testbed floorplan (\Cref{fig:non_los_environment}) and move the TX to different locations marked by the blue numbered circles. \Cref{fig:non_los_environment} only shows the basic floor plan/architecture of the office, and omits the presence of numerous large-sized objects, such as computers or lockers. At each predefined TX location, the TX transmits data packets with a randomly selected modulation, and the RX also randomly selected the RX bandwidth to receive the signal. The result of BER analysis is shown in \Cref{tab:non_los_evaluation}. It is clear that \system{} achieves significantly lower BER than any single branch in all measurements. For example, at location $\#4$, the measured BER of \system{} is $100$ times lower than branch 1 and $10$ times lower than branch 2. Meanwhile at location $\#13$, \system{}'s BER is approximately $30$ times lower than branch 1 and more than $100$ times lower than branch 2. These results demonstrate the universality of \system{} beamforming system, as it was trained on a single modulation (BPSK) and bandwidth (\SI{1}{\unit{\mega\hertz}}).

\section{Universal RF Beamforming-relay}\label{sec:relay}

\begin{figure}[t]
   \subcaptionbox{\label{fig:relay_map}}
   {
        \includegraphics[width=0.51\linewidth]{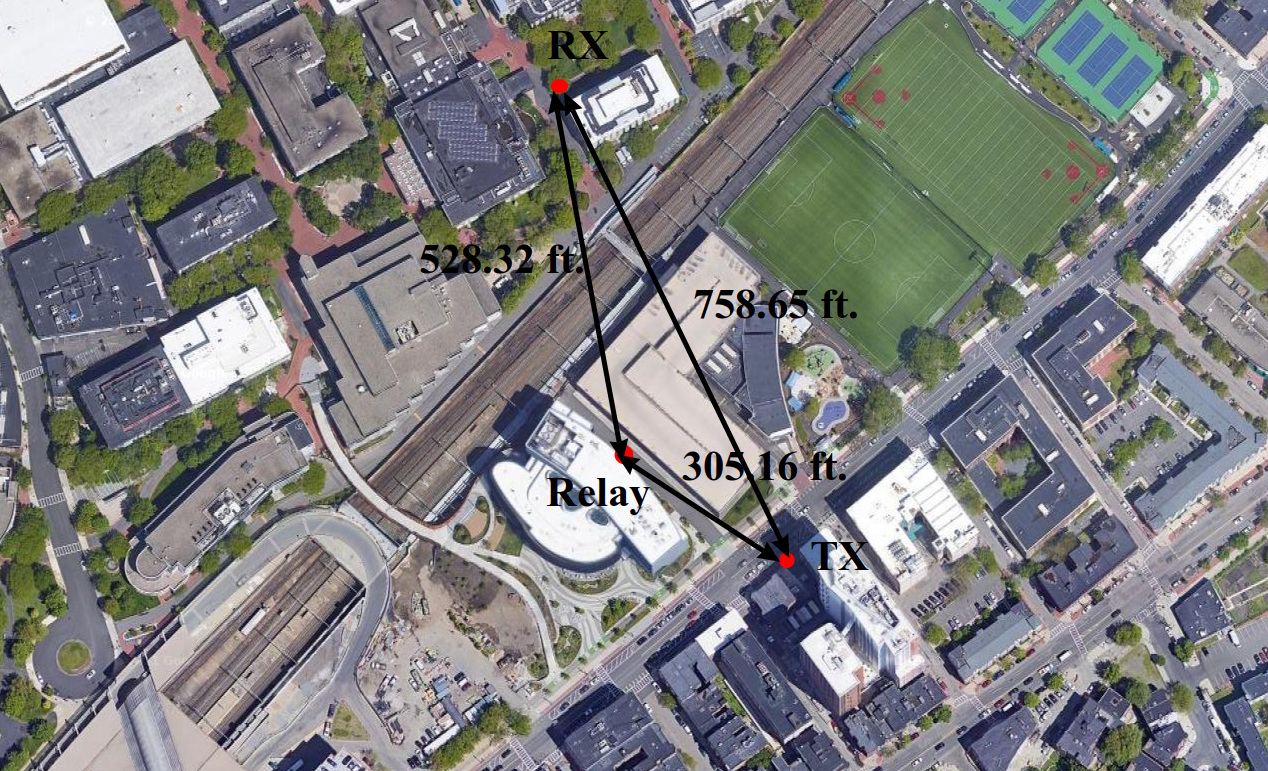}
   }
    \subcaptionbox{\label{fig:txrx_location}}
    {
        \includegraphics[height=3cm, width=0.21\linewidth]{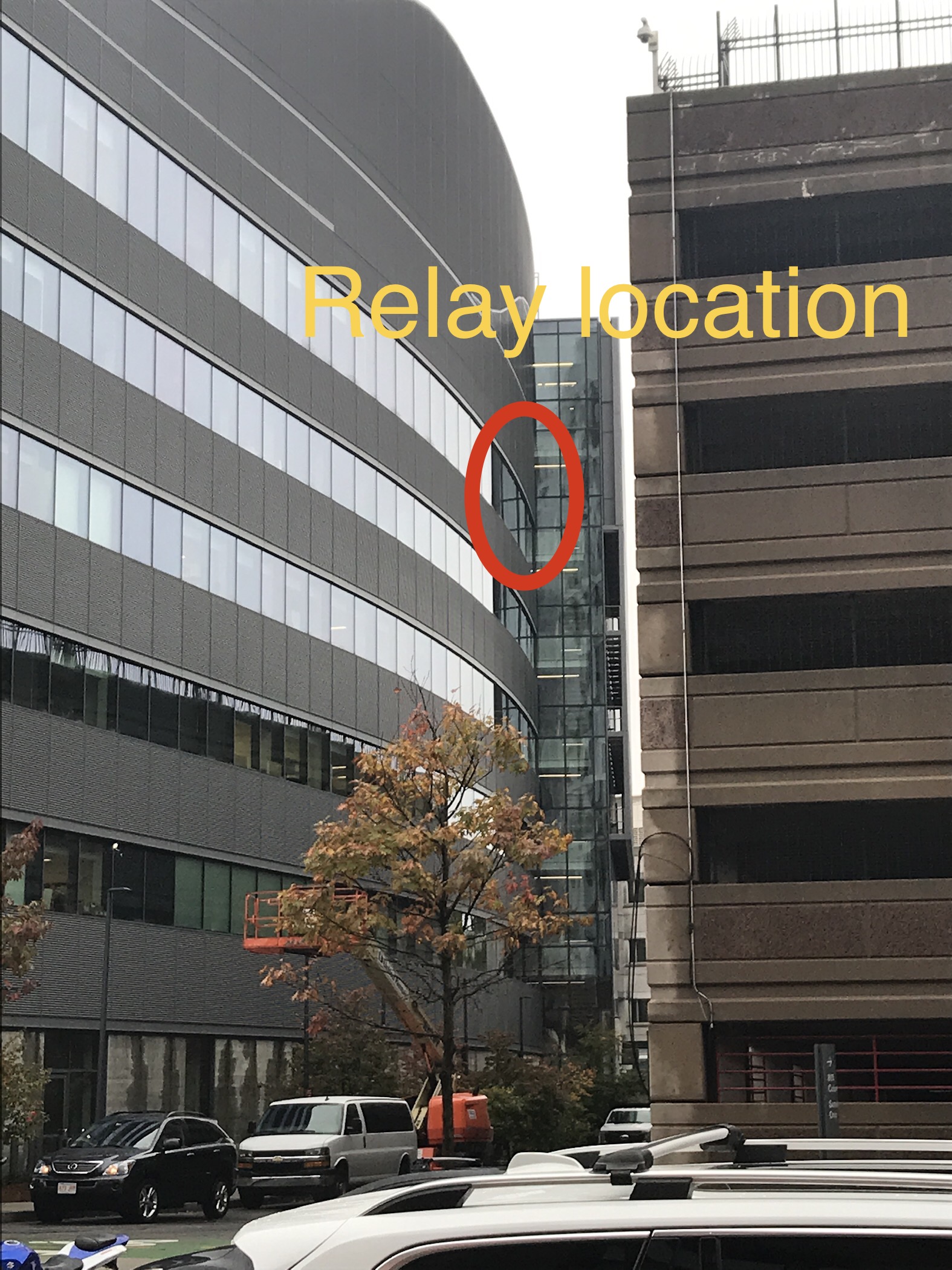}
        \hspace{0.5pt}
        \includegraphics[height=3cm, width=0.21\linewidth]{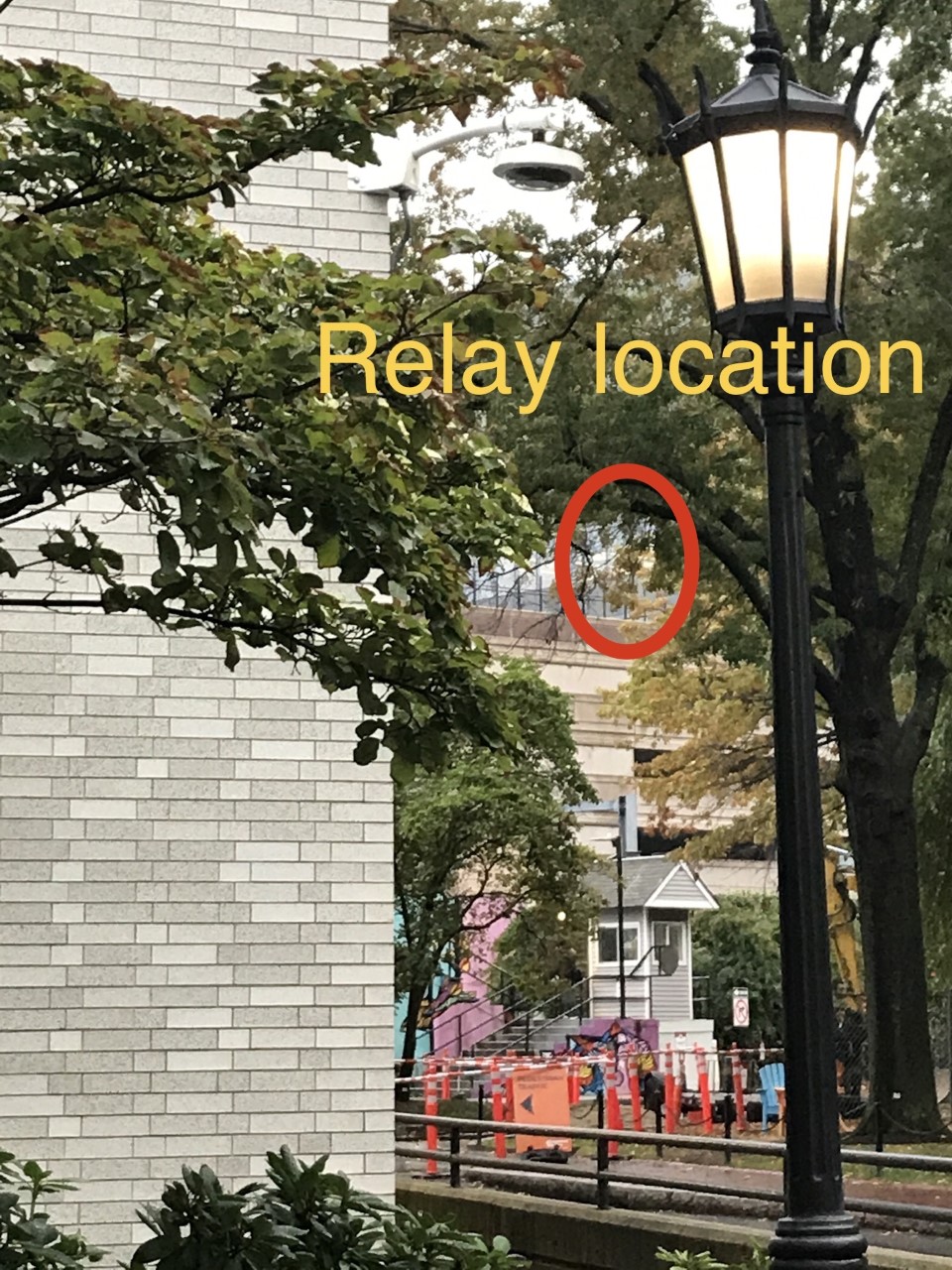}
    }
    \caption{Testbed environment for beamforming-relay experiments. TX and RX are mobile within a small range from the marked spots on the map. (a) Satellite view map. (b) Viewpoints of relay node's location from TX (left) and RX (right).}
    \label{fig:lora_environment}
\end{figure}

We demonstrate the universality of \system{} with the beamforming-relay application for LoRa~\cite{lora_semtech} and ZigBee~\cite{zigbee_csa} - two popular IoT technologies. In recent years, LoRa and ZigBee have seen significant rise in popularity in enabling the IoT communications thanks to the long range communication capability, low power requirement, and low cost. However, wireless channels are highly dynamic, and the communications can experience significant fading if the direct path is not present. An intermediate relay node that has the Line of Sight (LOS) to the transmitter and receiver can help maintain the communications. Motivated by this, we conduct experiments where \system{} is deployed as a universal beamforming-relay system when the direct LoRa/ZigBee communications are not possible. The relay system leverages \system{} with a modification where the combining signal \textit{is relayed to the TX chain instead of being sent to the decoder} (as in the original workflow in \Cref{fig:DEFORM_workflow}). In our setup, the location of relay node is fixed, while the TX/RX nodes are mobile within a small range around the locations that create a LOS with the relay node, as marked in~\Cref{fig:relay_map} and pictured in~\Cref{fig:txrx_location}.

First, we report on \system{} performance in relaying LoRa signal. We note that the original CNN model (trained for differential BPSK modulation) optimized by Temporal Smoothing (\Cref{sec:model_optim}) is maintained for all the experiments. \system{} beamforming system, is agnostic to RF signal characteristics, and its capability to improve the quality of the relay signal from LoRa, a chirp spread spectrum (CSS) modulation, is another strong indicator of its universality. We use Ettus USRP X310 for the relay node to achieve a better TX power, and Heltec ESP32 Development Kit for LoRa devices. It is noted that TX gain is always set to maximum and RX gain is fixed at the relay node. Also, to illustrate the system capability, LoRa's TX power is configured to the lowest level. We conduct 4 experiments, each consists of 4 measurements of 1 minute: The first measurement determines the Packet Loss Rate of direct TX-RX communications, in which we got 100\% for all experiments, confirming that the direct communications are not possible. The next two measurements determine the PLR of Amplify-and-Forward relay from a single antenna, and the last one provides the PLR for beamforming-relay using \system{}. The result of the measurements shown in~\Cref{tab:lora_plr} indicates that the beamforming-relay approach achieves less than $10\%$ Packet Loss Rate (PLR), significantly better compared to typical single-antenna Amplify-and-Forward relay of LoRa communications 
(\textit{PLR is up to 12 and 23 times lower than the stronger and weaker antennas}, respectively).

We repeat the same experiments for relaying ZigBee communications. We use two XBee-PRO 900HP equipped with the XBee Grove Development Boards for the TX and RX ZigBee nodes. It is noted that because ZigBee supports shorter communication range than LoRa, we set the TX power of ZigBee node to maximum. Due to such limited capabilities of ZigBee for long-range communications, we focus on the Packet Delivery Rate (PDR) of the direct and relay communications. We acquire measurements for 6 sets of 1-minute experiments in which the direct ZigBee communications are not feasible (with $0\%$ PDR). It is clear to see in the results (\Cref{tab:zigbee_pdr}) that the universal beamforming-relay still outperforms the single-antenna Amplify-and-Forward: With \system{}, We are able to achieve \textit{a successful packet reception as high as $193\%$ of the stronger antenna relay} (In experiment \#3, \system{} achieves $42.57\%$ successful reception while the stronger antenna only achieves $21.92\%$), \textit{and up to $858\%$ of the weaker antenna relay }(Experiment \#2).

\section{Discussion and Related work}
\ignore{While the Deep Learning model developed in \system{} is fast, it is still challenging to satisfy the real-time requirements for massive wideband RF data, as one RF sample captured over 1 MHz bandwidth is only \SI{1}{\unit{\micro\second}}  long. As shown in~\Cref{tab:benchmark}, the model prediction is the bottleneck that affects the overall runtime of the pipeline. However, in cases when channels are slowly varying (i.e., stable within a window of hundreds of milliseconds), it is acceptable to reuse the same prediction for a number of chunks of RF samples. For example, using a prediction for 5 chunks of 128 RF samples collected over \SI{1}{\unit{\mega\hertz}} bandwidth (which has a total duration of \SI{0.64}{\unit{\milli\second}}) allows a realtime operation of the system.}

It is clearly seen that the Deep Learning-based approach of \system{} can be extended to larger multi-antenna systems to achieve even higher than 3 dB gain. For a RF receiver system of $N$ receiving antenna elements, the deep learning architecture can be modified to have $2\times(N-1)$ outputs in which two estimations are made for each relative phase between the pre-selected antenna and one other antenna. The new requirement of such systems are time synchronization mechanisms for RX radios as typical wireless peripherals have a limited number of antennas (Most of Ettus's USRPs only support 2 simultaneous RX channels \cite{ettus}).

Our current system assumes that there is only one user in the observing spectrum at any given time. The problem becomes more challenging when multiple users are simultaneously transmitting on the same frequency band, which results in significant interference that damages the phase structure in the captured RF samples. In this case, the prediction could be accurate for only one user, or even not working for any users. It would be interesting to investigate the CNN's capability for multi-band spectrum and collision analysis in the future work.

\begin{table*}
\small
  \begin{tabular}{|c|c|c|c|c|c|c|c|c|c|c|c|c|c|c|c|}
    \hline
    \multirow{2}{5em}{} & \multicolumn{15}{c|}{\textbf{Measured BER (log scale) for different TX positions}}\\
    \cline{2-16}
    & \textbf{1} & \textbf{2} & \textbf{3} & \textbf{4} & \textbf{5} & \textbf{6} & \textbf{7} & \textbf{8} & \textbf{9} & \textbf{10} & \textbf{11} & \textbf{12} & \textbf{13} & \textbf{14} & \textbf{15}\\
    \hline
     \textbf{Branch 1} & -0.43 & -1.74 & -0.76 & -0.81 & -1.79 & -1.34 & -1.11 & -0.75 & NaN & -2.15 & 
-3.63 & -0.9 & -1.15 & -1.76 & -3.12\\
    \textbf{Branch 2} & -0.83 & -2.17 & -1.26 & -1.77 & -1.28 & -1.07 & -4.59 &     $\boldsymbol{\times}$ & NaN & -5.3 & -2.3 & -6 & -0.5 & -1.95 & -5.4\\
    \hline
    \textbf{\system{}} & -1.09 & -2.7 & -1.72 & -2.79 & -2.39 & -1.84 & -5.22 & -1.43 & NaN & -5.7 & -3.89 & NaN & -2.61 & -2.12 & -5.7\\
    
    \hline
  \end{tabular}
  \caption{BER analysis of non-LOS over-the-air experiment. The indexes corresponds to the numbered marks in ~\Cref{fig:non_los_environment}. Cross mark indicates no packets are detected resulted by insufficient SNR. NaN indicates a BER of zero.}
  \label{tab:non_los_evaluation}
\end{table*}

\begin{table}
\centering
\small
  \begin{tabularx}{0.93\linewidth}{|c|c|c|c|c|}
    \hline
    \multirow{2}{2em}{} & \multicolumn{4}{c|}{\textbf{Packet Loss Rate (\%)}}\\
    \cline{2-5}
    Index & Direct & Antenna 1-RL & Antenna 2-RL & \system{}-RL\\
    \hline
    1 & 100 & 22.33 & 11.65 & \textbf{0.97}\\
    2 & 100 & 4.85 & 20.45 & \textbf{2.91}\\
    3 & 100 & 20.39 & 12.91 & \textbf{8.74}\\
    4 & 100 & 16.67 & 7.84 & \textbf{0.98}\\
    \hline
\end{tabularx}
  \caption{Comparison of Packet Loss Rate for LoRa direct and relay (RL) communications.}
  \label{tab:lora_plr}
\end{table}

\begin{table}
\centering
\small
  \begin{tabularx}{0.93\linewidth}{|c|c|c|c|c|}
    \hline
    \multirow{2}{2em}{} & \multicolumn{4}{c|}{\textbf{Packet Delivery Rate (\%)}}\\
    \cline{2-5}
    Index & Direct & Antenna 1-RL & Antenna 2-RL & \system{}-RL\\
    \hline
    1 & 0 & 57.53 & 54.11 & \textbf{77.4}\\
    2 & 0 & 63.01 & 8.22 & \textbf{70.55}\\
    3 & 0 & 21.92 & 15.75 & \textbf{42.47}\\
    4 & 0 & 73.97 & 21.23 & \textbf{76.03}\\
    5 & 0 & 26.03 & 0 & \textbf{39.73}\\ 
    6 & 0 & 47.94 & 28.08 & \textbf{75.34}\\    
    
    \hline
\end{tabularx}
  \caption{Comparison of Packet Delivery Rate for ZigBee direct and relay (RL) communications.}
  \label{tab:zigbee_pdr}
\end{table}


Theoretical aspects of beamforming have been investigated in the literature \cite{cardoso_93, zhang_2017, vijay_2010, masouros_2015, wu_2019}, including some efforts to build the so-called \textit{blind beamformers} \cite{cardoso_93, himawan11, zhang_2017},  that estimates the channel phase offsets without explicit knowledge from the transmitter. The practicality of those methods and systematic deployment guidelines, however, are still open questions. \ignore{For instance, a blind beamformer for acoustic signals is built in \cite{himawan11} by analyzing the difference of measured Time Difference of Arrival (TDoA) on a subset of microphone arrays that is structurally positioned. While this approach may work in the environment with inherent Line of Sight and no multi-path, it is not effective where no LOS is present and the effects of multi-path is significant, which can create significant errors in the measurements.}Popular wireless communications technologies still utilize \textit{informed beamformer} approaches that estimate the channel using information from the transmitter such as pilot sequence \cite{wlan_pilot} in IEEE 802.11 or reference signal \cite{5g_dmrs} in 5G radios. As discussed, this approach typically requires significant communication overhead, and is limited to specific TX-RX setup and communication techniques.

The capabilities of phased-array systems \cite{van2004optimum} have been investigated for decades. Popular approaches using phase-array to efficiently extract the phase characteristics of signal in the presence of noise are MUSIC and ESPRIT \cite{van2004optimum}. Nonetheless, it is widely understood that these approaches are ineffective against multi-path effects, where coherent signals from multiple transmitting path distort the phase patterns at the receiver \cite{rahamim2004source}. Recently, several works in localization areas such as ArrayTrack \cite{arraytrack}, SpotFi \cite{kotaru2015spotfi}, or Ubicarse \cite{ubicarse} have investigated this problem, and achieved some improvements. However, those systems have certain limitations considering the goal of \textit{universal} beamforming. Firstly, they typically requires a lot of antennas \cite{arraytrack}, or other types sensing hardware such as gyroscope or accelerometer \cite{ubicarse}. Secondly, they rely on specific assumptions of radio mobility to easily distinguish between the direct path and reflection paths, and therefore are not universal. Finally, they are designed specially for localization, i.e. finding the Line-of-Sight direction of the transmitter, thus cannot find the optimal beamforming where the reflection paths are constructive and can improve the SNR of the direct path.  

Advances in Machine Learning and Deep Learning have emerged as a solution for critical problems in various areas, including wireless communications. In recent years, ML and DL approaches have been extensively utilized in various tasks such as modulation recognition \cite{oshea16}, RF technology identification~\cite{nguyen2021wideband}, or wireless localization \cite{wang2016csi}. There are also some efforts to enable ML-driven beamforming. For example, in \cite{ye_2017}, a Deep Neural Network (DNN) is developed for OFDM channel estimation. In \cite{huang2018unsupervised}, DNN is also utilized for downlink MIMO beamforming. Nonetheless, those works are limited by simulated data, and specific assumptions of channel model. Furthermore, they lack the explanation and evaluation for the impact of various RF characteristics, i.e. link modulations, bandwidths, and wireless channels. Compared to those, our work is unique in multiple aspects. First, our CNN-based DL model is trained with real RF data acquired by an efficient dataset collection process. Second, our system is agnostic to different RF settings of modulations, bandwidths, and channels. Despite being trained on a fixed, basic RF settings, we can still achieve the \textit{optimal} beamforming gain in complex, unseen settings. Third, \system{} is vital for the universal beamforming-relay application that achieves Packet Loss Rate of 23 times lower for LoRa and Packet Delivery Rate of 8 times higher for ZigBee compared to the conventional Amplify-and-Relay method. Hence, \system{} can be used as a universal RX beamforming module for existing multi-antenna RF receivers.


\balance


\bibliographystyle{ACM-Reference-Format}
\bibliography{./bib/hai,./bib/gn-beamforming}


\begin{thebibliography}{38}


\ifx \showCODEN    \undefined \def \showCODEN     #1{\unskip}     \fi
\ifx \showDOI      \undefined \def \showDOI       #1{#1}\fi
\ifx \showISBNx    \undefined \def \showISBNx     #1{\unskip}     \fi
\ifx \showISBNxiii \undefined \def \showISBNxiii  #1{\unskip}     \fi
\ifx \showISSN     \undefined \def \showISSN      #1{\unskip}     \fi
\ifx \showLCCN     \undefined \def \showLCCN      #1{\unskip}     \fi
\ifx \shownote     \undefined \def \shownote      #1{#1}          \fi
\ifx \showarticletitle \undefined \def \showarticletitle #1{#1}   \fi
\ifx \showURL      \undefined \def \showURL       {\relax}        \fi
\providecommand\bibfield[2]{#2}
\providecommand\bibinfo[2]{#2}
\providecommand\natexlab[1]{#1}
\providecommand\showeprint[2][]{arXiv:#2}

\bibitem[\protect\citeauthoryear{??}{gnu}{2021}]%
        {gnuradio}
 \bibinfo{year}{2021}\natexlab{}.
\newblock \bibinfo{title}{GNURadio}.
\newblock
\newblock
\urldef\tempurl%
\url{https://www.gnuradio.org}
\showURL{%
\tempurl}


\bibitem[\protect\citeauthoryear{??}{zig}{2021}]%
        {zigbee_csa}
 \bibinfo{year}{2021}\natexlab{}.
\newblock \bibinfo{title}{Zigbee}.
\newblock
\newblock
\urldef\tempurl%
\url{https://zigbeealliance.org/solution/zigbee/}
\showURL{%
\tempurl}


\bibitem[\protect\citeauthoryear{Abdel-Hamid et~al\mbox{.}}{Abdel-Hamid
  et~al\mbox{.}}{2014}]%
        {cnn_speech_recog}
\bibfield{author}{\bibinfo{person}{O. Abdel-Hamid} {et~al\mbox{.}}}
  \bibinfo{year}{2014}\natexlab{}.
\newblock \showarticletitle{Convolutional Neural Networks for Speech
  Recognition}.
\newblock \bibinfo{journal}{\emph{IEEE/ACM TASLP}} (\bibinfo{year}{2014}).
\newblock


\bibitem[\protect\citeauthoryear{Bishop}{Bishop}{2006}]%
        {bishop_prml}
\bibfield{author}{\bibinfo{person}{C.~M. Bishop}.}
  \bibinfo{year}{2006}\natexlab{}.
\newblock \bibinfo{booktitle}{\emph{Pattern Recognition and Machine Learning}}.
\newblock


\bibitem[\protect\citeauthoryear{Cardoso et~al\mbox{.}}{Cardoso
  et~al\mbox{.}}{1993}]%
        {cardoso_93}
\bibfield{author}{\bibinfo{person}{J.F. Cardoso} {et~al\mbox{.}}}
  \bibinfo{year}{1993}\natexlab{}.
\newblock \showarticletitle{Blind beamforming for non-gaussian signals}.
\newblock \bibinfo{journal}{\emph{IEEE Proceedings F}} (\bibinfo{year}{1993}).
\newblock


\bibitem[\protect\citeauthoryear{{CISCO Meraki}}{{CISCO Meraki}}{2018}]%
        {SNR}
\bibfield{author}{\bibinfo{person}{{CISCO Meraki}}.}
  \bibinfo{year}{2018}\natexlab{}.
\newblock \bibinfo{title}{{SNR and Wireless Signal Strength }}.
\newblock
\newblock


\bibitem[\protect\citeauthoryear{Glorot et~al\mbox{.}}{Glorot
  et~al\mbox{.}}{2011}]%
        {relu_2011}
\bibfield{author}{\bibinfo{person}{X. Glorot} {et~al\mbox{.}}}
  \bibinfo{year}{2011}\natexlab{}.
\newblock \showarticletitle{Deep Sparse Rectifier Neural Networks}. In
  \bibinfo{booktitle}{\emph{AISTATS'11}}.
\newblock


\bibitem[\protect\citeauthoryear{Goldsmith}{Goldsmith}{2005}]%
        {goldsmith_2005}
\bibfield{author}{\bibinfo{person}{Andrea Goldsmith}.}
  \bibinfo{year}{2005}\natexlab{}.
\newblock \bibinfo{booktitle}{\emph{Wireless Communications}}.
\newblock \bibinfo{publisher}{Cambridge University Press}.
\newblock


\bibitem[\protect\citeauthoryear{Goodfellow et~al\mbox{.}}{Goodfellow
  et~al\mbox{.}}{2016}]%
        {goodfellow_dl}
\bibfield{author}{\bibinfo{person}{I. Goodfellow} {et~al\mbox{.}}}
  \bibinfo{year}{2016}\natexlab{}.
\newblock \bibinfo{booktitle}{\emph{Deep Learning}}.
\newblock


\bibitem[\protect\citeauthoryear{He et~al\mbox{.}}{He et~al\mbox{.}}{2016}]%
        {resnet16}
\bibfield{author}{\bibinfo{person}{K. He} {et~al\mbox{.}}}
  \bibinfo{year}{2016}\natexlab{}.
\newblock \showarticletitle{Deep Residual Learning Image Recognition}. In
  \bibinfo{booktitle}{\emph{CVPR'16}}.
\newblock


\bibitem[\protect\citeauthoryear{Himawan et~al\mbox{.}}{Himawan
  et~al\mbox{.}}{2011}]%
        {himawan11}
\bibfield{author}{\bibinfo{person}{I. Himawan} {et~al\mbox{.}}}
  \bibinfo{year}{2011}\natexlab{}.
\newblock \showarticletitle{Clustered Blind Beamforming From Ad-Hoc Microphone
  Arrays}.
\newblock \bibinfo{journal}{\emph{IEEE TASLP}} (\bibinfo{year}{2011}).
\newblock


\bibitem[\protect\citeauthoryear{Huang et~al\mbox{.}}{Huang
  et~al\mbox{.}}{2018}]%
        {huang2018unsupervised}
\bibfield{author}{\bibinfo{person}{H. Huang} {et~al\mbox{.}}}
  \bibinfo{year}{2018}\natexlab{}.
\newblock \showarticletitle{Unsupervised learning-based fast beamforming design
  for downlink MIMO}.
\newblock \bibinfo{journal}{\emph{IEEE Access}} (\bibinfo{year}{2018}).
\newblock


\bibitem[\protect\citeauthoryear{Ioffe et~al\mbox{.}}{Ioffe
  et~al\mbox{.}}{2015}]%
        {batchnorm}
\bibfield{author}{\bibinfo{person}{S. Ioffe} {et~al\mbox{.}}}
  \bibinfo{year}{2015}\natexlab{}.
\newblock \showarticletitle{Batch Normalization: Accelerating Deep Network
  Training by Reducing Internal Covariate Shift}. In
  \bibinfo{booktitle}{\emph{ICML'15}}.
\newblock


\bibitem[\protect\citeauthoryear{Kalchbrenner et~al\mbox{.}}{Kalchbrenner
  et~al\mbox{.}}{2014}]%
        {dcnn_language}
\bibfield{author}{\bibinfo{person}{N. Kalchbrenner} {et~al\mbox{.}}}
  \bibinfo{year}{2014}\natexlab{}.
\newblock \bibinfo{title}{A Convolutional Neural Network for Modelling
  Sentences}.
\newblock
\newblock
\showeprint[arxiv]{1404.2188}


\bibitem[\protect\citeauthoryear{Kingma et~al\mbox{.}}{Kingma
  et~al\mbox{.}}{2017}]%
        {kingma2017adam}
\bibfield{author}{\bibinfo{person}{D. Kingma} {et~al\mbox{.}}}
  \bibinfo{year}{2017}\natexlab{}.
\newblock \bibinfo{title}{Adam: A Method for Stochastic Optimization}.
\newblock
\newblock
\showeprint[arxiv]{1412.6980}


\bibitem[\protect\citeauthoryear{Kotaru et~al\mbox{.}}{Kotaru
  et~al\mbox{.}}{2015}]%
        {kotaru2015spotfi}
\bibfield{author}{\bibinfo{person}{M. Kotaru} {et~al\mbox{.}}}
  \bibinfo{year}{2015}\natexlab{}.
\newblock \showarticletitle{Spotfi: Decimeter level localization using wifi}.
  In \bibinfo{booktitle}{\emph{ACM SIGCOMM'15}}.
\newblock


\bibitem[\protect\citeauthoryear{Kumar et~al\mbox{.}}{Kumar
  et~al\mbox{.}}{2014}]%
        {ubicarse}
\bibfield{author}{\bibinfo{person}{S. Kumar} {et~al\mbox{.}}}
  \bibinfo{year}{2014}\natexlab{}.
\newblock \showarticletitle{Accurate indoor localization with zero start-up
  cost}. In \bibinfo{booktitle}{\emph{ACM Mobicom'14}}.
\newblock


\bibitem[\protect\citeauthoryear{Kwon et~al\mbox{.}}{Kwon
  et~al\mbox{.}}{2019}]%
        {KwonLC2019}
\bibfield{author}{\bibinfo{person}{H.J. Kwon} {et~al\mbox{.}}}
  \bibinfo{year}{2019}\natexlab{}.
\newblock \showarticletitle{Machine Learning-Based Beamforming in Two-User MISO
  Interference Channels}. In \bibinfo{booktitle}{\emph{ICAIIC'19}}.
\newblock


\bibitem[\protect\citeauthoryear{Lin et~al\mbox{.}}{Lin et~al\mbox{.}}{2019}]%
        {5g_dmrs}
\bibfield{author}{\bibinfo{person}{X. Lin} {et~al\mbox{.}}}
  \bibinfo{year}{2019}\natexlab{}.
\newblock \showarticletitle{5G New Radio: Unveiling the Essentials of the Next
  Generation Wireless Access Technology}.
\newblock \bibinfo{journal}{\emph{IEEE CSM}} (\bibinfo{year}{2019}).
\newblock


\bibitem[\protect\citeauthoryear{Liu et~al\mbox{.}}{Liu et~al\mbox{.}}{2010}]%
        {liu_beamforming_2010}
\bibfield{author}{\bibinfo{person}{W. Liu} {et~al\mbox{.}}}
  \bibinfo{year}{2010}\natexlab{}.
\newblock \bibinfo{booktitle}{\emph{Wideband Beamforming: Concepts and
  Techniques}}.
\newblock


\bibitem[\protect\citeauthoryear{Masouros et~al\mbox{.}}{Masouros
  et~al\mbox{.}}{2015}]%
        {masouros_2015}
\bibfield{author}{\bibinfo{person}{C. Masouros} {et~al\mbox{.}}}
  \bibinfo{year}{2015}\natexlab{}.
\newblock \showarticletitle{Exploiting Known Interference as Green Signal Power
  for Downlink Beamforming Optimization}.
\newblock \bibinfo{journal}{\emph{IEEE TSP}} (\bibinfo{year}{2015}).
\newblock


\bibitem[\protect\citeauthoryear{Nguyen et~al\mbox{.}}{Nguyen
  et~al\mbox{.}}{2021}]%
        {nguyen2021wideband}
\bibfield{author}{\bibinfo{person}{H.~N. Nguyen} {et~al\mbox{.}}}
  \bibinfo{year}{2021}\natexlab{}.
\newblock \showarticletitle{Wideband, Real-time Spectro-Temporal RF
  Identification}. In \bibinfo{booktitle}{\emph{ACM MobiWac'21}}.
\newblock


\bibitem[\protect\citeauthoryear{O'Shea}{O'Shea}{2016}]%
        {oshea16}
\bibfield{author}{\bibinfo{person}{T. O'Shea}.}
  \bibinfo{year}{2016}\natexlab{}.
\newblock \showarticletitle{Convolutional Radio Modulation Recognition
  Networks}. In \bibinfo{booktitle}{\emph{Engineering Applications of Neural
  Networks}}.
\newblock


\bibitem[\protect\citeauthoryear{Paszke et~al\mbox{.}}{Paszke
  et~al\mbox{.}}{2019}]%
        {pytorch}
\bibfield{author}{\bibinfo{person}{A. Paszke} {et~al\mbox{.}}}
  \bibinfo{year}{2019}\natexlab{}.
\newblock \showarticletitle{PyTorch: An Imperative Style, High-Performance Deep
  Learning Library}.
\newblock In \bibinfo{booktitle}{\emph{NIPS'19}}.
\newblock


\bibitem[\protect\citeauthoryear{Rahamim et~al\mbox{.}}{Rahamim
  et~al\mbox{.}}{2004}]%
        {rahamim2004source}
\bibfield{author}{\bibinfo{person}{D. Rahamim} {et~al\mbox{.}}}
  \bibinfo{year}{2004}\natexlab{}.
\newblock \showarticletitle{Source localization using vector sensor array in a
  multipath environment}.
\newblock \bibinfo{journal}{\emph{IEEE TSP}} (\bibinfo{year}{2004}).
\newblock


\bibitem[\protect\citeauthoryear{Research}{Research}{2021}]%
        {ettus}
\bibfield{author}{\bibinfo{person}{Ettus Research}.}
  \bibinfo{year}{2021}\natexlab{}.
\newblock \bibinfo{title}{Products}.
\newblock
\newblock
\urldef\tempurl%
\url{https://www.ettus.com/products/}
\showURL{%
\tempurl}


\bibitem[\protect\citeauthoryear{Semtech}{Semtech}{2021}]%
        {lora_semtech}
\bibfield{author}{\bibinfo{person}{Semtech}.} \bibinfo{year}{2021}\natexlab{}.
\newblock \bibinfo{title}{What are LoRa and LoRaWAN?}
\newblock
\newblock
\urldef\tempurl%
\url{https://lora-developers.semtech.com/documentation/tech-papers-and-guides/lora-and-lorawan/}
\showURL{%
\tempurl}


\bibitem[\protect\citeauthoryear{Simonyan and Zisserman}{Simonyan and
  Zisserman}{2014}]%
        {VGG-2014}
\bibfield{author}{\bibinfo{person}{K. Simonyan} {and} \bibinfo{person}{A.
  Zisserman}.} \bibinfo{year}{2014}\natexlab{}.
\newblock \showarticletitle{Very deep convolutional networks for large-scale
  image recognition}. In \bibinfo{booktitle}{\emph{arXiv:1409.1556}}.
\newblock


\bibitem[\protect\citeauthoryear{Stuber et~al\mbox{.}}{Stuber
  et~al\mbox{.}}{2004}]%
        {stuber2004broadband}
\bibfield{author}{\bibinfo{person}{G. Stuber} {et~al\mbox{.}}}
  \bibinfo{year}{2004}\natexlab{}.
\newblock \showarticletitle{Broadband MIMO-OFDM wireless communications}.
\newblock \bibinfo{journal}{\emph{Proc. IEEE}} (\bibinfo{year}{2004}).
\newblock


\bibitem[\protect\citeauthoryear{Van~Trees}{Van~Trees}{2004}]%
        {van2004optimum}
\bibfield{author}{\bibinfo{person}{H.L. Van~Trees}.}
  \bibinfo{year}{2004}\natexlab{}.
\newblock \bibinfo{booktitle}{\emph{Optimum Array Processing: Part IV of
  Detection, Estimation, and Modulation Theory}}.
\newblock \bibinfo{publisher}{Wiley}.
\newblock
\showISBNx{9780471463832}


\bibitem[\protect\citeauthoryear{Venkateswaran et~al\mbox{.}}{Venkateswaran
  et~al\mbox{.}}{2010}]%
        {vijay_2010}
\bibfield{author}{\bibinfo{person}{V. Venkateswaran} {et~al\mbox{.}}}
  \bibinfo{year}{2010}\natexlab{}.
\newblock \showarticletitle{Analog Beamforming in MIMO Communications With
  Phase Shift Networks and Online Channel Estimation}.
\newblock \bibinfo{journal}{\emph{IEEE TSP}} (\bibinfo{year}{2010}).
\newblock


\bibitem[\protect\citeauthoryear{Wang et~al\mbox{.}}{Wang
  et~al\mbox{.}}{2016}]%
        {wang2016csi}
\bibfield{author}{\bibinfo{person}{X. Wang} {et~al\mbox{.}}}
  \bibinfo{year}{2016}\natexlab{}.
\newblock \showarticletitle{CSI-based fingerprinting for indoor localization: A
  deep learning approach}.
\newblock \bibinfo{journal}{\emph{IEEE TVT}} (\bibinfo{year}{2016}).
\newblock


\bibitem[\protect\citeauthoryear{Wu et~al\mbox{.}}{Wu et~al\mbox{.}}{2019}]%
        {wu_2019}
\bibfield{author}{\bibinfo{person}{Q. Wu} {et~al\mbox{.}}}
  \bibinfo{year}{2019}\natexlab{}.
\newblock \showarticletitle{Intelligent Reflecting Surface Enhanced Wireless
  Network via Joint Active and Passive Beamforming}.
\newblock \bibinfo{journal}{\emph{IEEE TWC}} (\bibinfo{year}{2019}).
\newblock


\bibitem[\protect\citeauthoryear{Xiong et~al\mbox{.}}{Xiong
  et~al\mbox{.}}{2013}]%
        {arraytrack}
\bibfield{author}{\bibinfo{person}{J. Xiong} {et~al\mbox{.}}}
  \bibinfo{year}{2013}\natexlab{}.
\newblock \showarticletitle{ArrayTrack: A Fine-Grained Indoor Location System}.
  In \bibinfo{booktitle}{\emph{NSDI'13}}.
\newblock


\bibitem[\protect\citeauthoryear{Ye}{Ye}{2018}]%
        {ye_2017}
\bibfield{author}{\bibinfo{person}{H. Ye}.} \bibinfo{year}{2018}\natexlab{}.
\newblock \showarticletitle{Power of Deep Learning for Channel Estimation and
  Signal Detection in OFDM Systems}.
\newblock \bibinfo{journal}{\emph{IEEE WCL}} (\bibinfo{year}{2018}).
\newblock


\bibitem[\protect\citeauthoryear{Zaheer}{Zaheer}{2018}]%
        {yogi}
\bibfield{author}{\bibinfo{person}{M. Zaheer}.}
  \bibinfo{year}{2018}\natexlab{}.
\newblock \showarticletitle{Adaptive Methods for Nonconvex Optimization}. In
  \bibinfo{booktitle}{\emph{NIPS'18}}.
\newblock


\bibitem[\protect\citeauthoryear{Zhang et~al\mbox{.}}{Zhang
  et~al\mbox{.}}{2017}]%
        {zhang_2017}
\bibfield{author}{\bibinfo{person}{L. Zhang} {et~al\mbox{.}}}
  \bibinfo{year}{2017}\natexlab{}.
\newblock \showarticletitle{An Eigendecomposition-Based Approach to Blind
  Beamforming in a Multipath Environment}.
\newblock \bibinfo{journal}{\emph{IEEE CL}} (\bibinfo{year}{2017}).
\newblock


\bibitem[\protect\citeauthoryear{Zhao et~al\mbox{.}}{Zhao
  et~al\mbox{.}}{2013}]%
        {wlan_pilot}
\bibfield{author}{\bibinfo{person}{Z. Zhao} {et~al\mbox{.}}}
  \bibinfo{year}{2013}\natexlab{}.
\newblock \showarticletitle{Channel Estimation Schemes for IEEE 802.11p
  Standard}.
\newblock \bibinfo{journal}{\emph{IEEE ITSM}} (\bibinfo{year}{2013}).
\newblock


\end{thebibliography}

\end{document}